\documentstyle[12pt,puthesis,chapterbib,psfig,wrapfig]{report}
\newcommand{\half}{\mbox{$\frac{1}{2}$}}
\newcommand{\svec}{{\bf S}}
\newcommand{\Del}{\hat{\Delta}}
\newcommand{\D}{\hat{D}}
\newcommand{\J}{{\bf J}}
\newcommand{\sig}{\mbox{\boldmath $\sigma$}}
\newcommand{\T}{{\bf T}}
\newcommand{\Tr}{{\rm Tr}}
\newcommand{\unity}{\mbox{1\hspace{-0.28em}l}}
\newfont{\double}{msbm10 scaled\magstep1}
\newcommand{\CC}{\mbox{\double C}}

\newcommand{\del}{\partial}
\newcommand{\bsigma}{\mbox{\boldmath$\sigma$}}
\newcommand{\hd}{\hat{\Delta}}
\newcommand{\hg}{\hat{\gamma}}
\newcommand{\htrans}{\hat{T}}
\newtheorem{rule1}{EMPERICAL RULE}
\newcommand{\GS}{{\rm GS}}
\newcommand{\epbar}{\overline{\epsilon}}
\newcommand{\oc}{{\cal O}}

\begin{document}

\title{\huge\bf Integrability and Applications of the 
Exactly-Solvable Haldane-Shastry One-Dimensional Quantum Spin Chain} 
\author{Johan Cornelis Talstra}
\dept{Physics}
\submitdate{November 1995}

\copyrighttrue
\copyrightyear{1995}
\signaturefalse
\figurespagetrue
\tablespagetrue

\beforepreface
\prefacesection{Abstract}
Recently, the one dimensional model of $N$ spins with $S=\frac{1}{2}$ on a
circle, interacting with an exchange that falls off with the inverse square of
the separation: \\ 
$H_{\rm ISE} =\sum_{i\neq j} \frac{1}{[\frac{N}{\pi}
\sin(\frac{i-j}{N}\pi)]^2}(\svec_i\cdot
\svec_j-\frac{1}{4})$, or ISE-model, has received ample attention.  This model
was introduced by Haldane and Shastry in 1988.  Its special features include:
relatively simple eigenfunctions, 
non-interacting elementary excitations that obey semionic 
statistics (spinons), and a
large ``quantum group'' symmetry algebra called the Yangian.
This model is fully
integrable, albeit in a slightly different sense than the more traditional
nearest neighbor exchange (NNE) Heisenberg chain. The simplicity of the
ISE-model warrants its paradigmatic
r\^{o}le as the semionic analog of the free boson
or fermion gas.

This thesis comes in two parts. Part I (chapters 
1
and
2)
 deals with the integrability  of
the model, and Part II (chapters 
3
and 
4)
discusses some of its applications.
Chapter 
1
introduces the model and presents the construction of a
subset of the eigenfunctions. The other eigenfunctions are shown to be
generated by the action of the  Yangian symmetry algebra of $H_{\rm ISE}$. 
The Yangian is derived from a transfer matrix, 
thus establishing a relation to more common exactly
solvable models. Chapter 
2
presents a method
to construct the
set of constants of the motion of the ISE-model. Normally they
derive from the trace of transfer matrix, but here they are obtained by
considering the deformation of the spin model into a dynamical version.

The ISE-model is tractable enough to obtain its zero-magnetic-field dynamical
structure factors.  Chapter 
3
attempts to extend this to a non-zero
magnetic field, where, due to the presence of spinons in the groundstate, more
complicated excitations contribute:  in low fields just up to two extra
spinons, for high fields two oppositely moving magnons.  We discuss the
relation to the more complicated NNE-model.  Finally chapter 
4
illustrates how a recently conjectured new form of Off Diagonal Long Range
Order in antiferromagnetic spin chains can be reinterpreted as a spinon
propagator in the ISE-model, and verified numerically.  We briefly comment on
its relevance to stabilizing superconductivity in the layered cuprates.

\prefacesection{Acknowledgements}

It is a great pleasure to thank my advisor, Duncan Haldane, for 
sharing his enthusiasm, his many ideas and his insights with me,  
and for making available the means to follow up on these.
I want to thank Steve Strong for the great time we had 
collaborating,
 of which Chapter \ref{chapter4} of this thesis is hopefully a testimony.

I also benefited greatly from the interaction with the other graduate
students  and faculty at Princeton. I acknowledge very useful comments from
a non-condensed matter perspective from Prof. Curt Callen. 
I want to thank Zach Ha and Chetan Nayak for discussions, 
and I  mention 
Eddy van de Wetering, since it is partially his example that brought me to
Princeton.
I especially thank Per Kraus, for making our office a place to learn
and to relax, scientifically and otherwise.

Finally, for many months of hospitality throughout the last 5
years as well as constant parental encouragement and support, 
I am deeply indebted to my family.

While at Princeton, I was supported by teaching assistantships from the
University, an IBM Graduate Fellowship, and lastly by NSF grant \# DMR
9224077.

\pagebreak
\begin{center}
{\large\em To my parents}
\end{center}
\prefacesection{Preface}
The work in this thesis is all based on previously published work. Chapter
\ref{chapter1} is an extension of work started in Phys.\ Rev.\ Lett.\ {\bf 69},
2021 (1992), by F.D.M. Haldane, Zach Ha, J.C. Talstra, Denis Bernard and Vincent
Pasquier. Chapter \ref{invar} was done in collaboration with Duncan Haldane
and is reported in J.\ Phys.\ A. Math.\ Gen.\ {\bf 28}, 2369 (1995). Chapter
\ref{dsf} on structure functions, also in collaboration with Duncan Haldane,
appeared in Phys.\ Rev.\ B {\bf 50}, 6889 (1994). Chapter \ref{chapter4} grew
out of work with Steve Strong and Phil Anderson. It has appeared
in Phys.\ Rev.\ Lett.\ {\bf 74}, 5256 (1995).

\pagebreak

\afterpreface
\part*{}
\renewcommand{\thechapter}{}
\def\chapvar{}
\chapter{Introduction}
\label{plan}
\markright{Introduction}

In this thesis, we will confine ourselves almost exclusively to one space
dimension, and mostly one specific model that hasn't been realized
experimentally yet. The experimental side of other models that we will
discuss is going to receive little attention either.  Given these serious
restrictions, one may wonder to what extent this work should belong to the
field of Condensed Matter Physics. We will try to address this concern in
three steps. 

Why one dimensional physics? The fabulous thing about one dimensional
Quantum Mechanics (or equivalently two dimensional statistical mechanics) is
that there exist interacting many-particle models that are {\em exactly}
solvable. Unlike in higher dimensions,  particles cannot evade
each other and have to scatter. For special choices of the interaction
potential, particles can be made to propagate in plane waves {\em between}
scattering events. During a collision they exchange their 
plane wave momenta. The coefficients of both sets of plane waves (before and
after) 
are related to each other by the phase shift, picked up by the two 
particles that
scattered. This heuristic attempt at formulating a wavefunction is called
the {\em Bethe Ansatz} \cite{Suth85,B31}. The Bose gas with delta-function
interactions is an example where the Ansatz works.

We use the word solvability in the sense that the corresponding Schr\"{o}dinger
differential/difference equation can be reduced to a system of algebraic
equations that can be solved, at least numerically \footnote{ E.g.\ for the
Bethe Ansatz, the plane wave momenta are related to each other and to the phase
shift via a set of nonlinear equations that can be solved numerically for
finite $N$ and analytically in the thermodynamic limit.}.  
For most models the latter is indeed all one can do analytically.  This brings
us to the second point:  why we discuss  one specific model, 
viz.\ the model with $N$
spin-$\half$ particles  distributed equidistantly on a circle, coupled with
an exchange interaction that decays proportional to the inverse squared
distance between the spins:  
$$ H_{\rm ISE} = \sum_{i\neq j}^N
\frac{1}{\left[(\frac{N}{\pi})^2 \sin (\frac{i-j}{N} \pi)\right]^2}
\svec_i\cdot\svec_j.  
$$ 
It is also called the Inverse Square Exchange or ISE-model.  In this model
solvability means a great deal more:  not only are a lot of the physically
relevant eigenstates known
analytically (and the others can be generated via the action of ladder
operators), they are also of a
relatively simple form, so one can actually calculate things like structure
factors in closed form.  These are averages of local operators in a certain
state.  To date this is still impossible in the Bethe Ansatz style models.

The benefits don't stop here; the elementary excitations of this model,
so-called {\em spinons}, are spin-$\half$ particles that obey semion
statistics, that is to say, for every two spinons we create, there is one less
orbital available for the next spinon.  For fermions this would be 2 and for
bosons this would be 0.  This is an example of the {\em exclusion} statistics
interpretation of fractional statistics \cite{H91b}, as opposed to the usual
{\em
exchange} statistics  of the two dimensional fractional quantum hall effect
(in which case both formulations can be  shown to be consistent).
The long range spin model is not the only place where one finds fractional
excitations, but contrary to spin models such as the traditional Nearest
Neighbor Exchange (NNE) 
Heisenberg model, the ISE-model describes a {\em free} gas of these
spinons. 

In the continuum limit there exists another solvable 
interacting model, the Sutherland model
\cite{Suth71}, of $M$ bosonic  particles on a circle of circumference $N$, 
repelling each other via a
$\frac{1}{r^2}$ interaction, which has fractional excitations. 
$$
H_{\rm Suth} = -\sum_{i}^M \del^2_{x_i} +\lambda(\lambda-1)\sum_{i\neq j}^M
\frac{1}{\left[ (\frac{N}{\pi}\sin(\frac{x_i-x_j}{N}\pi)\right]^2}
$$
As a matter of
fact, one can select their statistics, by varying the coupling constant in the
potential energy.  For a particular value of that coupling $\lambda=2$, 
(corresponding to
exactly semion statistics) we can set up a mapping to the ISE spin model.  The
mapping identifies a number of low lying states in the continuum with a number
of maximal $S^z$ states in the spin model.  The correspondence is not trivial
in that all other (missing) spin states are obtained by the action of some
simple symmetry like $SU(2)$ or $SU(2)\times SU(2)$ (as in the NNE and 1D
Hubbard models respectively).  Recently it was found that the ``missing''
states are generated by acting on the ``continuum'' 
wavefunctions with the much larger symmetry algebra called the Yangian,
$Y(sl_2)$, a quantum deformation of the $SU(2)$ Lie group.
Unlike the case of the Quantum Affine Algebra extension of $SU(2)$:
$U_q(\widehat{sl}_2)$, the Yangian is not obtained by deforming the Lie group
commutation relations, but rather by adding new generators. 
  The Yangian algebra
is well studied in the mathematics literature and was first introduced into
physics in quantum field theory \cite{B91}.  The spin model is the first finite
system which serves as a representation of this algebra.  It can be studied in
a hands-on manner, for instance on a computer, and one can construct the
ladder operators explicitly, just as for SU(2) spins.

At this point we comment on the third issue: the relation to experiments.
The utility of the ISE-model and its Yangian doesn't lie so much in the fact
that it clarifies experiments, waiting for an explanation, but  rather, it
presents a  new paradigm, a language. We mean this in the same
sense that  free boson and fermion models are paradigms that do not occur in 
nature, but a class of more
realistic models only differ from them by the addition of irrelevant 
perturbations; these models reside in the same universality class, 
and have essentially the the same long distance physics as the free fixed
points. Similarly, the nearest neighbor Heisenberg
model is  in the same universality class as the ISE model, but its semions
are weakly interacting, as we will see in chapter \ref{dsf}.

The remainder of this thesis is organized as follows. The first part,
chapters \ref{chapter1} and \ref{invar}, will be concerned with the ``exactly
solvable'' qualifier of the ISE-model, illustrating the richness of the
underlying integrability structure. Chapters \ref{dsf} and \ref{chapter4}
are more applied and stress the paradigmatic aspects of the model. 
Chapter \ref{chapter1} will review the relevant facts about the solutions of
the ISE spin-model, and show how its Yangian symmetry emerges. In other
words, we will try to construct the {\em transfer matrix} of this
model. Chapter \ref{invar} is a bit more technical in nature. In general,
all exactly solvable models will possess a set of commuting operators of
which the Hamiltonian is one: the constants of the motion (in classical
physics, the conserved quantities). The usual derivation of these
quantities does not work for the  ISE model, for the precise reason that
the Yangian is an {\em exact} symmetry of the ISE Hamiltonian. By deforming the
model through allowing the particles to leave their lattice sites, via a
kinetic energy term, and taking the limit where the spin potential energy
dominates the kinetic energy we can recover these constants of the motion.

Chapter \ref{dsf} describes the {\em dynamical} correlation functions of the
ISE-model in a non-zero magnetic field, where some knowledge of the
underlying Yangian structure---the division of the Hilbert space
into Yangian multiplets---is actually required. It builds on recent work
based on various methods, such as supersymmetric matrix functional
integrals
\cite{als93,hz93} and Jack polynomial techniques \cite{Ha94} from mathematics, which
obtained these structure functions (or Greens functions in the corresponding
quantum field theory), exactly, at at zero external field. 
We investigate the case $h\neq 0$, in which case the $h$-field tunes
the number of fractional quasiparticles in the groundstate.

Finally, chapter \ref{chapter4} show how a new order parameter for
antiferromagnetic spin models \cite{RVB} can be analyzed in the free spinon
ISE-model. The order parameter involves inserting two sites into a spin
chain and putting two spins in a singlet configuration on them.
 Whereas only numerical and circumstantial evidence exists for
the Long Range Order (LRO) of this order parameter in the NNE-model, it can
be approached analytically in the ISE-model, once we interpret the
order parameter as a spinon propagator. Some analogies to the $\nu=\half$
bosonic FQHE are drawn, and we suggest some implications for stabilizing
superconductivity in layered cuprates via singlet pair hopping.

\setcounter{part}{0}
\renewcommand{\thepart}{\Roman{part}}
\part{Integrability Structure of the $\frac{1}{r^2}$-Spin model.}
\label{part1}
\setcounter{chapter}{0}
\renewcommand{\thechapter}{\arabic{chapter}}
\def\chapvar{Chapter}
\chapter{The Inverse Square Exchange Spin Model and its Yangian.}
\label{chapter1}

\section{Introduction}
\label{introduction1}

Solvable models play an interesting r\^{o}le in many-body physics and not
just because of the mathematical satisfaction of being able to obtain the
exact solution to a particular
problem. In a world of uncontrolled approximations, they can be the
touchstone and foundation of theories that describe ``real'' systems more
realistically than the mathematical toy models themselves. The 1-D Hubbard
model of spin-1/2 fermions on a circle illustrates this eminently
\cite{RVBbook}.

Although nowadays scores of models have been solved exactly, many of them
still await practical application. This is not surprising considering the
mathematical {\em tour de force} that is often required. A consequence
is that solving a model (i.e., knowing the eigenfunctions of that
particular Hamiltonian) does not mean that we can actually compute physical
properties if those wavefunctions are of a complicated nature. Such is the
case
for the Nearest Neighbor Exchange $S=1/2$ Heisenberg-chain, for example.
Its Hamiltonian is given by \cite{bethe31}:
\begin{equation}
H_{\rm NNE}=J\sum_{i=1}^{N} {\bf S}_i\cdot{\bf S}_{i+1},
\label{NNEdef}
\end{equation}
where ${\bf S}_{N+1}={\bf S}_1$. 

A system that doesn't suffer from this problem is the model of spin-$\half$
particles on a one-dimensional ring, interacting antiferromagnetically
through an Inverse Square Exchange interaction (ISE-model for short). It was
introduced in 1988 by Haldane and Shastry \cite{H88,S88}. The Hamiltonian is
as follows:
\begin{eqnarray}
H_{\rm ISE} &=& \frac{J}{4}\sum_{i\neq j=1}^N d(i-j)^{-2} (P_{ij}-1)\nonumber\\
&=& \frac{v_s}{\pi}\sum_{i\neq j=1}^N d(i-j)^{-2} ({\bf S}_i\cdot {\bf S}_j
-\frac{1}{4}),
\label{ISEdef}
\end{eqnarray}
\begin{wrapfigure}{r}{2.5in}
\framebox[2.5in]{\centerline{\psfig{file=circle.epsi,width=2.3in}}}
\end{wrapfigure}
where $P_{ij}$ permutes the spins on sites $i,j$ and 
$d(i-j)$ represents the chord distance between two points $i,j$ on a
circle that has been subdivided into equidistant sites: $d(n)=|\frac{N}{\pi}
\sin^2(\frac{n\pi}{N})|^2 $. Periodic boundary conditions are assumed.

Despite the long range of the force between the spins, this model is
solvable to a much larger extent than its short-ranged cousins. The key
feature is the non-transcendental structure of the eigenfunctions of
(\ref{ISEdef}), and the simplicity of the spectrum. It may seem that
the particular exchange integral in eq.\ (\ref{ISEdef}) makes $H_{\rm ISE}$
less suited as a paradigm for a family of physically applicable models (but
see \cite{Vacek93} for exceptions). However, the fact that the model's
solution is simple, in combinations with the  fractional statistics nature of
the elementary excitations in its spectrum, make it an appropriate candidate
for an ideal gas of {\em anyons}. 

We will start with a short introduction to the ISE-model and its integrability
structure; precise details can be found in \cite{H94,BL95}.  Section\
\ref{ISEdiag} will explain how $H_{\rm ISE}$ was solved originally. 
Section\ \ref{ISEyang} will analyze the symmetries of the Hamiltonian and
provide an explanation for the exact degeneracies of its spectrum.  The
methods on which section \ref{ISEyang} are based, are familiar from quantum
inverse scattering and the Yang-Baxter equation literature---thereby 
elucidating the
relation to more conventional solvable models.

\section{The ISE-Hamiltonian and its Spectrum}
\label{ISEdiag}

To solve the Schr\"{o}dinger equation for $H_{\rm ISE}$  let us first
transcribe it into a language of hard-core bosons on a lattice, 
where down-spins are
considered  particles and up spins represent the absence of these
particles. The hard-core constraint refers to projecting out the states
in the Hilbert space
with more than one particle per site.
With $M$ particles present, $S^z=\half N -M$. The usual
separation of the $\svec_i\cdot\svec_j$ term into a kinetic energy part
($S^+_i S^-_j + S^-_i S^+_j$) and a potential energy part $S^z_i S^z_j$, 
reduces  eq.\
(\ref{ISEdef}) to \cite{Fradkin91}:
\begin{equation}
\label{ISEbose}
\sum_{k=1}^N \epsilon_0(k)\hat{\rho}_k + \frac{v_s}{\pi}\sum_{i\neq j=1}^{N}  
d(i-j)^{-2}  n_i n_j,
\end{equation}
where the discrete Fourier transform of the exchange is $\epsilon_0(k)=
\frac{v_s}{2 \pi} k(k-2\pi)$, $k=\frac{2\pi}{N}m,\:  m=0\ldots N-1$.
$S^-_j=c^\dagger_j$, $n_j=c^\dagger_j c_j=\half-S^z_j$, and
$\hat{\rho}_k=\hat{c}^\dagger_k \hat{c}_k$
is the usual number operator for the extended orbital with momentum $k$. 
From now on we will measure energy in units of $(\frac{2\pi}{N})^2 
\frac{v_s}{\pi}$. 
Notice that
the kinetic energy is strictly quadratic (albeit periodically repeated in
successive Brillouin zones),
despite the fact that the particles hop on a lattice. This peculiar
connection with the kinetic energy in the continuum is a special property of
the $1/\sin^2$ potential that people have attempted to  employ 
before in the context of
fermion lattice gauge theory \cite{Drell76}. This phenomenon raises the
suspicion
that the relation between the lattice model and some continuum model of
particles on a ring may be exploited. Haldane and Shastry \cite{H88,S88,H91}
found indeed
that the set of eigenfunctions of the  continuum Hamiltonian of the
so-called Sutherland model \cite{suth70}:
\begin{equation}
\label{SUTHdef}
H=-\sum_i^M  (\del_i)^2 +
\sum_{i<j}^M\frac{2\lambda(\lambda-1)}{(\frac{N}{\pi})^2
\sin^2(\frac{x_i-x_j}{N}\pi)},
\end{equation}
of $M$ particles on a circle of circumference $N$ 
are also eigen\-functions of $H_{\rm ISE}$ via the par\-ticle/down-spin analogy.
Let us repeat this derivation since it involves a few transformations that we
will run into frequently. 

Sutherland found the (unnormalized) groundstate
for his {\em continuum} model to be of the simple Jastrow form:
\begin{equation}
\label{psi0}
\Psi_0(n_1,\ldots,n_M)=\prod_{i<j}\left|\sin
(\frac{x_i-x_j}{N}\pi)\right|^\lambda.
\end{equation}
This is the case of bosons; for fermions we adopt (\ref{psi0}) for
$x_1<x_2\ldots <x_M$ and multiply with the appropriate sign for other
orderings of the particles.

As a notational aside, in the spin model on the circle---with a lattice 
spacing that we have chosen to be 1---let us parametrize
the sites with the complex numbers $\left\{ z_j=e^{\frac{2\pi i}{N}j}
\right\}_{j=0}^{N-1}$.
The locations of the particles (down-spins) $\{ n_i \}_{i=1}^M$ in a given 
spin-configuration denoted by the ket
$|n_1,n_2,\ldots,n_M\rangle$, are given by
${z_{n_j},\: j=1,\ldots,M}$ and $0\leq n_j<N$. As an example, in this complex 
notation the
``interaction potential'' between down-spins becomes: 
$(\frac{2\pi}{N})^2 \cdot \frac{-z_{n_i}
z_{n_j}}{(z_{n_i}-z_{n_j})^2}$. 
Haldane and Shastry showed that (\ref{psi0}) is an eigenstate of spin
Hamiltonian 
(\ref{ISEdef}) as well if we restrict the $x_i$ to the lattice, i.e.\ to be
integers $\{ n_i\}$, and pick $\lambda=2$. In other words, the spin
groundstate is a member of the family of wavefunctions (in complex notation):
\begin{equation}
\label{psi0compl}
\psi_0(z_{n_1},\ldots,z_{n_M})=\prod_{i<j}^{M}(z_{n_i}-z_{n_j})^2
\prod_{i=1}^M z_{n_i}^{J} =
\prod_{i<j}\sin^2(\frac{n_i-n_j}{N}\pi)
\prod_ie^{\frac{2\pi i}{N}(J+2(M-1))n_i}.
\end{equation}
In the spin model, the absolute groundstate as  function of $M$ occurs 
when $J=1,\, M=N/2$, 
and 
$S^z=0$, (at least for $N$ even, which is what we will assume in the rest 
of this chapter). In the sector $M\neq \frac{N}{2}$ we have to choose 
$J=1+\frac{N}{2}-M$ to
find the state with lowest energy for that value of $S^z$. 
For what follows we choose $J=1$.
Here, and in the following we adhere to the convention that uppercase greek
letters refer to wavefunction of the $\{ n_i \}$ and lowercase greek letters
to functions of the $\{ z_{n_i} \}$. That is: $\Psi_0$ is related to 
$\psi_0$ as follows:
\begin{equation}
\Psi_0(n_1,\ldots,n_M)\equiv \psi_0(z_{n_1},\ldots,z_{n_M}).
\end{equation}

As we shall see in section \ref{suthsolve}, 
the excited states in the continuum are obtained by
multiplying the groundstate wavefunction by a symmetric set of 
plane-wavefunctions---$\Psi=\Psi_0(x_1\ldots x_M)\Phi(x_1\dots x_M)$, 
with momenta that satisfy simple Bethe-Ansatz type 
equations.  The rationale is that the strong correlations are described well
by the ``groundstate'' wavefunction and small fluctuations are added by the
multiplying factor.
In the spin case we can try the same: we write a spin wavefunction as the 
product of the wavefunction (\ref{psi0compl}) and some
symmetric polynomial $\phi$ in the ${z_{n_i}}$:~\footnote{$(z_n)^m$ is a plane
wave of momentum $\frac{2\pi}{N}m$.}
$\Psi(n_1,\ldots,n_M)=\Psi_0(n_1,\ldots,n_M) \Phi(n_1,\ldots,n_M)$. 
The crux the mapping of $H_{\rm ISE}$ onto the continuum Sutherland
Hamiltonian is the following: the spin Schr\"{o}dinger equation for
$\Psi_0\Phi$ which is really a difference equation for $\Phi$, can be
rewritten as a {\em differential} equation in $\Phi$, if the latter is of a
polynomial (i.e.\ plane wave) form. This differential equation is identical
to the continuum Schr\"{o}dinger equation that $\Phi(x_1\ldots
x_M)=\frac{\Psi^{\rm cont}(x_1\ldots x_M)}{\Psi^{\rm cont}_0(x_1\ldots x_M)}$ 
satisfies (the superscript ``cont'' indicates that the $x_i$ are
considered continuum variables). This equation has
been solved in \cite{suth70}. Let us first illustrate the 
difference-to-differential transformation. 

\subsection{From the Lattice to the Continuum}

In complex notation:
\begin{equation}
\label{ISEsplit}
H_{\rm ISE}=H_1+H_2=\sum_{i\neq j}\frac{-z_i z_j}{(z_i-z_j)^2} S^z_i S^z_j +
\sum_{i\neq j} \frac{-z_i z_j}{(z_i-z_j)^2} \half (S^+_iS^-_j+S^-_iS^+_j)
\end{equation}
We have dropped the additive constant part in $H_{\rm ISE}$ for now, and we
will put it back in at the end, by requiring that the fully polarized state
$|\uparrow\cdots\uparrow\rangle$ has zero energy.
Now $\langle n_1,\ldots,n_M|H_1|\Phi\Psi_0\rangle=
\Phi(n_1,\ldots,n_M)\langle n_1,\ldots,n_M|H_1|\Psi_0\rangle$ 
since $H_1$ is diagonal in
the particle number occupation basis. Furthermore
\begin{eqnarray}
\langle n_1,\ldots,n_M| &H_2& |\Phi\Psi_0\rangle =
\sum_{j=1}^M \sum_{i\neq n_j}^N -\frac{ z_i z_{n_j}}{(z_i-z_{n_j})^2}
\times\nonumber\\
&&\Psi_0(n_1,\ldots,n_{j-1},i,n_{j+1},\ldots)
\Phi(n_1,\ldots,n_{j-1},i,n_{j+1},\ldots).
\label{H2def}
\end{eqnarray}
We can write this as a ``continuum'' expression with the help of the fact
that the Fourier transform of $1/\sin^2$ is quadratic:
\begin{equation}
\label{SLACder}
\sum_{n(\neq m)}^N \frac{z_n z_m}{(z_n - z_m)^2} P(z_n)=
\left\{-\half (z_m\del_{z_m})^2 + \frac{N}{2} z_m\del_{z_m}
-\frac{1}{12}(N^2-1)\right\}P(z_m)
\end{equation}
if $P(z)$ is a polynomial of degree $<N$ in every variable. Since $\psi_0$
is of has powers of $z_{n_i}$ in the range $0,\ldots,2(M-1)+1$,
 this means that we can only apply eq.\
(\ref{SLACder}) if $\phi$ has a degree in the range
$0,\ldots,N-2M$ in all its entries.  This
somewhat arbitrary constraint from the point of view of the spin model
 will become clearer later on. Then,
with the Leibnitz rule for differentiation:
 the  $H_2$ matrix elements (\ref{H2def}) become, 
\begin{eqnarray}
\lefteqn{-\Phi(n_1,\ldots,n_M)
\left( \sum_{j=1}^M
     \left[-\frac{1}{12}(N^2-1)+ \frac{N}{2} (z_{n_j}\del_{n_j})-
             \half (z_{n_j}\del_{n_j})^2
     \right] \psi_0
\right)}
\nonumber\\
\lefteqn{+ -\sum_{j=1}^M\left(-\half\cdot
2\cdot\left[(z_{n_j}\del_{n_j})\psi_0\right]+\frac{N}{2}\psi_0
-\half\psi_0 (z_{n_j}\del_{n_j})\right)(z_{n_j}\del_{n_j})\phi\nonumber}\\
&&\equiv\Phi(n_1,\ldots,n_M)\langle n_1,\ldots,n_M|H_2|\Psi_0\rangle +
\Psi_0(n_1,\ldots,n_M)\langle n_1,\ldots,n_M|H_2'|\Phi\rangle \nonumber\\
&&
\label{H2reduce}
\end{eqnarray}
where $H_2'=\sum_{j=1}^{M}\half(z_{n_j} \del_{n_j})^2+\sum_{j\neq k}^M 
w_{n_j n_k} z_{n_j} \del_{n_j} - {\rm deg}(\Phi)(\frac{N}{2}-M)$.
$w_{ij}=\frac{z_i+z_j}{z_i-z_j}=-i\cot(\frac{i-j}{N}\pi)$. 
Remember that $\phi$ is a symmetric
polynomial and therefore homogeneous of degree ${\rm deg}(\phi)$ and
satisfies Euler's theorem. Now using the Schr\"{o}dinger equation: $H(\Psi_0
\Phi)=E\cdot(\Phi\Psi_0)$ and $H\Psi_0=E_0(M) \Psi_0$, we have 
$H_2'\Phi=(E-E_0(M))
\Phi$. This reduced Schr\"{o}dinger equation for $\Phi$
is identical to the equation obtained when $\Psi_0\cdot\Phi$ 
would
have been considered a continuum expression and be plugged 
into the continuum limit Schr\"{o}dinger equation
(\ref{SUTHdef}), {\em provided} $\lambda=2$ \cite{suth70}.
For good measure: $E_0(M) = \frac{1}{6}M(4M^2-1)-\half M^2 N $.
The case $\lambda\neq 2$ would correspond to an XXZ generalization of the 
isotropic spin model.  We follow the solution of Sutherland for
diagonalizing $H_2'$ in the next section.

\subsection{Eigenstates of the Sutherland Hamiltonian}
\label{suthsolve}

The (non-hermitean) Schr\"{o}dinger equation for $\Phi$ can be solved
by choosing a basis in the space of the symmetric polynomials, such that
$H_2'$ becomes triangular.  The eigenvalues can then be read off from the
diagonal.  The obvious basis for symmetric polynomials are the functions
$\phi^{|m|}=\sum_{P\in S_M} \prod_{i=1}^M z_{n_i}^{m_{P(i)}}$, where $0\leq
m_1 \leq\ldots\leq m_M\leq N-2M$.  Notice:  ${\rm deg}(\phi^{|m|})=\sum_i
m_i$.  The only degree of freedom left to make $H_2'$ triangular is to order
the different partitions $|m|$.  The off-diagonal 
action of $H_2'$ on a basis state $\phi^{|m|}$ is to
produce a linear combination of other basis
states that are labeled by partitions obtained from {\em repeatedly} 
``squeezing'' partition $|m|$ \cite{suth70}.  We say 
that
$|n|$ is produced by squeezing $|m|$ if $m_k=n_k, k\neq i,j$ and if for some
$i<j$: $n_i=m_i+1$, $n_j=m_j-1$. This is illustrated in the next figure:
\[
\begin{array}{ccccc}
m_1 & m_2 & m_3 & m_4 & m_5 \\
1   &  3  &  4  &  6  &  9 \\
    &\rule{.8em}{0em}\uparrow_{+1} &&\rule{0.8em}{0em}\uparrow_{-1}&\\
\end{array} 
\rule{2em}{0em}\Longrightarrow\rule{2em}{0em}
\begin{array}{ccccc}
n_1 & n_2 & n_3 & n_4 & n_5 \\
1   & 4   &  4  &  5  &  9 \\
    &     &     &     &
\end{array}
\]
Thus, if we order the states according to whether they can or cannot obtained
from each other by repeatedly squeezing, $H_2'$ is by definition diagonal.
This is possible since this partial ordering is transitive.
The eigenfunctions of $H_2'$ are then labeled by a unique partition $|m|$ and
consist of a linear combination of the corresponding basis state plus all
its squeezed descendants. They are known in the mathematics literature as
{\em Jack polynomials} \cite{forrester93,stanley90,zncha94}. The energy of
such a state is then \cite{suth70}: 
\begin{equation}
E_0(M)+\half\{\sum_j (n_j^2 - 2(M+1-2j)n_j \} -
(\frac{N}{2}-M)\sum_j n_j=\half\sum_{j=1}^M m_j (m_j- N) ,
\label{energydef}
\end{equation}
where the $m_j=n_j+2j-1$ 
are a
more convenient set of variables  in the range $0<m_j<N$. These variables,
 which are distinct
and cannot be consecutive either, are called {\em pseudomomenta}.
  The ${m_j}$ satisfy an equation that
makes contact with Asymptotic Bethe Ansatz methods:
\begin{equation}
e^{\frac{2\pi i}{N} m_i\cdot N} = \prod_{i<j} e^{\Theta(m_i-m_j)}
\label{ABAdef}
\end{equation}
and $\Theta(m_i-m_j)=\pi {\rm sgn} (m_i - m_j)$. $\Theta(m)$ is the quantum
mechanical phase shift for a scattering process  of two particles,
with relative momentum $\frac{2\pi m}{N}$, that interact via a $\lambda
(\lambda -1)/\sin^2(x)$ potential, at $\lambda=2$.

\subsection{Missing States and their Origin}

The states we have found thus far are all at the top of their spin
multiplets: $S=S^z$, which can be seen easily from the fact that $\psi_0$ is
a polynomial which vanishes at $z_i=0$ (see Appendix A).  By acting
with $S^-$ we can recover the rest. This gives us a total of (for $N$ even)
$\left(\rule{0em}{.8em}^{N-M}_{\rule{0.6em}{0em} N}\right)$
spin-multiplets with
$S=\frac{N-2M}{2}$ (we have to choose $M$ $n_k$'s in the range
${0,\ldots,N-2M}$). We know that there ought to be 
$\left(\rule{0em}{.8em}^N_M\right)-
\left(\rule{0em}{.8em}^{\rule{0.6em}{0em} N}_{M-1}\right) >
\left(\rule{0em}{.8em}^{N-M}_{\rule{0.6em}{0em} N}\right)$
\cite{bloch29}, and the missing multiplets must be of non-polynomial form,
that is, they correspond to continuum eigenfunctions with support beyond the
first Brillouin z\^{o}ne \cite{suth88}, which gets mapped back to
$[-\pi,\pi]$ because of Umklapp processes 
in the lattice case. Thus the trick (\ref{SLACder}) ceases to be
valid.

However, by diagonalizing small systems numerically, Haldane \cite{H88} finds
that all these ``missing'' states have energies already contained in the set
of energies of poly\-nomial-type states.  I.e., the energy levels have enormous
degeneracies beyond the regular $SU(2)$ symmetry of the model (see table
\ref{spectrum} later on).  This hints at
the fact that there exists some hidden symmetry generated by operators
commuting with the Hamiltonian (but not amongst each other).  Since the
introduction of the model various operators that commute with $H_{\rm ISE}$
were suggested by Inozemtsev \cite{Inoz90}, but not with the intention of
reproducing  the degeneracies---all these operators commuted
amongst themselves.  Such an algebra should {\em not} commute with the $SU(2)$
Casimir $\svec^2$.  The two Inozemtsev invariants of the third order can be
linearly combined into an operator that commutes with $\svec^2$:
\begin{equation}
\label{H3def}
H_3=\sum_{ijk}\rule{0in}{1.3ex}^{'}\frac{z_i z_j z_k}{z_{ij} z_{jk} z_{ki}}
 \svec_i\cdot\svec_j\times\svec_k
\end{equation}
where $z_{ij}=z_i-z_j$ and the sum $\sum\rule{0in}{1.3ex}^{'}$ excludes 
terms where any of the
dummy indices coincide. We will revisit this operator in chapter \ref{invar}.
The other linear combination:
\begin{eqnarray}
I_3'& =& \sum_{ijk}\rule{0in}{1.3ex}^{'}
(w_{ij}+w_{jk}+w_{ki})\svec_i\cdot\svec_j\times\svec_k
\nonumber\\
&\propto&\svec\cdot \left(\sum_{i\neq j} w_{ij} \svec_i\times\svec_j\right) 
\equiv \svec\cdot {\bf \Lambda}
\label{lambdadef}
\end{eqnarray}
also commutes with $\svec$ and with $H_{\rm ISE}$, but ${\bf \Lambda}$ only
commutes with $H_{\rm ISE}$ and {\em
doesn't} commute with $\svec^2$!
${\bf \Lambda}$ is called the {\em rapidity operator}, since $\Lambda^z$
measures the total rapidity of one of the polynomial type eigenstates, i.e.,
$\sum_j m_j$ (not to be confused with the translation/momentum operator
which determines $\prod_j e^{\frac{2\pi i}{N} m_j}$ or $\sum_j m_j\,\,{\rm
mod}\, N$). Now in Bethe's traditional model with nearest neighbor exchange
(NNE) interaction this operator has an analogue: ${\bf \Lambda}_{\rm
NNE}=\sum_{i\neq j} \theta(i-j) \svec_i\times \svec_j$, where $\theta(n)$ is
the Heavyside step function =
$\lim_{\kappa\rightarrow\infty}\coth(\kappa(i-j))$. So the NNE rapidity
operator is seen to be a particular limit of the analytic (hyperbolic)
continuation of the one in the trigonometric model with ISE, i.e.,
$z_j=e^{\frac{2\pi i}{N}j}\rightarrow e^{\kappa j}$. 

The algebra generated by $\svec_{\rm Tot}$ and ${\bf \Lambda}_{\rm NNE}$ has 
been
studied in the mathematics literature and is known as a representation of
the {\em Yangian algebra} $Y(sl_2)$ ($sl_2$ is isomorphic to $su_2$).  The
Yangian, a Hopf algebra, was introduced in 1985 by Drinfel'd \cite{drinfeld85}  in the
context of the Yang-Baxter equation (YBE), and its representation theory 
has been
worked out subsequently by Tarasov, Kulish et al. \cite{tarasov85} and 
Chari and Pressley \cite{CP1990}. The set of
Yangian operators does not commute with the Hamiltonian in the NNE model,
except when $N\rightarrow \infty$. However in the long-range spin model
$\svec$ and ${\bf \Lambda}_{\rm ISE}$ do commute with $H_{\rm ISE}$ for any
{\em finite} $N$. The
entire representation of the Yangian that they generate in this case must also
commute with $H$, i.e., for the ISE case $Y(sl_2)$ is a genuine symmetry, 
giving rise to additional degeneracies. In other words: this
particular spin model is a reducible representation of the Yangian! 
From the Yangian
representation theory we can also understand the special r\^{o}le of the
polynomial wavefunctions: they are the Yangian Highest Weight States (YHWS)
of different irreducible Yangian representations. 
Every one of them generates an irreducible representation space by letting 
$\svec$ and 
${\bf \Lambda}$ act on them
repeatedly. All the resulting states will have the same energy as their
parent state, but won't in general be polynomials. 

Every  irreducible $SU(2)$ representation is parametrized by one integer,
since there is only one element in the Cartan algebra---for $SU(N)$ there are
$N-1$ integers. For $Y(sl_2)$ we characterize each irrep with a
polynomial, the so-called Drinfel'd polynomial (a polynomial is
determined by the prefactor of the highest power and the positions of its
roots; here we
only care about the  relative positions of the roots)\footnote{In a $SU(N)$
spin chain a $Y(sl_N)$ representation is characterized by 
$N-1$ polynomials, see chapter \protect\ref{invar}.}. 
In the spin-chain
case the degree of this polynomial is $N-2M$ for a YHWS with $M$ down-spins.
The locations of its roots, which are spaced by integer multiples of some
unit called $h$, are determined by the pseudomomenta $\{ m_i \}_{i=1}^M$ and
vice versa.

One of the things the representation theory of the Yangian algebra can teach
us is the degeneracy of the multiplets:  i.e.,  exactly which spin-multiplets
occur in a given irrep, belonging to a particular energy level.
  Haldane \cite{H91} has identified a rule that
gives the spin-content of a particular energy level, 
but in the next section, we will go through the ``Yangian
derivation'' \cite{BGHP93,BPS93}.  We will do this by constructing another
irreducible representation in a Hilbert space of which we know the $SU(2)$ spin
content in section \ref{spincontent}.  
This representation is equivalent to one based on
one of the polynomial states, as they are made to have 
the same Drinfel'd polynomial. 

In order to understand the representation theory we should first discuss the
abstract algebraic structure underlying the Yangian, the {\em transfer matrix}
and the Yang-Baxter equation, in \ref{yangrep}.  In the case of the NNE-model
the NNE representation of the transfer matrix (which contains $\svec_{\rm
Tot}$, ${\bf \Lambda}_{\rm NNE}$ and all other elements of the Yangian as
coefficients in a power series) emerges naturally if one calculates its
partition function ${\rm Tr}(e^{-\beta H})$.  We construct it specifically in
\ref{yangexample}, as well as for the ISE-model.  The transfer matrix for the
latter, unfortunately doesn't have such a nice physical interpretation, but
its representation is reducible, explaining the multiplet structure of
$H_{\rm ISE}$.

\section{The Transfer Matrix, its Yangian, and the Algebraic Bethe Ansatz}
\label{ISEyang}

Historically the Yangian emerged from a study of the algebra of the 
transfer matrix.  Almost all exactly solvable 1D models owe their
integrability to the existence of a transfer matrix which encodes the
consistency rules on scattering processes that lead to the Bethe Ansatz
equations \cite{suth85}.  The transfer matrix $\T(u)$ is an operator-valued 
matrix function of a spectral parameter $u$.  
$u\in\CC$ and $\T$ is a $2\times 2$
matrix in our $SU(2)$ case (for $SU(N):  \; N\times N$).  Its $n$th
Taylor-expansion coefficients in $\frac{1}{u}$ contains essentially the
operators generated by commuting $n$ $\svec$'s and ${\bf \Lambda}$'s with
each other.
Specific to $SU(2)$ the transfer matrix is defined by 
\begin{eqnarray}
T^{ab}(u) &=&\sum_{\mu =0}^3\sigma^{ab}_\mu T^\mu(u)\nonumber\\ 
&=& \delta ^{ab}+\sum_{n=0}^\infty \frac{1}{u^{n+1}}\sum_{\mu=0}^3 
\sigma^{ab}_\mu
J_n^\mu, \,\,\,\, a,b=1,2 \label{tdef} 
\end{eqnarray} 
$\{ \sigma_1, \sigma_2 ,\sigma_3\}$, the three Pauli matrices, and 
$\sigma_0=\unity$, satisfy 
$\Tr(\sigma_\alpha \sigma_\beta)=2\delta_{\alpha\beta}$. 
These Pauli matrices do {\em not} act on  the local spins, but are just a
convenient way to label different linear combinations of the four components
of $\T$.
The $J^\mu_n$ are
abstract operators, but in the spin-chain representation they are
combinations of the local $\svec_i$.  To follow the
upcoming discussion it may be helpful to take the following into account:
\begin{eqnarray} 
&& {\bf J}_0 = \svec_{\rm Tot}\nonumber\\
 & & {\bf J}_1\propto {\bf \Lambda} 
\label{Jhelp} 
\end{eqnarray} 
The Yangian defining commutation relations---the algebra of $\svec_{\rm
Tot}$ and ${\bf \Lambda}$ and the operators generated  by commuting these with
each other---are 
encoded in the Fundamental Commutation Relations (FCR) for $\T$:
\begin{equation} 
R(u-v) (\unity\otimes\T(u))(\T(v)\otimes\unity)=
(\T(v)\otimes\unity)(\unity\otimes\T (u)) R(u-v) 
\label{FCRdef} 
\end{equation}
where $R$ is a 4x4 matrix acting on the
$a,b$ indices of both $\T$'s but not in the space in which their matrix
elements act, ie\ $R$ commutes with the $J^\mu_n$.  
Here $R$ is Yang's ``rational'' 
solution (whence the name Yangian) of the
{\em Yang-Baxter equation}:  
\begin{equation} 
\label{Rdef} 
R=u\unity+hP_{12}.
\end{equation} 
$P_{12}$ is the operator that permutes vectors in the tensor product
$\CC^2\otimes\CC^2$:  $P_{12} (x\otimes y)=y\otimes x$.  
$h$ is some complex number, called the {\em quantum parameter}, 
which is fixed for a particular
model.
For the ISE-model, the Drinfel'd
polynomial 
will turn out to have real roots, and we can normalize $h$ to 1. For the
NNE-model, complex roots occur in a pattern symmetric around the imaginary 
axis,
and it is customary to choose $h=i$.
From now on we will use
the summation convention with a sum over Greek indices running from $0\ldots
3$ and Latin indices running from $1\ldots 3$.  The FCR (\ref{FCRdef}) can be
written out more transparently in components:  
\begin{equation}
(u-v)[T^{ab}(u),T^{cd}(v)]=h(T^{cb}(v)T^{ad}(u)-T^{cb}(u)T^{ad}(v))
\label{FCRcomp} 
\end{equation}
 This holds in general for a chain based on for $sl_n$ spins  as well.
Notice that for $h\rightarrow 0$ the transfer matrix commutes with itself;
this is referred to as the {\em classical limit}.  For the particular case of a
spin-$\half$ model we rewrite (\ref{FCRcomp}) in basis of $\sigma$-matrices:
\begin{equation}
\begin{array}{rclr}
(u-v)[T^0(u),T^i(v)]&=&\frac{h}{2}i\epsilon_{ijk}T^j(v),T^k(u)\},
&(a)\\
(u-v)[T^i(u),T^j(v)]&=&\frac{h}{2}i\epsilon_{ijk}(T^k(v)T^0(u)-T^k(u)T^0(v))
&(b)\\
&=&\frac{h}{2}i\epsilon_{ijk}(\{T^0(u)T^k(v)-T^0(v)T^k(u))&
\end{array}
\label{Tcompcomm}
\end{equation}
$\{ ,\}$ denotes the anticommutator.
Expanded in modes this becomes:
\begin{equation}
\begin{array}{rclr}
 \left[ J^0_{m+1},J^i_n\right] -\left[ J^0_m,J^i_{n+1}\right] &=&
\frac{h}{2}i\epsilon_{ijk}
\{ J^j_n,J^k_m\} &(a)\\
\left[ J^{i}_{m+1},J^{j}_n\right] -\left[J^i_m,J^j_{n+1}\right] &=&
\frac{h}{1}i\epsilon_{ijk} (J^k_n J^0_m-J^k_m J^0_n) &(b)\\
&=&\frac{h}{2}i\epsilon_{ijk} (J^0_m J^k_n -J^0_n J^k_m) &
\end{array}
\label{modescomm}
\end{equation}
This is to be supplied with the ``boundary condition'' $J^0_{-1}=\unity$; ${\bf
J}_{-1}=0$. In deriving (\ref{Tcompcomm}, \ref{modescomm}) 
we have used the fact
that $[T^\mu(u),T^\nu(v)]=[T^\mu(v),T^\nu(u)]$, which follows from adding
the FCR and the one with $u$ and $v$ interchanged. An important consequence
of this last relation with $\mu=\nu$ is that the components of the
transfer matrix commute with themselves for different values of the spectral
parameter. In particular, if $\mu=\nu=0$ we have $[\Tr(\T(u)),\Tr(\T(v))]=0$. 
If we expand this result in powers of $u,v$ we find $[J^0_m,J^0_n]=0$ 
$\forall_{m,n}$, 
i.e., we have a set of commuting operators. For the NNE model, the
Hamiltonian is a member of this family, and thus we obtain a set of
constants of motion. However $[\Tr(\T),T^{ab}]\equiv 0$ only for
$N\rightarrow\infty$, so the Yangian doesn't provide any extra degeneracies. 
Unfortunately, for the ISE model $H_{\rm ISE}$ cannot be a member of that 
family since
$H_{\rm ISE}$ commutes with $\T$. We will return to this issue in chapter
\ref{invar}. 

Let us look at eq.\ (\ref{modescomm})  for a few particular cases.  When we
choose $m=-1$ in (\ref{modescomm}$^a$) we find: $[J^0_0,{\bf J}_n]=0$ so
$J^0_0$ commutes with the algebra and is scalar. Setting $m=-1$ in
(\ref{modescomm}$^b$): $[J^i_0,J^j_n]=hi\epsilon_{ijk} J^k_n$, i.e., the $\J$
are $SU(2)$ vector operators. Similarly with $m=0$ in (\ref{modescomm}$^a$)
we find that the $J^0_n$ are scalars.
With $m=0$ in (\ref{modescomm}$^b$) we have:
\begin{equation}
i\epsilon_{ijk}[J^j_1,J^k_n]=-2J^i_{n+1}-2h(J^0_0+1)J^i_n-2hJ^i_0(J^0_n).
\label{recursion}
\end{equation}
This last equation tells us that we can generate all the vectors $\J_n, n>1$
once we specify $\J_0$ and $\J_1$, plus the trace operators $J^0_m,\: m<n$. We
will get back to the latter shortly. 

In order for the ensuing $\{ \J_n\}$ to satisfy the FCR, $\J_0$ and $\J_1$ have
to obey a set of consistency relations, known as the deformed Serre
relations.  Our candidates for $\J_0$ and $\J_1$ viz.  $\svec$ and ${\bf
\Lambda}$ turn out to satisfy these relations (for both ISE and NNE model)
\cite{HHTBP92}, but since we will explicitly 
construct a transfer matrix that satisfies
(\ref{FCRdef}) and has $\svec$ and ${\bf \Lambda}$ as its first moments we do
not have worry about these constraints.

To fix the set $\{ J^0_n\}$ we require some additional information beyond the
FCR. We need to specify the so-called {\em Quantum Determinant}:
\begin{equation}
\label{Qdetdef}
{\rm Det}_q = T^{11}(u) T^{22}(u-h)-T^{12}(u) T^{21}(u-h)
\end{equation}
It can be shown with the FCR that $[T^{ab}(u), {\rm Det}_q \T(v)]=0$ (see
\cite{korepin81}). The quantum determinant commutes with the Yangian
algebra. In the particular representations of $Y(sl_2)$ in the NNE and ISE
spin chains, it is actually a c-number function times the identity operator
(but in general it doesn't
need to be, see chapter \ref{invar}). From eq.\ (\ref{Qdetdef}) we see that
${\rm Det}_q \left(\phi (u)\T (u)\right)=\phi(u)\phi(u-h){\rm Det}_q \T(u)$ 
so by
multiplying the transfer matrix by a scalar function we can choose the 
normalization  of the quantum determinant. After an expansion in powers of
$\frac{1}{u}$ this normalization fixes the constant (non-operator) part of
the $J^0_n$. We should also note that ${\rm Det}_q (\T_1 \T_2) =
{\rm Det}_q (\T_1) {\rm Det}_q (\T_2)$.

\section{Representations of the Yangian}
\label{yangrep}

Like in the case of Lie algebras, finite dimensional irreducible
representations  of the Yangian are built on a Highest Weight State which is
annihilated by a set of raising operators: $\{ J^+_n\equiv J^x_n+iJ^y_n\}$.
Furthermore the YHWS  are eigenvectors of $J^z_n$ and $J^0_n$. The action of
the lowering operators $\{ J^-_n\equiv J^x_n-i J^y_n\}$ generates the
other states that span
 the representation space. This information can be encoded in the following
equation:
\begin{equation}
\T(u)|{\rm YHWS}\rangle =\left(
\begin{array}{cc} t^{11}(u) & 0\\ T^{21}(u) &
t^{22}(u)
\end{array}
\right) |{\rm YHWS}\rangle.
\label{HWSeval}
\end{equation}
$t^{11}(u)$ and $ t^{22}(u)$  are complex
functions of $u$ that label the representation. Just like $S^z$ acting on a
maximal $S^z$ HWS determines the spin of an $SU(2)$ irrep, the diagonal
elements of $\T$, $t^{11}$ and $t^{22}$, 
determine the particular $Y(sl_2)$ irrep
that we are dealing with. Actually, only their ratio does, since the quantum
determinant constraints their product. {\em If} the representation is finite 
dimensional and irreducible, this ratio can be written as follows:
$\frac{t^{11}(u)}{t^{22}(u)}=\frac{P(u+h)}{P(u)}$, where $P$ is a
polynomial.
$P(u)$ is the Drinfel'd polynomial. 

The other states in the Hilbert space are generated by the repeated action of $T^-(u)=T^{21}(u)$:
\begin{equation}
\label{ABEdef}
|\lambda_1,\ldots,\lambda_r\rangle = \prod_{i=1}^{r} T^-(\lambda_i)|{\rm
YHWS}\rangle
\end{equation}
The $\{\lambda_i\}$ aren't arbitrary numbers: we want to choose them such
that the states that they generate are an orthonormal basis for the
representation. 
It turns out that choosing them
to be eigenvectors of the trace of the transfer matrix
$T^0(u)$ does just that, since $T^0$ is hermitean\footnote{At least the
spin chain realization is.} and consequently, its eigenvectors
are orthogonal (just 
like $(S^-)^n|S^z=S_{\rm Tot}\rangle$ must be an $S^z$ eigenstate)\footnote{
See chapter \protect\ref{invar} for a more specific discussion.}. With the FCR
(\ref{Tcompcomm}) this translates to the following set of Bethe Ansatz equations
for the $\{\lambda_j\}$ \cite{FaddeevTakt81}:
\begin{equation}
\label{ABAeq}
\frac{t^{11}(\lambda_n)}{t^{22}(\lambda_n)}\cdot
\prod_{j(\neq n)}^r \frac{\lambda_n-\lambda_j +h}{\lambda_n-\lambda_j -h}=1,
\;\,n=1,\ldots,r
\end{equation}
In the spin models this procedure yields maximal $S^z$ states; 
the remaining states follow by acting with
$S^-=J^-_0$. Those states in the same multiplet have the same
$T^0(u)$-eigenvalue since $S^-$ commutes with the trace of the transfer matrix.
This procedure of constructing a representation space of the
Yangian is called the {\em Algebraic Bethe Ansatz} \cite{FaddeevTakt81}.

\section{Specific examples: NNE and ISE chains}
\label{yangexample}

An important property of the solution of the FCR eq.\ (\ref{FCRdef}) is that
if $\T_1 (u)$ and $\T_2 (u)$ satisfy it (and matrix elements of $\T_1$
commute with those of $\T_2$)  then so will $\T_1\otimes\T_2$. 
The simplest solution to the FCR with R-matrix
(\ref{Rdef}) is:
\begin{equation}
\label{EVdef}
\T(u)=(1+\frac{h}{2u})\sigma_0 +\frac{h}{u} \sig\cdot\svec =
\unity +\frac{h}{u} P_{01}
\end{equation}
It acts in the Hilbertspace of one spin $S$. 
The second step is only valid for $S=\half$. 
$P_{01}$ is the operator that
permutes the ``real'' spin with the ``auxiliary'' spin defined by the matrix
indices of $\T$ i.e., \mbox{$P_{01}=\half+ \sig\cdot\svec$}. 
This representation of $\T$ is called
the {\em Fundamental Representation}, but is referred to in the mathematics
literature as the ``evaluation homomorphism''. 
The Hilbert space in which it acts is
just that of a single spin $S$. 
One can simply verify that its quantum determinant is
given by $\frac{u+h}{h}$, for $S=\half$	and
$\frac{u^2-(h/2)^2-h^2 S(S+1)}{u(u-h)}$ for general $S$. Furthermore: 
$t^{11}(u) =
1+\frac{h}{2u}+\frac{h}{u}S;\; t^{22}(u)=1+\frac{h}{2u}-\frac{h}{u}S$ and 
$\frac{P(u+h)}{P(u)}=\frac{u+\frac{h}{2}+hS}{u+\frac{h}{2}-hS}$ which has
the unique solution: 
\begin{equation}
P(u)=\prod_{j=1}^{2S}\left(u+\frac{h}{2}+hS -hj\right)
\label{STRdef}
\end{equation}
We see that Drinfel'd polynomial $P(u)$ has a string of $2S$ zeroes, 
equally spaced, and centered
around 0.

From it we can construct a transfer matrix for $N$ spins by tensoring it $N$
times, for $N$ different spins:
\begin{equation}
\label{TNNE}
\T_{\rm NNE}(u)=\prod_{i=1}^N\left[\unity+\frac{h}{u}\left( \half+
\sig\cdot\svec_i\right)\right]=
\prod_{i=1}^N\left(\unity+\frac{h}{u}P_{0i}\right).
\end{equation}
Again, the second equality only holds for $S_i=\half$. This is exactly the
transfer matrix of the NNE model. Since it doesn't commute with $H_{\rm
NNE}$, 
it suggests that this Yangian representation 
is {\em irreducible}, which, as we shall see later, it is in fact. Its YHWS is
just the completely ferromagnetic state. We can
expand it in powers of $\frac{1}{u}$ to make contact with the Yangian
operators. We specialize to $sl_2$.
\begin{eqnarray}
&&\T_{\rm NNE}(u)=\unity + h\sum_{n=0}^\infty \frac{1}{u^{n+1}}
\left(\sum_{i,j=1}^N (\half  + \sig\cdot\svec_i) 
(L^n)_{ij}\right)\nonumber\\
&&L_{ij}=h(1-\delta_{ij})\theta(i-j) P_{ij}
\label{NNEyang}
\end{eqnarray}
where $\theta(n)$ is the step function, that has been introduced to keep the
dummy indices in the expansion of (\ref{TNNE}) in powers of $\frac{1}{u}$
 ordered. $P_{ij}$ permutes
spins on sites $i$ and $j$. This result follows from the fact that
\begin{equation}
P_{0i_1}\cdots P_{0i_m}= (\half+\sig\cdot\svec_{i_m}) P_{i_m i_{m-1}}
P_{i_{m-1} i_{m-2}} \cdots P_{i_2 i_1}
\label{permtrick}
\end{equation}
From eq.\ (\ref{NNEyang}) we find by explicit calculation:
$\J_0=h\sum_{i}\svec_i=h\svec_{\rm Tot}$ and $\J_1=h^2\frac{N-1}{2}\svec
-\frac{ih^2}{2}\sum_{j<i} \svec_i \times \svec_j$, as expected. 

For the ISE model there isn't such an obvious candidate for a transfer
matrix.  However, if we compare the form of the $\J_1$ Yangian generator for
the NNE model with what we want in the ISE case: $\J_1\propto \sum_{i<j}
\frac{z_i}{z_i-z_j} \svec_i\times\svec_j$ (where we have slightly rewritten the
expression for ${\bf \Lambda}_{\rm ISE}$ (\ref{lambdadef})), we see they only
differ in that $\theta(i-j)\rightarrow\theta_{ij}\equiv\frac{z_i}{z_i-z_j}$.
Using this replacement as an Ansatz in eq.\ (\ref{TNNE}), we find that the
resulting $\T$ with $L$ matrix
\begin{eqnarray}
\T&=&\unity+\sum_{ij}\left(\half+\sig\cdot\svec_i\right)\left(\frac{1}{u-L}
\right)_{ij}\nonumber\\
L_{ij}&=&(1-\delta_{ij}) \theta_{ij}
\label{Ldef}
\end{eqnarray}
 still satisfies the FCR---for arbitrary $\{z_i\}$ as a matter of 
fact \cite{BGHP93}.  The corresponding product form (\ref{TNNE}) isn't so
simple to find, but was eventually traced after the introduction of the
so-called Dunkl operators (see chapter \ref{invar} for a more general
discussion):
\begin{eqnarray}
\hat{\gamma}_i &= & h\sum_{j(\neq i)} \left[\theta_{ij}\theta(j-i)-
\theta_{ji}\theta(i-j) \right] K_{ij}\nonumber\\
&=& \frac{h}{2}\sum_{j(\neq i)}(w_{ij}-{\rm sgn}(i-j))K_{ij}
\label{dunkldef}
\end{eqnarray}
where $K_{ij}$ permutes $z$-labels $K_{ij}z_j=z_i K_{ij}$, but not the spins!
These Dunkl operators
commute with one another: $[\hat{\gamma}_i,\hat{\gamma}_j]=0$. Therefore, 
$\T=\unity + \frac{h}{u-\hat{\gamma}_i}P_{0i}$ and thus also
\begin{equation}
\label{TISE}
\tilde{\T}=
\prod_{i=1}^N\left(\unity + \frac{h}{u-\hat{\gamma}_i}P_{0i}\right)
\end{equation}
still satisfy the FCR (remember: $\hat{\gamma}_i$ only acts on $z_i$'s, not
on spins). We turn $\tilde{\T}$ into a spin operator by replacing
$K_{i,i+1}$ by $P_{i,i+1}$, once ordered to the right of an expression
(imagine the particles having {\em dynamical} $z$ co-ordinates, as well as
spins, then the projection is equivalent to just acting on states that are
symmetric in simultaneous permutations of spin and $z$ co-ordinate). Ref.
\cite{BGHP93} shows that the projected version of $\tilde{\T}$ 
preserves integrability and reproduces the asymptotic series 
(\ref{Ldef}).  Again it is trivial to compute $\J_0=h\sum_{i}\svec_i$ and
$\J_1=-\frac{ih^2}{2}\sum_{i\neq j} w_{ij}\svec_i\times\svec_j + 
h^2\frac{N-1}{2}
\sum_i\svec_i$. At this point the $z_i$ are still arbitrary complex number,
but the requirement that $\J_{0,1}$ commute with $H_{\rm ISE}$, and thus
generate its symmetry algebra, restricts
them to be equidistant on a circle. 

Representations of this version of $Y(sl_2)$ {\em are} in fact reducible:
we know that $H_{\rm ISE}$ commutes with $\T$, and we will find
that there is more than one YHWS! We will now
elucidate the decomposition. It was mentioned earlier that all the polynomial
states translated from the Sutherland model are  YHWS in the spin model. We can
check this via eq.\ (\ref{HWSeval}): those states should be annihilated by
$J^+_0$ and $J^+_1$ and therefore also by $T^+(u)$. This is verified in
Appendix A. They should also be eigenvectors of $T^{11}(u)$ and
$T^{22}(u)$ and we will compute their eigenvalues. Potentially there could
be non-polynomial YHWS, but since we will find that the polynomial ones and
their descendants  exhaust the Hilbert space, we can rule this
option out. 

Because of the special polynomial character of the YHWSs we can write the
action of $T^{11}(u)$ and $T^{22}(u)$ on them as {\em differential
operators}, just as we did in diagonalizing $H_{\rm ISE}$ eq.\
(\ref{H2reduce}). It suffices to find $t^{11}(u)$, as $t^{22}(u)$ will follow
from ${\rm Det}_q \T(u) |{\rm YHWS}\rangle=t^{11}(u)t^{22}(u-h)|{\rm YHWS}\rangle$. 
In
general the {\em allowed} 
set of rational functions $t^{11}(u),t^{22}(u)$ can be obtained
from just knowing the quantum determinant  and the fact that
$\frac{t^{11}}{t^{22}}$ is of the restrictive form 
$\frac{P(u+h)}{P(u)}$, with polynomial $P$. But to identify specific Yangian
multiplets with particular eigenvalues of $H_{\rm ISE}$ we will consider the
action of $\T(u)$ explicitly in Appendix B. 
	
There we find:
\begin{equation}
\label{t22def}
T^{22}(u)|{\rm YHWS}\rangle=
\prod_{i=1}^M\left(1+\frac{h}{u-h(\hat{D}_i-\frac{N+1}{2})}\right)|{\rm
YHWS}\rangle
\end{equation}
where $M=\frac{N}{2}-S^z$ is the number of down-spins in the YHWS. The
$\D_i$ are the {\em dynamical} Dunkl operators: $\D_p=z_p \del_p
+\frac{1}{h}\hat{\gamma}_p$, obeying the {\em same} algebra 
as the $\hat{\gamma}_i$. We
will revisit these operators in chapter \ref{invar}. We can solve the
eigenvalues problem of the $\D_i$ very much along the same lines as the way
$H_{\rm ISE}$ was (partially) diagonalized. This is done in Appendix
C. What we find there is that the eigenfunctions  of the dynamical
Dunkl eigenfunctions are
most easily described in a space spanned by polynomials of the form
$\prod_{i<j}(z_i-z_j)\prod_{i}z_i^{n_i}$. The eigenvalues  are organized
in blocks of partitions $|n|={n_1\geq n_2\ldots \geq n_M}$ and are labeled
within the blocks by permutations of that partition
$\{ P_1,\ldots ,P_r | P_i\in S_M \},\; r\leq M!$. I.e., for every $P$ there is an
eigenfunction $\psi^{|n|}_P$ such that its eigenvalues with respect to 
the $\{ \D_i\}$ are: 
$d_i=n_{P(i)}-P(i)+M$. Any symmetric function of the $\D_i$---such as
$T^{22}(u)$---will be insensitive to the particular $P$ and  must have an
$r!$-fold degeneracy in that block of states. 
Now we have to impose two ``physical''
constraints: 1) that the eigenfunctions are {\em symmetric}---down-spins are
hard-core {\em bosons}---, and 
2) they have degree in every one
of the $z_i$ less than $N$---the hard-core constraint 3) that they vanish 
when two 
co-ordinates coincide is taken care of by the antisymmetric prefactor. 
1) implies that of $r!$ degenerate eigenfunctions
we retain only one: their {\em anti}symmetric combination (to cancel the
antisymmetry of the Jastrow prefactor). A consequence is that two $n_i$ 
can never coincide and (because of the $M-i$ factor) two $d_i$ cannot 
be adjacent. 2) restricts the $n_i$ to be in the range $0,\ldots,N-M$. 
Actually, since the
eigenfunctions have to be annihilated  by $T^+(u)$, which contains $S^+$, they
have to vanish if one of  the $z_i=0$ (see Appendix A), so we have
rather: $1\leq n_i\leq N-M$ and thus $1\leq d_i \leq N-1$.

With these additional constraints the eigenfunctions are effectively in a
smaller subspace spanned by $\prod_{i<j}(z_i-z_j)^2\prod_i z_i\cdot
\Phi(z_1,\ldots,z_M)$, with $\Phi$ a symmetric polynomial of order $\leq
N-2M$.
We recognize this form from solving eq.\ (\ref{H2reduce}).
The second factor $\prod(z_i-z_j)$ comes from the fact that we only retained
the antisymmetric combination, which contains such a factor.

We finally find that 
\begin{eqnarray}
t^{22}(u)&=&\prod_{i=1}^M \left(1+\frac{h}{u-h(d_i-\frac{N+1}{2})}\right)
\nonumber\\
&=& \frac{f_2(u+h)}{f_2(u)};\,\,\,
f_2(u)=\prod_{i=1}^M\left(u-h(d_i-\frac{N+1}{2})\right)
\label{t22ev}
\end{eqnarray}
To obtain $t^{11}(u)$, whose ratio to $t^{22}(u)$ determines the Drinfel'd
polynomial and thus the particular representation, we need the 
quantum determinant.
Since it is a scalar, we can obtain it by having it act on any vector, for
instance on the YHWS $|\uparrow\uparrow\ldots\uparrow \rangle$, corresponding to
$M=0, \psi=1$. In that case $t^{22}(u)=1$, so ${\rm Det}_q \T(u) = 1\cdot
t^{11}(u) =1+\sum_{i,j}\left(\frac{1}{u-h\theta}\right)_{ij} = 1+
\langle 1|\frac{1}{u-h\theta} |1\rangle$, where $|1\rangle$ 
is the vector of length $M$:
$(1,1,\ldots,1)$ and $\theta$ the matrix with elements $\theta_{ij}, j\neq
i$ and 0 otherwise. In the case $z_j=e^{\frac{2\pi i}{N}j}$, 
$|1\rangle$ happens to be an eigenvector of $\theta$ with
eigenvalue $\frac{N-1}{2}$.
Then
\begin{equation}
\label{detqise}
{\rm Det}_q \T(u)=\frac{u+h\frac{N+1}{2}}{u-h\frac{N-1}{2}}=
\frac{\Delta(u+h)}{\Delta(u)}.
\end{equation} 
$\Delta(u) =\prod_{i=1}^N\left(u+h(\frac{N+1}{2}-j)\right)$ is the ratio
of two {\em strings}  (a set zeroes of a polynomial, spaced by $h$) of length
$N$. With $t^{11}(u)t^{22}(u-h)=\frac{\Delta(u+h)}{\Delta(u)}$,
$t^{22}(u)=\frac{f_2(u+h)}{f_2(u)}$, and the definition of the Drinfel'd
polynomial $P(u)$ we have:
\begin{equation}
\label{drinsolve}
\frac{\Delta(u+h)}{\Delta(u)}=\frac{P(u+h)f_2(u+h)f_2(u)}{P(u) f_2(u)
f_2(u-h)} \equiv \frac{g(u+h)}{g(u)}.
\end{equation}
Since $\Delta(u)$ has an $N$-string of zeroes, it is easy to see that the
only way to satisfy eq.\ (\ref{drinsolve}) for a polynomial $g(u)$  is 
$\Delta(u)=g(u)=f_2(u) P(u) f_2(u-h)$. 
Note that this only admits solutions for $P(u)$ if  the zeroes of $f_2$ are
not adjacent, which we know from the explicit calculation of the $\{ d_i\}$
they are not.  The
order of the Drinfel'd polynomial is thus $N-2M$.  A graphical example of
eq.\ (\ref{drinsolve}) is shown in fig.\ (\ref{figmotif}). We have
absorbed the factor of $\frac{N+1}{2}$ into $u$ for now. 
\begin{figure}[htb]
\framebox[\hsize]{\centerline{\psfig{file=motif.epsi,width=5.8in}}}
\vspace{.2truein}
\caption{A graphical solution of eq.\ (\protect\ref{drinsolve}), with
\mbox{$d_1\!\!=\!\! 1,\, d_2\!\!=\!\! 3,\,
d_3\!\!=\!\! 7,\,  
d_4\!\!=\!\! 12\, $}  \& \protect\linebreak
$d_5\!\!=\!\! 15$}
\label{figmotif}
\end{figure}
A picture like this figure is called a {\em motif} and, as we shall see
shortly, the spin multiplets present in the Yangian action on the
corresponding YHWS can be read off quite easily. 

The rule is as follows:
A sequence of $Q$ consecutive roots in the Drinfel'd polynomial is to be
identified with an $SU(2)$ spin $S=\frac{Q}{2}$. In the example motif we
have therefore: ${\scriptstyle (S=1)\otimes ( S=\frac{3}{2})\otimes
(S=\half) = ( S=3)\oplus (S=2^2)
\oplus (S=1^2) \oplus (S=0)}$; thus a  $7+2\cdot 5 + 2\cdot 3+ 1=24$-dimensional
representation, a 24-fold degeracy of this energy level. 

Haldane \cite{H91} had already identified this state
counting rule based on numerical studies. 
He introduced a nice physical
description in terms of {\em spinons}, the elementary excitations, where
their number is  $N_{sp}=N-2M$. Once we identify the $d_i$ with the
pseudomomenta $m_i$ later on, we will see that the absolute groundstate of the
model,  which occurs for $M=\half N ,S^z=0$, has 0 spinons. The lowest
excited states have $S^z=1$ and thus contain 2 spinons\footnote{The ones 
that don't,
derive from YHWS with a higher $S^z$ and thus higher energy; see expression
for $E_0(M)$ in section \protect\ref{ISEdiag}.}: spinons are
created in pairs! In the motif picture, a sequence of $Q$ consecutive
Drinfel'd zeroes is to be identified with $Q$ spinons ``in the same orbital''.
To have the total spin come out right, the spinons have to have
spin-$\half$, and be in a totally symmetric state. This would make them
bosons, except that the number of ``states'', $M+1$, varies with $N_{sp}$
(which allows the identification as semions by the generalized exclusion
principle, see section \ref{semions}).

So every degenerate energy level has a YHWS (with say $M$ down-spins), that
corresponds to a sequence of $N-2M$ integers in the range $0\ldots N$. There
is a spinon for every integer, and $Q$ consecutive integers refer to $Q$
spinons in a maximal $S^z$ state. 
We will now prove this statement.

\subsection{The Spin Content of the Yangian Irreducible Representations.}
\label{spincontent}

We will go back to the simple fundamental transfer matrix of the NNE :
$\T_i (u-hp_i)=(u-hp_i+\frac{h}{2})+h\sig\cdot\svec_i$ 
\footnote{We have changed the normalization of the fundamental $\T(u)$ by
multiplying it by $(u-hp_i)$ to simplify the algebra.} where the spectral
parameter has been shifted by an amount $hp_i$. We will take the tensor
product of $N-2M$ copies of this transfer matrix, one for every spinon, 
each copy with a different
{\em shift} $p_i$. We pick the $p_i$ as the roots of the Drinfel'd
polynomial corresponding to a particular 
YHWS in the ISE-model (so {\em here} the $p_i$ do
not correspond to the crosses in fig.\ (\ref{figmotif}) but rather the
circles). All $\svec_i$ are chosen to be spin-$\half$. 
\begin{equation}
\T(u)=\bigotimes_{i=1}^{N-2M}\left[
u-hp_i+\frac{h}{2}+h\sig\cdot\svec_i\right]
\label{evaltensor}
\end{equation}
By having $\T$ act on the state $|\uparrow\uparrow\ldots\uparrow \rangle$
(this is {\em not} a state in the spin model, but represents all $N-2M$
{\em spinons} pointing up, the YHWS).
This is obviously the---or at least one of the---highest weight state(s)  
of this 
representation) we see that it  has the form of eq.\ (\ref{HWSeval})---just
add an auxiliary spin and read off the action of $\T$.
$t^{11}(u)=\prod_i (u-hp_i),\, t^{22}(u)=\prod_i (u-hp_i+h)$, so that
the Drinfel'd polynomial reads: $\frac{P(u+h)}{P(u)}=\prod_i
\frac{u-hp_i+h}{u-hp_i}$ i.e., $P(u)=\prod_i (u-hp_i)$ which is the same
Drinfel'd polynomial as in the ISE model for the YHWS characterized by the
same $p_i$. The two representations must be isomorphic, 
provided that they are irreducible \cite{CP1990}. 
We will assume the latter is true and
concern ourselves with obtaining the $SU(2)$ content of eq.\
(\ref{evaltensor}) and thus (\ref{t22def}).

Now we need to show that consecutive shifts, or spinons in the same orbital,
only occur in  their symmetric, maximal spin, combination.
To that end we need to show the following fact about the tensor
product:\newline
$\T(u)\equiv (u-hp_1+\frac{h}{2}+h\sig\cdot\svec_1)\otimes
(u-hp_2+\frac{h}{2}+h\sig\cdot\svec_2)$ with arbitrary values of spins
$S_1,S_2$ is reducible, such that the subrepresentation contains
the highest spin $S_1+S_2$ multiplet, {\em only} if
$p_2-p_1=S_1+S_2$. (For $p_1-p_2=S_1+S_2$ we get only the lowest spin
$|S_1-S_2|$-multiplet, which we don't care about now). 
A new feature of Yangian
representations is  that reducibility doesn't imply {\em full} reducibility,
that is: in matrix form the representations might look like:\\
\begin{figure}[htb]
\psfig{file=reducible.epsi,width=\hsize}
\label{reducible}
\end{figure}

The span of vectors in block $A$ is invariant under the action of the
representation, but it's ortho-complement might not be. If $p_2-p_1=S_1+S_2$,
$A$ corresponds to just the ``highest component'' of $S_1\otimes S_2$, that
is the $S_{\rm Tot}=S_1+S_2$ states. Let us show this by letting $\T$ act in
this highest spin sector:
\begin{eqnarray}
&&\T(u)= T_0+\sig\cdot\vec{T} = 
\left[(u-hp_1+\frac{h}{2})(u-hp_2+\frac{h}{2})+h^2\svec_1\cdot\svec_2
\right]\nonumber\\
&&+ 
\sig\cdot\left[(\half(u-hp_1)+\half(u-hp_2)+1)\svec + (\half (u-hp_2)-
\half (u-hp_1)){\bf \Delta} + i\svec_1\times\svec_2\right]\nonumber\\
&&
\label{T2}
\end{eqnarray}
where ${\bf \Delta}\equiv \svec_1-\svec_2$. Using the fact that
$\svec_1\cdot\svec_2=S_1 S_2$ {\em on a highest} $S_{\rm Tot}$ state, 
and the identity:
$i\svec_1\times\svec_2=-\frac{1}{2}[\svec_1\cdot\svec_2,{\bf \Delta}]$, the
vector part of $\T$ becomes:
\begin{equation}
\label{Treduce}
\vec{T}=\half(1+ u-hp_1+u-hp_2)\svec + \half(S_1 S_2 + u-hp_2 -(u-hp_1)
-\svec_1\cdot\svec_2){\bf \Delta}
\end{equation}
($\svec\equiv\svec_{\rm Tot}$).
Since ${\bf \Delta}$ is a vector operator, it acting on an maximal
$S_{\rm Tot}$  state produces only $S_{\rm Tot}=S_1+S_2$ or $S_1+S_2-1$ states.
To eliminate the latter and make $\T$ reducible, we arrange the prefactor of
${\bf \Delta}$  to annihilate non-highest spin states. Those states have
$\svec_1\cdot\svec_2 = S_1 S_2 -(S_1+S_2)$, so we must choose:
$p_2-p_1=S_1+S_2$.

The resulting transfer matrix is again of the {\em fundamental} type, but now 
with $S=S_1+S_2$. We will need its shift as well. To find it we must
evaluate (\ref{Treduce}) for the special choice of $p_2-p_1$. The only
non-trivial step is that:
\begin{equation}
P_{S=S_1+S_2}{\bf \Delta} P_{S=S_1+S_2} = \left(\frac{S_1-S_2}{S_1+S_2}
\svec\right) P_{S=S_1+S_2},
\label{projection}
\end{equation}
where $P_{S=S_1+S_2}$ is the projector onto the highest component. This is
easily proved by evaluating it on states of the type
$(S^-)^n|S^z_1=S_1,S^z_2=S_2\rangle$ and using that for $n=0$
$\Delta^z=\frac{S_1-S_2}{S_1+S_2}$ and the fact that ${\bf \Delta}$ is a vector
operator. Substituting this back into the expression for the combined
transfer matrix we have
\begin{equation}
\label{Tred}
\T_{S_1+S_2}=(u-hp_1-hS_1+\frac{h}{2})\left[(u-hp_1-hS_2)+
h(\frac{1}{2}+\sig\cdot\svec)\right].
\end{equation}
The prefactor just changes the normalization. The new shift is $p_1 + S_2$.
Note that this is not symmetric in $(p_1,p_2)$, a fundamental property of
the Yangian, where 
\mbox{$A\otimes B\, \mbox{$\simeq \hspace{-0.14truein} /$}\, B\otimes A$}
in general.
For our specific case $S_1=\half$, $S_2=\half$, $p_2-p_1=1$, we thus build a
representation containing just $S=1$ states, {\em without} any $S=0$ states
connected to them. The new shift resides
between $p_1$ and $p_2$. We can add a third spin-$\half$  by multiplying
$\T_1 \T_2 $ form the left by $\T_3$, to create a $S=\frac{3}{2}$ fundamental
representation, provided $(p_1+S_2)-p_3=(S_1+S_2)+S_3$, i.e., $p_3=p_1-1=p_2
-2$: a string of 3 points. We can repeat this $Q$ times to create a
$S=\frac{Q}{2}$ fundamental transfer matrix if the $p_1\ldots p_Q$ are all
equally spaced, and increase from the left to the right in the tensor
product. 
In other words, $\T$ acts in the space of $Q$ spins-$\half$, but leaves the
subspace with $S_{\rm Tot}=\frac{Q}{2}$ invariant, and in that subspace it
is isomorphic to a {\em single} spin $S=\frac{Q}{2}$ fundamental transfer 
matrix (\ref{evaltensor}).
The shift of the total spin $\frac{Q}{2}$ is in the center of mass of
this string of $\{ p_i\}$. All of this is in accord with what we would expect
from the spin $S$ fundamental representation, in eq.\ (\ref{STRdef}).

Going back to (\ref{evaltensor}), we can thus conclude that any string of
$Q$ consecutive Drinfel'd zeroes indeed represents a $S=\frac{Q}{2}$
multiplet. The total spin content of the multiplet is just the tensor
product of these ``string-spins''.
 The only remaining potential obstacle is that the transfer matrix 
(\ref{evaltensor}) can be reduced further.
However, Yangian representation
theory tells us that for it to be reducible the strings of $p$'s have to be
in {\em special} positions \cite{CP1990}. Two strings are in such a special
position if their union is again a string (they are collinear and ``touch'')
or if one string is a {\em proper} subset of the other. This is obviously
not the case since all strings are separated by at least two zeroes of the
polynomials $f_2(u)$ and $f_2(u-h)$---the crosses in fig.\ (\ref{figmotif}).

Similarly, we see that the NNE Yangian representation (\ref{TNNE}) is
irreducible since all its $2S$ strings have the same shifts, coincide, and
are therefore {\em not} proper subsets of each other.

\subsection{The Yangian, the Hamiltonian and Completeness.}
\label{semions}

With the knowledge that the spinon description is correct we can now show that
it is also complete \cite{H91}.  For a given $M$ (number of down-spins), we
have $N_{orb}\equiv M+1$ orbitals available for the $(N-2M)$ bosonic spinons
(see fig.\ (\ref{figmotif})), which can have spin up or down, resulting in
$\left(\rule{0em}{0.8em}^{N-2M+2(M+1)-1}_{\rule{1.9em}{0em}N-2M}\right)$
states.
Summing over $M=0,\ldots,\frac{N}{2}$ ($N$ even) we recover indeed $2^N$
states.  The fact that the number of orbitals available to the spinons changes
with $M$ and therefore with the number of spinons $N_{sp}$ itself, gives rise
to the interesting issue of one dimensional fractional statistics.  Since
the number of orbitals available to the spinons goes down by 
{\em one} if we add {\em two} spinons ($\Delta N_{sp}=-2\Delta(M+1)$) we see that
by this token, spinons interpolate halfway between bosons and fermions
and are therefore called {\em semions} \cite{H91b}.
This form of statistics, which is based on state-counting rather than
exchanging particles (as is customary in the two dimensional quantum Hall
effect), can be applied in any number of dimensions.
The fact that the Yangian algebra is the raising/lowering algebra of the
semionic spinons, means that with it we have at our disposal a
well-developed mathematical tool for probing fractional statistics
\cite{schoutens92}.

To connect every Yangian representation with a degenerate eigenvalue of
$H_{\rm ISE}$ we need to rewrite the Hamiltonian's action on YHWS 
 in terms of the Dunkl operators $\D_i$. 
Using the trick (\ref{SLACder}) to convert a convolution with $1/\sin^2$ 
into derivatives we have (in units of $\frac{v_s}{\pi})$:
\begin{eqnarray}
H_{\rm ISE} &=& \half\sum_{i\neq j} \frac{-4 z_i z_j}{(z_i-z_j)^2}(P_{ij}-1)=
\sum_i \half \D_i(\D_i-N)\nonumber\\
&=& \sum_i\half  d_i (d_i -N) \hspace{2em}d_i\in \{1,\ldots,N-1\},  
\hspace{1em} i=1,\ldots,M
\label{HISEdunkl}
\end{eqnarray} 
which reproduces the result from the Sutherland model diagonalization if we
identify the  $ d_i$ with the pseudomomenta $m_i$.  
Knowing the spectrum of $H_{\rm ISE}$ as well as the degeneracies of the
energy levels, one can obtain the thermodynamics \cite{H91}. 
In table \ref{spectrum} we have listed the entire spectrum and its
degeneracies for $N=6$ spins. 
\begin{table}
\caption{The spectrum of $H_{\rm ISE}$ for 6 spins. The notation for the
motifs in the first column is identical to that in fig. 
\protect\ref{figmotif}. The $d_i$ in column 2 are the same as the $m_i$ of
eq. (\protect\ref{energydef}), which is also used to compute the energy in
the last column.}
\vspace{1em}
\psfig{file=table1.epsi,width=\hsize}
\samepage
\comment{
\begin{tabular}{cccccc}
\hline \hline
Motif & Pseudomomenta & Drinfel'd Roots & Spin Content & Degeneracy & 
$E\, \left( (\frac{2\pi}{N})^2 \frac{v_s}{\pi}\right)$ \\
\nopagebreak
123456& $\{d_i \}$ & $\{ p_i \}$ & & $g$ & \\
\hline
XxXxXx & 1,3,5 & --- & 0 & 1 & -9.5\\
\nopagebreak
OXxXxO & 2,4 & 1,6 & $\half\otimes\half=0\oplus 1$ & 4 & -8.0\\
\nopagebreak
OOXxXx & 3,5 & 1,2 & 1 & 3 & -7.0\\
\nopagebreak
XxXxOO & 1,3 & 5,6 & 1 & 3 & -7.0\\
\nopagebreak
OXxOXx & 2,5 & 1,4 &  $\half\otimes\half=1\oplus 0$ & 4 & -6.5 \\
\nopagebreak
XxOXxO & 1,4 & 3,6 & $\half\otimes\half=1\oplus 0$ & 4 & -6.5 \\
\nopagebreak
XxOOXx & 1,5 & 3,4 & 1 & 3 & -5 \\
\nopagebreak
OOXxOO & 3 & 1,2,5,6 & $1\otimes1=2\oplus 1\oplus 0$ & 9 & -4.5\\
\nopagebreak
OXxOOO & 2 & 1,4,5,6 & $\half\otimes\frac{3}{2}=2\oplus 1$ & 8 & -4.0 \\
\nopagebreak
OOOXxO & 4 & 1,2,3,6 & $\frac{3}{2}\otimes\half=2\oplus 1$ & 8 & -4.0 \\
\nopagebreak
XxOOOO & 1 & 3,4,5,6 & 2 & 5 & -2.5 \\
\nopagebreak
OOOOXx & 5 & 1,2,3,4 & 2 & 5 & -2.5 \\
\nopagebreak
OOOOOO & --- & 1,2,3,4,5,6 & 3 & 7 & 0.0\\
\nopagebreak
       &     &             &   & $\rule{0em}{0.85em}^+ \overline{
              \rule{1.0em}{0em}64\rule{1.0em}{0em}}\rule{.6em}{0em} $ & \\
\hline
\end{tabular}
}
\label{spectrum}
\end{table}

\section{Conclusion}
We have found that in the ISE model we can separate the
Hilbertspace of $2^N$ states into groups labeled by a sequence of M integers
$\{d_i \}$  in the range $1,\ldots,N-1$ and $M=0,\ldots,\frac{N}{2}$. For
every sequence or set of pseudomomenta there is {\em one}  spin state which
is of a special polynomial character, called a YHWS. The energy of such a
state is (in units of $(\frac{2\pi}{N})\frac{v_s}{\pi}$)  $\half \sum_i d_i (d_i-N)$. The
rest of the states in that group are generated by the repeated action of
the Yangian lowering operator $T^-(u)$ on it. Those states are all
degenerate with their YHWS. The motif picture of the pseudomomenta
interprets the Yangian multiplet as $N-2M$ bosonic spin-$\half$ spinons  in
$M+1$ orbitals. In the YHWS, these spinons are all fully polarized (maximal
spin), and the subsequent action of the Yangian rotates these spins and
lowers the polarization.

For the NNE Heisenberg model, there is only one YHWS for finite $N$:
\mbox{$|\uparrow\uparrow\ldots\uparrow \rangle$} and the spinons picture 
is only approximately correct (see chapter \ref{dsf}).

\newpage

\section{Appendix}
\subsection{Appendix A: Sutherland wavefunctions are YHWS.}
\label{appA}

In this Appendix we will show that the subset of eigenstates of $H_{\rm
ISE}$ that derive from the continuum wavefunctions of the Sutherland model
are indeed YHWS. That is to say: they have to be annihilated by both
$S^+=J^+_0$ and $J^+_1$ or equivalently $\Lambda^+_{\rm ISE}$. The
Sutherland states are of the form
$\Psi(n_1,\ldots,n_M)\equiv\psi(z_{n_1},\ldots,z_{n_M})=\prod_{i<j}^M
(z_{n_i}-z_{n_j})^2\prod_{i=1}^M z_{n_i}\cdot
 \phi(z_{n_1},\ldots,z_{n_M})$, where
$\phi$ is a symmetric polynomial, such that $\psi$ is of degree less than
$N$ in every one of the $z_{n_i}$. Then the action of $S^+$ (which reduces $M$
by one)  is, using the symmetry of the wavefunction:
\begin{eqnarray}
&&(S^+\Psi)(n_1,\ldots,n_{M-1})=\sum_{j=1}^N\Psi(n_1,\ldots,n_{M-1},j)=
\sum_{j=1}^N\psi(z_{n_1},\ldots,z_{n_{M-1}},e^{\frac{2\pi ij}{N}})\nonumber\\
&&= \psi(z_{n_1},\ldots,z_{n_{M-1}},0)\equiv  0 .
\label{Spluseq}
\end{eqnarray}
The last step holds because the sum over $j$ is just a Fourier transform
at momentum 0 on the last entry of $\psi$, which picks out the (vanishing) 
constant part in $\psi$.

The second Yangian raising operator is $\Lambda^+\propto\sum_{i\neq j}w_{ij}
 S^+_i S^z_j$. In a basis of down-spins $S^z_i$ is given by:
$\half- \sum_{j=1}^M \delta_{i,n_j}$. Then: 
\begin{eqnarray}
(\Lambda^+\Psi)(n_1,\ldots,n_{M-1})&=&\half\sum_{i\neq \{n_p\} }^N
\left(\sum_{j(\neq i)}^N w_{ij}\right) \Psi(n_1,\ldots,n_{M-1},i)\nonumber\\
&&-\sum_{k=1}^M\sum_{i\neq \{ n_p\} }^N w_{in_k}
\psi(z_{n_1},\ldots,z_{n_{M-1}},
z_i)
\label{szeq}
\end{eqnarray}
The first term vanishes because the the sum over the odd function $w_{ij}$ 
is zero. We can simplify the second term, involving the convolution with 
$w_{ij}$,  with the following identity, inspired by eq.\ (\ref{SLACder}):
\begin{equation}
\label{cotconv}
\sum_{i=1}^N w_{ij} P(z_i) = N P(z_j) -2z_j P'(z_j)-NP(0)
\end{equation}
($P'$ is the derivative of $P$) if $P$ is a polynomial of degree less than
$N$. The $z=0$ term vanishes as with $S^+\Psi$, and  the other two
are zero as well since $\Psi$ has a {\em double} zero when two of its arguments
coincide. 

\subsection{Appendix B: $T^{22}(u)$ Acting on a YHWS.}
\label{appB}

We will follow the derivation in \cite{BPS93}. In a basis of states $\{
|n_1,\ldots ,n_M \rangle\}$ 
where the $n_i$ label the positions of the down-spins,
we consider the action of $T^{22}(u)$ (we set $h=1$ during the calculation):
\begin{equation}
T^{22}(u)=\unity + \sum_{n=0}^\infty \frac{1}{u^{n+1}}\sum_{i,j=1}^N
\left( L^n\right)_{ij} X^{22}_j
\label{t22action}
\end{equation}
with $L_{kl}=h(1-\delta_{kl})\theta_{kl} P_{kl}$;
$\theta_{kl}=\frac{z_k}{z_k-z_l}$, and the $z$'s lie on equidistant points on
a circle. $X^{22}_j$ is the operator  that projects out the states with a
down-spin on site $j$. We let $T^{22}(u)$ act on a polynomial type state
$\psi(z_{n_1},\ldots,z_{n_M})$ of which the only constraint is that it is 
symmetric
and vanishes
when one of the $z_i$ become identical or equal to one of the other $z_j$. 
Thus the powers of the $z_i$ in
every term are at least one. This corresponds to $\psi$ being a maximal
$S^z$ state (see Appendix \ref{appA}). Under these circumstances we have
(eq.\ \ref{cotconv}): $\sum_{k(\neq n)}\theta_{kn}P(z_k)=z_k
P'(z_k)-\frac{N+1}{2} P(z_k)$, if $P$ is of degree $\leq N-1$.
Evaluating the $n=0$ term in (\ref{t22action}) is trivial. The $n=1$ term
for {\em fixed} $i$ $\sum_{j}'\theta_{ij} P_{ij} X^{22}_j$ gives:
\begin{eqnarray}
\Psi^{(1)}(i|n_2,\ldots,n_M) &=&
\langle i,n_2,\ldots,n_M|\sum_j \rule{0in}{1.3ex}^{'}
\theta_{ij}P_{ij}X^{22}_j|\psi\rangle\nonumber\\
&=& \sum_{j=1}^N \rule{0in}{1.3ex}^{'}\theta_{ij}
\Psi(j,n_2,\ldots,n_M)+\sum_j \theta_{in_j}
\Psi(i,n_2,\ldots,n_M)\nonumber\\
&=&(z_i\del_i -\frac{N+1}{2})\psi(z_i|z_{n_2},\ldots,z_{n_M}) +
\sum_j \rule{0in}{1.3ex}^{'} \theta_{in_j}\Psi(i,n_2,\ldots,n_M)\nonumber\\
&&
\label{L1}
\end{eqnarray}
We have used the fact that $\psi$ is symmetric in its entries. The prime on
the summation symbols indicates that the terms with diagonal elements of
$\theta$ are absent. The second term comes from the exchange from two down-spins 
and the first is the result of an up- and a down-spin trading places.
Notice that the wavefunction $\Psi^1$ is {\em only} symmetric in entries 2
through $M$, and has one down-spin fixed at the (parametric) site $i$. 

For the $n=2$ term we have to convolve this set of functions with $L$
resulting in :
\begin{eqnarray}
\Psi^{(2)}(i|n_2,\ldots,n_M)&=& \langle i,n_2,\ldots,n_M|\sum_{jk} 
\rule{0in}{1.3ex}^{'}L_{ij}
L_{jk}X^{22}_k|\psi \rangle \nonumber\\
&=& h\sum_j\rule{0in}{1.3ex}^{'}
 \theta_{ij}\Psi^{(1)}(j|n_2,\ldots,n_M)+h\sum_{l=2}^M\theta_{in_l}
\Psi^{(1)}(n_l|n_2,..i..,n_M)\nonumber\\
& \equiv & h\left( (D_1-\frac{N+1}{2})\Psi^{(1)}\right) (i|n_2,\ldots,n_M).
\label{L2}
\end{eqnarray}
$D_j =z_j \del_i +\sum_{k(\neq j)}K_{jk}$. The $\{K_{ij}\}$ are 
the same operators that are used in the definition of the Dunkl operators
(\ref{dunkldef}). 
$\Psi^{(2)}$ is obviously symmetric in entries $2,\ldots,M$ as well 
and we can now do all orders inductively. 
Finally, we have to do the sum over all locations
of the spin ``$i$'' in (\ref{t22action}). Since the down-spins are
indistinguishable and now unrestricted, we are allowed to symmetrize
$\Psi^{(n)}(n_1,\ldots,n_M)$ (calling final summation index $i$ now $n_1$). 
This is done by applying $\sum_{j=1}^M K_{1j}$ where $K_{11}=\unity$.
It is easy to verify that $K_{ij}D_j = D_i K_{ij}$ so with
$K_{ij}\Psi^{(0)}=\Psi^{(0)}$:
\begin{equation}
T^{22}(u)=1+\sum_{n=0}^N \frac{h^n}{u^{n+1}}\sum_{j=1}^M D_j^n=
1+\sum_{j=1}^M \frac{1}{u-hD_j}
\label{t22sum}
\end{equation}
It can be shown that the summation (\ref{t22sum}) can be brought into one
fraction by exploiting a special case of the  Calogero-Sutherland model with
spin (see chapter \ref{invar}, eq.\ (\ref{equiva}))
by acting on a state in which
all particles have their spin up. 
\begin{equation}
T^{22}(u)=\prod_{j=1}^{M}\left(1+\frac{1}{u-h(\D_j-\frac{N+1}{2})}\right)
\label{t22product}
\end{equation}
with 
\begin{eqnarray}
\D_j&=&z_j\del_j + \sum_k \rule{0in}{1.3ex}^{'}
\theta_{jk}K_{jk}-\sum_{k(<j)}K_{jk}\nonumber\\
&=&z_j\del_j +\half\sum_k\left(w_{jk}-{\rm sgn}(j-k)\right)K_{jk}
\label{dunkl2def}
\end{eqnarray}

\subsection{Appendix C: Dunkl Operators and their
Eigenfunctions.}
\label{appC}

The Dunkl operators from eqs.\ (\ref{t22def}, \ref{dunkl2def}) are given for
arbitrary {\em odd} $\lambda$ by $\D_i=z_i\del_i +
\frac{\lambda}{2}\sum_{j(\neq
i)}\left(w_{ij}-{\rm sgn}(i-j)\right)K_{ij}$. We have introduced the free
parameter $\lambda$ so that we can accommodate $\lambda=1$ in chapter 
\ref{introduction1}, and $\lambda\rightarrow\infty$ in chapter \ref{invar}.
Even with arbitrary $\lambda$ the $\D_i$ still satisfy the Hecke algebra
relations (see chapter \ref{invar}). We will follow the derivation of
\cite{BGHP93}.

We expect the eigenstates to
be polynomials, but we have to select some subset of polynomials in order
for the $\frac{z_i+z_j}{z_i-z_j}$ term  in $\D_i$ to still produce a
polynomial.  Therefore we choose the form $\psi(z_1,\ldots,z_M)=
\psi_0(z_1,\ldots,z_M)\phi(z_1,\ldots,z_M)$ where
$\psi_0=\prod_{i<j}(z_i-z_j)^\lambda$ to cancel the pole when two $z$'s
come together, and $\phi(z_1,\ldots,z_M)=\prod_{i=1}^M (z_i)^{n_i}$. The
action of $\D_i$ on $\phi$ is then 
\begin{eqnarray}
\psi^{-1}_0\cdot\D_i (\psi_0\phi)&=& \left(z_i\del_i
-\frac{\lambda}{2}\sum_{j(\neq i)} w_{ij} (K_{ij}-1)\right.\nonumber\\
&&+\left. \frac{\lambda}{2}\sum_{j(\neq i)}{\rm sgn}(i-j) (K_{ij}-1) + 
\frac{M-1}{2} + i -\frac{M+1}{2}\right)\phi\nonumber\\
&\equiv & \D_i' \phi
\label{dgauge}
\end{eqnarray}
since $\half\sum_j \rule{0in}{1.3ex}^{'}
{\rm sgn}(i-j) = i-\frac{M+1}{2}$. Now, for $n_i\neq n_j$
(if $n_i=n_j$ the result is obviously 0):
\begin{eqnarray}
&&\left(\sum_{j(\neq i)}w_{ij} (K_{ij}-1)\right)z_i^{n_i} 
z_j^{n_j}=\nonumber\\
&& {\rm sgn}(i-j)\left( z_i^{n_i}z_j^{n_j} +z_i^{n_j}z_j^{n_i} + 
2\left\{\begin{array}{l}\sum_{k=1}^{n_i-n_j-1}z_i^{n_i-k}z_j^{n_j+k};\,\,
n_i >n_j\\
\sum_{k=1}^{n_j-n_i-1}z_i^{n_i+k}z_j^{n_j-k};\,\, n_i<n_j
\end{array}\right.\right),
\label{squeezing}
\end{eqnarray}
we see that the action of one of the non-diagonal terms is to produce a
series of squeezed polynomials, in the sense that the degree remains the
same, but the powers get closer (the high ones decrease, and the low ones
increase). Inspired by the solution of the Sutherland model we can attempt
to define a hierarchy on the partitions $|n|=\{ n_1\geq n_2\ldots\geq n_M\}$
such that $\D_i$ becomes triangular, and we can read off its eigenvalues
from the diagonal. In anticipation of what is to come, we group the
polynomials into blocks which have the same set of powers $\{n_i \}$  
but differ in
their ordering. For two partitions $|n|$ and $|m|$ we say that $|n|>|m|$ if,
counting from the left, for the first unequal pair $(n_i,m_i)$ holds
$n_i>m_i$, e.g.\ $\{7,4,3,2\} > \{6,5,5,5\}$. 
\begin{wrapfigure}{r}{2in}
\framebox[2in]{\centerline{\psfig{file=blocktriang.epsi,width=1.8in}}}
\end{wrapfigure}
``Squeezing'' obviously only
produces states in lower partitions. The ${\rm sgn}(i-j)(K_{ij}-1) $ term just
connects states in the same partition. 
So $\D_i$ is block-triangular in this
basis, and we only need to make the blocks on the diagonal, which contain only
matrix elements of states of the  same partition, triangular.
In such a block, states are all labeled by permutations $P\in S_M$. Using
(\ref{dgauge}) and (\ref{squeezing}) and forgetting terms with the wrong
partition, we have

\begin{eqnarray}
\D_i'z_i^{n_{P(i)}}z_j^{n_{P(j)}}&=&z_i^{n_{P(i)}}z_j^{n_{P(j)}}\left(
n_{P(i)}+
{\scriptstyle \frac{M+1}{2}+i-\frac{M+1}{2}}\right) \nonumber\\
&&+ z_i^{n_{P(i)}}z_j^{n_{P(j)}} [{\rm sgn}(n_{P(i)}-n_{P(j)})-{\rm sgn}(i-j)]
\nonumber\\
&&+ z_i^{n_{P(j)}}z_j^{n_{P(i)}}[{\rm sgn}(n_{P(i)}-n_{P(j)})+{\rm sgn}(i-j)]
\label{dpart}
\end{eqnarray}
The third, non-diagonal, term is non-zero only when $i>j$ and 
$n_{P(i)}<n_{P(j)}$ or $i<j$ and $n_{P(i)}>n_{P(j)}$; in either case, when
ordered by $i,j$ the power on the right is always goes {\em down} in the
application of $\D_i'$. So if the permutations are ordered by looking at the
{\em last} non-identical pair $n_{Q(i)}>n_{P(i)}$, implying that 
$|n_Q|>|n_P|$, $D_i'$ is also triangular inside the blocks. The eigenvalues
on the diagonal read:
\begin{eqnarray}
d_i^{|n|,P}&=&n_{P(i)}-\frac{\lambda}{2}\sum_{j(\neq i)}{\rm
sgn}(n_{P(i)}-n_{P(j)}) -\frac{\lambda}{2}\sum_{j(\neq i)} {\rm sgn}(i-j) +
\lambda{\scriptstyle \left(\frac{M+1}{2}+i-\frac{M-1}{2}\right)}\nonumber\\
&=& n_{P(i)}+\lambda (M-P(i)).
\label{dev}
\end{eqnarray}
To evaluate the sum over $n_{P(i)}$ we first pick $P=\unity$ and then apply
the permutation $P$, which is trivial since $\sum_j\rule{0in}{1.3ex}^{'}$ is invariant under
permutations. Notice that the set of eigenvalues $\{ d_i^{|n|,P}\}_{i=1}^M$
for fixed $M,P\in S_M$ is the same for {\em all} $\D_i$. 

If two of the powers in the expression for $\phi$ are identical (\ref{dev})
is no longer valid, but in our case $\phi$ will be taken to be antisymmetric
so this will not cause problems. Note that the highest power occurring in the
{\em actual} eigenstate labeled by partition $|n|$ is max$\{ n_i\} + M-1$.

\def\bsigma{\mbox{\boldmath$\sigma$}}
\def\half{\mbox{$\frac{1}{2}$}}
\def\hd{\hat{\Delta}}
\def\hg{\hat{\gamma}}
\def\htrans{\hat{T}}

\chapter{Integrals of motion of the ISE Model}
\label{invar}

\section{Introduction}
\label{intro2}

In the previous chapter we presented two different spin-model
representations of the
Yangian Algebra, $\T_{\rm NNE}(u)$ , eq.\
(\ref{TNNE}), and $\T_{\rm ISE}(u)$, eqs.\ (\ref{Ldef}), (\ref{TISE}).
Via the algebraic Bethe Ansatz eq.\ (\ref{ABEdef}) we can construct a
$2^N$-dimensional basis of the Hilbert space. For $\T_{\rm NNE}$ we need
only one HWS: $| \uparrow\uparrow\cdots\uparrow\rangle$, but for
$\T_{\rm ISE}$, being reducible, we needed several YHWS, viz. the set of
polynomial type states (\ref{HWSeval}), (\ref{t22def}) (which are 
are also, miraculously, eigenstates of $H_{\rm ISE}$).

In general,
in choosing a different basis for the Hilbert space we haven't said
anything about the action of a particular Hamiltonian in
that space. This is where the power of the transfer matrix method comes in:
the specific way in which the ABA basis states were constructed was by
forcing them to be of the form (\ref{ABEdef}), with $\lambda_i$'s such
that they were eigenstates of $T^0(u)$ $\forall_{u}$. This automatically
leads to Bethe Ansatz eqs.\ (\ref{ABAeq}). 
This somewhat unintuitive constraint becomes
{\em imperative} once we realize that $H_{\rm NNE}$ is ``contained'' in
$T^0(u)$ in the following sense: as elucidated in the previous chapter
$[T^0(u),T^0(v)]=0\Leftrightarrow [J^0_m,J^0_n]=0\: \forall_{m,n}$. The
 (hermitean) $J^0_n$ form a set of commuting operators. One
particular combination of them: $\left.\frac{d}{du}\left\{ 
\ln(T^0(u))\right\}\right|_{u=0} \propto H_{\rm NNE}$. The ABA basis is
thus a set of eigenvectors that simultaneously diagonalize $H_{\rm NNE}$
{\em and}  the other (combinations of the) $J^0_n$. Since all $J^0_n$
commute with $H_{\rm NNE}$ they are conserved and therefore called {\em
constants of motion} or {\em invariants}. For $\T_{\rm NNE}$ total spin,
$\svec^2$, is in that set, as is is $P_{123\cdots N}$, the
momentum/translation operator, as expected. In short, diagonalizing the
NNE Hamiltonian is
taken care of by eq.\ (\ref{ABAeq}) with the added bonus of obtaining the
constants of the motion and their eigenvalues.

For the ISE model we can still construct the ABA basis states, which are
also orthogonal, but from the point of view of diagonalizing $H_{\rm
ISE}$ this achieves little since $[H_{\rm ISE},\T_{\rm ISE}(u)]=0$. Since
it commutes with $\T_{\rm ISE}$, $H_{\rm ISE}$ cannot be obtained from
${\rm Tr}(T_{\rm ISE}(u))$ since $[{\rm Tr}(\T(u)),\T(v)]$ $\neq 0$!
$H_{\rm ISE}$ must have a different origin. Also the question of
additional constants of motion in the ISE model isn't resolved.  We
already mentioned one invariant: $H_3$ in eq.\ (\ref{H3def}), conjectured
in \cite{Ino90}, which, as it turns out, also commutes with $H_{\rm ISE}$.
Based on the same work, an
additional invariant $H_4$ was conjectured \cite{HHTBP92}, but this was not
the result of a systematic algorithm.

Minahan and Fowler \cite{MF93} and Sutherland and Shastry \cite{SuthSh93} 
introduced sets of invariants that
commute
with the Hamiltonian, based on operators introduced by Polychronakos
\cite{Poly92}.  However, the generating functions for these sets are
essentially
the trace of the transfer matrix and thus contain only elements of
the
Yangian algebra\footnote{The authors of Ref.\protect\cite{MF93}
claim to have
found the Hamiltonian $H_2$ in their third order invariant, but in
reproducing
their algebra we did not find any such term; in fact we only
recovered Yangian
operators.}.

In this chapter we will construct the rest of the set $\{ H_n\}$ of extensive
operators that
commute among themselves and with the Yangian.  In order to do this
we have to
consider a more general {\em dynamical} model in which the particles
are
allowed to move along the ring:  the Calogero Sutherland model (CSM)
with an
internal degree of freedom.  This model, which has been studied in
\cite{Poly92,HH92} has the following Hamiltonian:

\begin{equation}
H_{\rm CSM}=\sum_{j=1}^{N} \hbar^2 \left(z_j \frac{\partial}{\partial
z_j}\right)^{2} -
\sum_{i\neq j} h (h +P_{ij} ) \frac{z_i z_j}{(z_i -z_j
)^2}.
\label{dynmodeldef}
\end{equation}
$P_{ij}$ permutes the spin of {\em particles} rather than sites $i$ and $j$.
When $h\rightarrow\infty$ or equivalently $\hbar\rightarrow 0$,
the
particles `freeze' into their classical equilibrium positions, and
barring some
subtleties we recover the spin Hamiltonian $H_2$.

The reason for this diversion through the dynamical model to obtain
the
constants of motion is the following:  the so-called {\em quantum
determinant}
of the transfer matrix, which commutes with the Yangian
algebra and
therefore a natural candidate for the generating function of the
constants of
motion, happens to be {\em c-number} in the spin model (i.e.\ when
$\hbar\rightarrow 0$), as we will verify again below.  But in the general
dynamical
model, this is not the case, and by carefully taking the limit
$h\rightarrow\infty$ we can isolate a generating function for
the $\{
H_n\}$.

We will first review briefly the integrability structure of the CSM which 
bears great resemblance to that of the ISE model.  Then in section
\ref{Hn} we will construct the constants of motion. Finally section
\ref{Hnev} will show us how to obtain the eigenvalues of all the constants
of motion.  Unless noted otherwise, we will use general $SU(p)$, $p\geq
2$ spins (i.e.\ spins that can have $p$ values rather than 2), since most
results in this chapter generalize away form the $SU(2)$ case quite
easily.

\section[Integrability in the Calogero  Sutherland Model]{
Integrability in the Calogero  Sutherland \protect\linebreak
 Model}
\label{intCSM}

Let us first review the r\^{o}le of the Yangian algebra in the
{\em dynamical} model.
A more extensive treatment can be found in \cite{H94,BGHP93}.  The
integrability of the CSM is also based on the existence of a
transfer matrix
$T^{ab}(u)$ that commutes with the CSM-Hamiltonian:

\begin{eqnarray}
T^{ab}(u) &=& \delta^{ab} + \sum_{n=0}^{\infty}
\frac{h}{u^{n+1}}
T^{ab}_n \nonumber\\ T^{ab}_{n} &=& \sum_{i,j=1}^{N}X^{ab}_i (L^n
)_{ij}\nonumber\\ L_{ij}
&=&
\delta_{ij}z_j\partial_{z_j}+(1-\delta_{ij})h\theta_{ij}P_{ij},
\label{tdyndef}
\end{eqnarray}
where $X^{ab}_j$, $a,b=1,\ldots ,p$ acts as $|a\rangle\langle b|$ on
the spin
of particle $i$, and $\theta_{ij}=\frac{z_i}{z_i-z_j}$, as usual.
This transfer matrix satisfies the Yang-Baxter equation as well:

\begin{equation}
R_{00'}(u-v) \T^{0}(u) \T^{0'}(v) = \T^{0'}(v) \T^{0}(u) R_{00'}(u-v),
\label{YBdef}
\end{equation}
with $R_{00'}(u)=u+h P_{00'}$ and $T^0 (u)=T(u)\otimes 1$ and
$T^{0'}(u)=1\otimes T(u)$.  For the purposes of this chapter we will
deal with
another form of the same transfer matrix.  We reiterate the 
representation of the so-called {\em Dunkl}-operators of eq.
(\ref{t22def}):
\begin{eqnarray}
\hat{D}_i &\equiv&\hbar z_i\partial_{z_i} + \hat{\gamma}_i = \hbar
z_i\partial_{z_i} + \frac{h}{2}\sum_{j(\neq i)}
(w_{ij}-{\rm sgn}(i-j))K_{ij} \nonumber\\
w_{ij} & = & \frac{z_i+z_j}{z_i-z_j}
\end{eqnarray}
We have made the scale of $\hat{D}_i$ explicit by putting a factor
$\hbar\neq 0$ in front of $z_p \del_p$. The idea is that eventually
$\hbar\rightarrow 0$. 
$K_{ij}$ is the operator that permutes the {\em spatial} co-ordinates
of
particles
$i$ and $j$. These Dunkl-operators commute:

\begin{equation}
[\hat{D}_i ,\hat{D}_j ] = 0,
\end{equation}
but are not covariant under permutations:

\begin{eqnarray}
&& [K_{i,i+1},\hat{D}_k] = 0 \;\;\;{\rm if}\; k\neq i,i+1\nonumber\\
&& K_{i,i+1}\hat{D}_i - \hat{D}_{i+1} K_{i,i+1} = h,
\label{dunklalg}
\end{eqnarray}
defining a so-called {\em degenerate affine Hecke Algebra}.
In terms of these Dunkl operators we can define a transfer matrix
that also
obeys the Yang-Baxter equation:

\begin{equation}
\hat{\T}^0 (u) = \left(1+\frac{h
P_{01}}{u-\hat{D}_1}\right)\cdots
\left(1+\frac{h P_{0N}}{u-\hat{D}_N}\right).
\label{equiva}
\end{equation}
It satisfies eq.\ (\ref{YBdef}) trivially since the $\{ \hat{D}_i \}$
commute
amongst themselves and commute with the $P_{0j}$, since the latter
only act on
spin degrees of freedom; furthermore
$\left(1+\frac{h P_{0i}}{u-\hat{D}_i}\right)$ is the elementary
transfer
matrix with spectral parameter $u-\hat{D}_i$.  To retrieve $\T^0 (u)$
from
$\hat{\T}^0 (u)$ we have to apply a projection $\Pi$, familiar from
chapter \ref{chapter1} to  $\hat{\T}^0$.
It
replaces every occurrence of $K_{i,i+1}$ by $P_{i,i+1}$ once ordered
to the
right of an expression (this is equivalent to having the unprojected
operator
act on wavefunctions that are symmetric under simultaneous
permutations of
spin- and spatial co-ordinates) \cite{BGHP93}. 
It is now possible to validate eq.\ (\ref{t22product}) in Appendix B of
chapter \ref{chapter1}: 
it follows from (\ref{equiva}) and (\ref{tdyndef}) 
by having it act on a state where all
dynamical particles have the same spin (say up).
We will drop the ``0''
superscript on $\hat{\T}(u)$ from here on.

\section{Construction of the Constants of Motion}
\label{Hn}

It was shown in \cite{BGHP93}, that this dynamical transfer 
matrix commutes with
the CSM Hamiltonian. 
As does the transfer matrix, the quantum determinant must commute
with $H_{\rm ISE}$. It is given in its full $SU(p)$ glory by:
\begin{equation}
{\rm Det}_q (\T(u)) = \sum_{\sigma\in S_p}\epsilon(\sigma )
T_{1\sigma_1}(u)T_{2\sigma_2}(u-h )\cdots
T_{p\sigma_p}(u-h(p-1)).
\end{equation}
It satisfies $[\T(u),{\rm Det}_q(\T(u)) ]=0$ and has been computed in
\cite{BGHP93} as:

\begin{equation}
{\rm Det}_q (T(u)) = \Pi\: {\rm Det}_q (\hat{T}(u))\Pi = \Pi\:\left(
\frac{\hat{\Delta}(u+h,\hbar)}{\hat{\Delta}
(u,\hbar)}\right)\Pi.
\label{tdelta}
\end{equation}
where (making the dependence on $\hbar$ explicit):

\begin{equation}
\hd (u,\hbar )=\prod_{i} (u-\hat{D}_i(\hbar )).
\label{deltadef}
\end{equation}
Now obviously:

\begin{equation}
[\hd (u,\hbar),\hat{T} (v,\hbar) ] = 0.
\label{dtcommut}
\end{equation}
This holds since the $\hat{D}_i$'s commute with each other and the
$P_{0j}$'s .
The projector doesn't get in the way since a product of projections
is the
projection of the product---both $\htrans(v)$ and $\hd(u)$ are
symmetric under
simultaneous permutation of spin- and spatial co-ordinates
\cite{BGHP93}.  Since we know the eigenvalues of the commuting set $\hat{D}_i$ 
form Appendix C of chapter \ref{chapter1}, the
eigenvalues of $\hd(u)$ are also known:  for every partition $|n|$
there is an
eigenvalue:

\begin{equation}
\delta^{|n|}(u) = \prod_{j=1}^{N}(u-\hbar n_j -h
(j-\frac{N+1}{2}))
\label{deltaev}
\end{equation}
We notice that as $\hbar\rightarrow 0$, i.e.\ in the limit of the
ISE model, all
eigenvalues become identical and $\hd(u,0)$ is a {\em multiple of the
identity operator} (compare eq.\ (\ref{detqise}).  Thus no
non-trivial constants of motion are contained in $\hd(u,0)$.
Nevertheless let
us study (\ref{dtcommut}) for small $\hbar$.  Writing
$\hat{\T}(v,\hbar)=\sum_{n}\hbar^n \hat{\T}_n (v)$;
$\hd(u,\hbar)=\sum_{n}\hbar^n
\hd_n(u)$:

\begin{eqnarray}
0 &=& [\hat{\T}(v,\hbar),\hd(u,\hbar)]  \nonumber\\
&=&[\hat{\T}_0(v),\hd_0(u)]+\hbar\left([\hat{\T}_0(v),\hd_1(u)]
+[\hat{\T}_1(v),\hd_0(u)]\right) +
{\cal O}(\hbar^2).
\label{hbarexp}
\end{eqnarray}
The ${\cal O}(\hbar ^0)$ term is trivially 0.  The rest of this
letter will
focus on the vanishing of the ${\cal O}(\hbar)$ term. As we shall
show below:
$[\hat{\T}_1(v),\hd_0(u)]=0$.  Therefore we have the important result:
$[\hat{\T}_0(v),\hd_1(u)]=0$, i.e.\ the ${\cal O}(\hbar)$ term in
$\hd(u,\hbar)$
commutes with the transfer matrix (and therefore the Yangian) of the
ISE spin model.  Furthermore it will also become
apparent that

\begin{equation}
[ \hd_1(u),\hd_1(u') ] =0 .
\end{equation}
So we can take $\hd_1(u)$ to be the generating function of the
constants of
motion of the ISE model!  In order to establish these results we first need
to prove
the following corollary:  $z_i\partial_{z_i} \hd_0(u)=0$ when
evaluated with the particles at their equilibrium positions, i.e.\
$z_j=\exp \left( \frac{2\pi i}{N} j\right)$.

{}From eq.\ (\ref{deltadef}) we have $\hd_0(u)=\prod_{i} (u-\hg_i)$.
Since we
know that $\hd_0(u)$ is scalar we can evaluate it by having it act on
any
convenient state, e.g.\ the one where all particles have identical
values for
their internal degree of freedom (for $p=2$ we would say:  all spins
pointing
up).  I.e.\ the permutations reduce to 1.  This has been done in
\cite{BGHP93}:

\begin{eqnarray}
\hd_0(u)& = & {\rm det}(u- \Theta)\nonumber\\
\Theta_{ij} & = & \frac{h z_i}{z_i -z_j} (1-\delta_{ij}).
\end{eqnarray}
Then using $\partial_x {\rm det}A(x) = {\rm Tr}[A^{-1}\partial_x
A(x)] {\rm
det}[A(x)]$ we have:

\begin{eqnarray}
-\partial_{z_i}\hd_0(u) &=&
{\rm Tr}\left[\frac{1}{u-\Theta}\partial_{z_i}\Theta\right]{\rm det}
[u-\Theta]
\nonumber\\
&=& \sum_{n=0}^{\infty} u^{-(n+1)}
{\rm Tr}\left[\Theta^n\partial_{z_i}\Theta\right] {\rm det}[u-\Theta
].
\end{eqnarray}
Now evaluate the trace in a basis where $\Theta$ is diagonal. This
can clearly
be done for $\hbar\rightarrow 0$ and $z_j = \exp \left( \frac{2\pi
i}{N}
j\right)$. $\Theta$ has eigenvectors $\psi_n$ , where $(\psi_n)_j =
\frac{1}{\sqrt{N}}\exp \left(\frac{2\pi i}{N} jn\right)$, with
eigenvalue
$h\left(\frac{N+1}{2}-n\right)$. Then

\begin{equation}
{\rm Tr}[\Theta^m \partial_{z_i}\Theta ] =
\sum_{n=1}^{N}\langle\psi_n|\partial_{z_i}\Theta|\psi_n\rangle
h^m \left(\frac{N+1}{2}-n\right)^m.
\end{equation}
Using that
\begin{equation}
(\partial_{z_i}\Theta )_{jk} =  \left\{
\begin{array}{ll}
\frac{h}{(z_i -z_j)^2}(z_j\delta_{ki}-z_k\delta_{ij}) & j\neq k
\\
0 & \mbox{otherwise}
\end{array}
\right.
\end{equation}
we find:

\begin{eqnarray}
\langle\psi_n |\partial_{z_i}\Theta |\psi_n\rangle &=&
\sum_{p(\neq i )}\frac{h}{N}
\frac{e^{\frac{2\pi i}{N}(i-p)n} - e^{\frac{2\pi i}{N}(p-i)n}}{\left(
e^{\frac{2\pi i}{N} p}-e^{\frac{2\pi i}{N} i}\right)^2} e^{\frac{2\pi
i}{N} p}\nonumber\\
&=&\frac{ih}{2N}\sum_{p=1}^{N-1}\frac{\sin\left(\frac{2\pi
pn}{N}\right)}{\sin^2\left( \frac{\pi p}{N}\right)} \equiv 0.
\end{eqnarray}
Therefore $\partial_{z_i}\hd_0(u) = 0$ at $z_j=\exp \left( \frac{2\pi
i}{N}j\right)$.

{}From expanding $\hd(u,\hbar)$ to ${\cal O}(\hbar)$ in
(\ref{deltadef}) we have

\begin{equation}
 \hd_1(u) = \sum_{i=1}^{N}\left(\prod_{j=1}^{i-1}(u-\hg_j)\right)
\: z_i\partial_{z_i} \: \left(\prod_{j=i+1}^{N}(u-\hg_j)\right).
\end{equation}
Then, with the corollary and the fact that $[\hg_i ,\hg_j ]=0$ for
all $i,j$
(the Dunkl algebra (\ref{dunklalg}) is satisfied for $\hbar =0$ as
well):

\begin{eqnarray}
[\hd_1(u),\hd_0(u)]&=&\sum_{i=1}^{N}\left(\prod_{j=1}^{i-1}
(u-\hg_j)\right)
[z_i\partial_{z_i},\hd_{0}(u)]\left(\prod_{j=i+1}^{N}(u-\hg_j)\right)
\nonumber\\
&=&\sum_{i=1}^{N}\left(\prod_{j=1}^{i-1}(u-\hg_j)\right)
\left\{ z_i\partial_{z_i} \hd_0(u)\right\} \left(\prod_{j=i+1}^{N}
(u-\hg_i)\right) = 0
\label{delzeroone}
\end{eqnarray}
With this result we find:

\begin{eqnarray}
\frac{1}{\hd(u,\hbar)}&=&\frac{1}{\hd_0(u)}-\hbar
\frac{1}{\hd_0(u)}\hd_1(u)
\frac{1}{\hd_0(u)} + {\cal O}(\hbar^2) \nonumber\\
&=& \frac{1}{\hd_0(u)}-\hbar
\frac{1}{\left(\hd_0(u)\right)^2}\hd_1(u) +
{\cal O}(\hbar^2).
\end{eqnarray}
This can be checked by multiplying the LHS and RHS by $\hd(u,\hbar)$.
Now let
us expand $\hat{\T}(v,\hbar)$ to ${\cal O}(\hbar)$:

\begin{eqnarray}
\hat{\T}(v,\hbar)&=& \prod_{i=1}^{N} \left( 1+\frac{h
P_{0i}}{v-\hat{D}_i}
\right) =
\frac{1}{\hd(v,\hbar)}\prod_{i=1}^{N} (v-\hat{D}_i +h P_{0i})\nonumber\\
&=& \frac{1}{\hd_0(v)}\prod_{i=1}^{N}(u-\hg_i +h P_{0i}) +
\nonumber\\
&& +
\hbar\left\{-\left(\frac{1}{\hd_0(v)}\right)^2
\hd_1(v)\prod_{i=1}^{N}\left( v-\hg_i
+h P_{0i}\right)\right. \nonumber\\
&& \left. -\frac{1}{\hd_0(v)}
\sum_{i=1}^{N}\left(\prod_{j=1}^{i-1}(v-\hg_j +h P_{0j})\right)
z_i\partial_{z_i} \left(\prod_{j=i+1}^{N}(v-\hg_j +h
P_{0j})\right)
\right\}+{\cal O}(\hbar^2)\nonumber\\
&\equiv& \hat{\T}_0(v)+\hbar\hat{\T}_1(v) + {\cal O}(\hbar^2)
\label{trmexpand}
\end{eqnarray}
It is now obvious how $[\hat{\T}_1(v),\hd_0(u)]=0$.  The contribution
from the
first term in curly brackets in (\ref{trmexpand}) vanishes by virtue
of eq.\
(\ref{delzeroone}).  As far as the second term is concerned:  the
$\hg_i$
commute amongst each other and with the $P_{0j}$, and
$[z_i\partial_{z_i},\hd_0(u)]=0$ as we showed before.

So far we have established that $\hd_1(u)$ respects the Yangian
symmetry, but
we also need to show that it is a {\em good} generator of constants
of motion
in that it commutes with itself at a different value of the parameter
$u$:
$[\hd_1(u),\hd_1(u')]=0$.  It will be enough to prove this on the
space spanned
by the Yangian Highest Weight states, the highest weight states of a 
representation of the Yangian
algebra.
All other states in the model are generated by acting on these YHWS
with the
elements of the Yangian algebra, i.e.\ the transfer matrix.  Since
$\hd_1(u)$
commutes with $\htrans_0(v)$ for any $u$, $[\hd_1(u),\hd_1(u')]=0$
will
therefore also hold on the non-highest weight states.  First of all
we should
note that $\hd_1(u)$ and $\hd_1(u')$ leave the space of YHWS
invariant.  This
follows from the fact that all such states $|\Gamma\rangle$ are
annihilated by
$\htrans_0^{ab}(v)$ with $a<b$ (see ref.  \cite{BGHP93}).  But since
$\htrans_0^{ab}(v)$ commutes with $\hd_1(u)$,
$\htrans_0^{ab}\left(\hd_1(u)|\Gamma\rangle\right)$ will also be 0
for $a<b$.

The proof that $\hd_1(u)$and $\hd_1(u')$ commute hinges on the
existence of an
operator in the $SU(p)$ ISE model 
that commutes with both these $\hd_1$'s and is {\em
non-degenerate} in the space of YHWS.
To find such an operator we need to review quickly how the action of the
transfer matrix on a YHWS generalizes from $SU(2)$, eq.\ (\ref{HWSeval})
to $SU(p)$. In that case \cite{CP91,BGHP93}:
\begin{equation}
\label{thwssup}
\hat{T}(u) |{\rm YHWS}\rangle=
\left(
\begin{array}{ccc} t^{11}(u) & & \rule{0em}{1.5em}_{\mbox{\huge 0}}\\
  & \ddots & \\
\mbox{\huge *}&& t^{pp}(u)
\end{array}\right)  |{\rm YHWS}\rangle,
\end{equation}
where $*$ is the set of lowering operators in the lower triangle of
$\hat{\T}$ $\{ \hat{T}^{ij}(u)\}_{i>j}$, and $t^{aa}(u),\: a=1,\ldots,p$
are scalar functions. Rather than one Drinfel'd polynomial, the $p>2$ case
has $p-1$ polynomials $P_k(u)$ describing the HWS, for the $p-1$ elements of the
Cartan algebra. They are defined through:
\begin{equation}
\prod_{k=j}^{p-1}\frac{P_k(u+h)}{P_k(u)}=\frac{t^{jj}(u)}{t^{pp}(u)}l;\;
k=1,\ldots,p-1,
\label{drinsup}
\end{equation}
and $t^{pp}(u)$ is again of the form $\frac{f_p(u+h)}{f_p(u)}$ as in eq.\
(\ref{t22ev}). The fact that the eigenvalues of $\hat{\T}$ on a YHWS have
to be of the form (\ref{drinsup}) severely restrict where the roots of
the polynomials can lie. E.g.\ given that the eigenvalue of the quantum 
determinant belonging to
 this YHWS is given by $\frac{\Delta(u+h)}{\Delta(u)}$
(\ref{detqise}),
with:
\begin{eqnarray}
\Delta(u)&=&\prod_{j=1}^{p}t^{jj}(u-h(j-1))\nonumber\\
&\equiv & \prod_{k=1}^N \left(u+h({\scriptstyle \frac{N+1}{2}}-k)\right)
\label{qdetdrin}
\end{eqnarray}
we find that the set of integers from 1 to $N$
(modulo the factor $\frac{N+1}{2}$)---the zeroes of $\Delta(u)$---must 
be subdivided into strings of length no greater than $p$.
Every root of the Drinfel'd polynomial $P_k(u)$ labels the {\em
left}most point in a string of $k$ zeroes of $\Delta(u)$. These zeroes
come from  $P_k(u),\, P_k(u-h),\ldots$ and $P_k(u-h(k-1))$ respectively.
This is most easily explained in a figure (see fig.\ \ref{stringfig}).
\begin{figure}
\framebox[\hsize]{\centerline{\psfig{file=string.epsi,width=5.8in}}}
\caption{Example of a string motif for $SU(3)$, for $N=16$ spins.}
\label{stringfig}
\end{figure}

It is clear from knowledge of the action of $\hat{\T}$ on a particular
YHWS, i.e.\ the set $\{ t^{kk}(u)\}_{k=1}^{p-1}$, we can uniquely
determine the appropriate set of Drinfel'd polynomials, since the solution
of an equation of the type $\frac{f(x)}{g(x)}=\frac{P(x+h)}{P(x)}$, for
polynomials $f,g$ and $P$ is
unique (up to a constant). As a matter of fact, just knowing
$\frac{t^{11}(u)}{t^{pp}(u)}=\frac{P_1(u+h)\cdots P_{p-1}(u+h)}{
P_1(u)\cdots P_{p-1}(u)}$ we can reconstruct all $P_i(u)$ since every root
of $f_p(u)$ has to cause $p$ consecutive zeroes in $\Delta(u)$.\footnote{
For the sake of completeness: to relate a particular set of Drinfel'd zeroes to
the pseudomomenta $m_i$ that go into the expression for energies eq.
(\ref{Hexpl}) and (\ref{HISEdunkl}): they are the roots of the polynomial
$\frac{\Delta(u)}{\prod_{j=1}^p P_j(u-h(j-1))}$. This is essentially
$\Delta(u)$ with
the last root in every string left out.}
So if $\frac{t^{11}(u)}{t^{pp}(u)}$ uniquely determines the YHWS,
$T^{11}$ is the nondegenerate operator that we are looking for. 

We can now prove that $[\hat{\Delta}(u),\hat{\Delta}(u')]=0$.
If $|\Gamma\rangle$ is a
YHWS with motif $\Gamma$ then $\hd_1(u)|\Gamma\rangle$ and
$\hd_1(u')|\Gamma\rangle$ both are scalar multiples of
$|\Gamma\rangle$
since they have the same $T^{11}_0(v)$-eigenvalue
($[T^{11}_0(v),\hd_1(u)]=
[T^{11}_0(v),\hd_1(u')]=0$).  So in this
YHWS-space both $\hd_1(u)$ and $\hd_1(u')$ are diagonal and thus
commute.

In the remaining part we will reproduce the integrals of motion that
have
already been found \cite{Ino90,HHTBP92}, and point out some
subtleties in their
construction.  As is customary for the Heisenberg model with nearest
neighbor
exchange we take $\Gamma_{1}(u)=\frac{d}{du} \ln(\hd_1(u) )$ rather
than
$\hd_1(u)$ to be the generating function for the integrals of motion.
In the NNE model this need arises, in order for $H_{\rm NNE}$ to be the
{\em local}, i.e.\ to just couple nearest neighbors. An added advantage
is that the spectrum of these operators will  be additive (in
both NNE and ISE models), reaffirming that the parameters that
characterize the eigenstates---be it the pseudomomenta in the ISE model or
the rapidities in the NNE case---describe {\em single} quasiparticle
excitations.

When expanded in
powers of $u$, $\Gamma_1(u)$
reads:
\begin{eqnarray}
\Gamma_1(u) &=& \Pi \sum_{i=1}^{N}\frac{1}{u-\hbar z_i\partial_{z_i}
-\hg_i}
\Pi\;\; {\rm  to}\;{\cal O}(\hbar) \nonumber\\
 & = & \sum_{n=0}^{\infty}u^{-(n+1)}\left\{
\Pi\sum_{i=1}^{N}\sum_{p=0}^{n-1}
(\hg_i)^p z_i\partial_{z_i} (\hg_i)^{n-p-1} \Pi\right\}\nonumber\\
&\equiv &\sum_{n=0}^{\infty}u^{-(n+1)} H_n,
\label{Hgendef}
\end{eqnarray}
where we have reinserted the projection operator that turns
$K_{ij}\rightarrow
P_{ij}$, when ordered to the right of an expression.
We have worked out the first few $H$'s. With $z_{ij} = z_i -z_j$ we
have:

\begin{eqnarray}
H_1 &=& \sum_{i=1}^{N} z_i\partial_{z_i} \equiv P\nonumber\\
H_2 &=& -\sum_{i,j}\rule{0in}{1.3ex}^{'}\,
\frac{z_i z_j}{z_{ij}^2} (P_{ij}-1)\nonumber\\
H_3 &=& \sum_{ijk}\rule{0in}{1.3ex}^{'}\,
\frac{z_i z_j z_k}{z_{ij}z_{jk}z_{ki}} P_{ijk} +
\frac{3}{4} \sum_{ij}\rule{0in}{1.3ex}^{'}\,
(1-(w_{ij})^2)z_i\partial_{z_i} = \nonumber\\
&=& \sum_{ijk}\rule{0in}{1.3ex}^{'}\,
\frac{z_i z_j z_k}{z_{ij}z_{jk}z_{ki}} P_{ijk} + 
{\scriptstyle \frac{N^2 -1}{4}} P \nonumber\\
H_4&=& \sum_{ijkl}\rule{0in}{1.3ex}^{'}\,
\frac{z_i z_j z_k z_l}{z_{ij}z_{jk}z_{kl}z_{li}}(P_{ijkl}-1)
-2 \sum_{ij}\rule{0in}{1.3ex}^{'}\,
\left(\frac{z_i z_j}{z_{ij} z_{ji}}\right)^2 (P_{ij}-1)
+{\scriptstyle \frac{N^2-1}{3}} H_2
\label{Hexpl}
\end{eqnarray}
A prime on the summation symbol indicates that the sum should be
restricted
to distinct summation-indices.  To compute the previous expressions,
we normal
ordered the $z_i \partial_{z_i}$ to the right in eq.\ (\ref{Hgendef})
and {\em
then} put $z_j = \exp\left(\frac{2\pi i}{N} j\right)$.  The identity
$w_{ij}w_{jk}+w_{jk}w_{ki}+w_{ki}w_{ij}=-1$ which lies at the heart
of the
integrability of these $\frac{1}{r^2}$-models is very useful in the
reduction
of these expressions.  $P$ indicates the total momentum, or degree of the
polynomial YHWS, and we will discuss
its
interpretation later on.  Notice that fortunately $H_{\rm ISE}$ is a member
of this set, as $H_2$. 
Notice the absence of Yangian operators, as
well as
terms containing both permutations {\em and} derivatives.
Although one would expect this, based on the fact that, physically, the
dynamical degrees of freedom freeze out when $h\rightarrow 0$, it 
still seems somewhat accidental that they cancel in an explicit
computation.
The expressions in eq.\
(\ref{Hexpl}) can be seen to coincide with those reported previously
\cite{HHTBP92},\footnote{We should point out a correction in eq.\ (7)
of
Ref.\ \protect\cite{HHTBP92} where $-\frac{1}{3} H_2$ should be
replaced by
$+\frac{1}{6} H_2$.  This changes the invariant in a harmless manner,
but this is relevant for
comparing the eigenvalues of the operators in that article and the
ones
that we will find later
on.}  lending credibility to this way of deriving the integrals of
motion.
Unfortunately for large $n$ it becomes prohibitively complicated to
compute
$H_n$.

\section{The Spectrum of the Invariants.}
\label{Hnev}

We have an alternative way to verify the validity of these constants
of motion
as well. We will proceed to compute the eigenvalues of the operators
$H_n$  and
compare these to the `rapidity' description of the eigenvalues in
\cite{HHTBP92}. We will constrain ourselves to the $SU(2)$ case to
simplify the
algebra.
The method that we have in mind is to produce the eigenvalues of
$\hat{\Delta}(u,\hbar)$ of the dynamical model and take the appropriate
limit $\hbar\rightarrow 0$. 
This method, called the Asymptotic Bethe Ansatz is designed to 
work in the low density limit. In that case, between collisions, the
particles are in free plane wave states. Then, by taking one particle and
bringing it all the way around the circle and scattering it through the
other $N-1$ particles, we can constrain their allowed ``free'' momenta via
the phase shifts picked up in every collision process. The spectrum of
the model then follows from adding the free kinetic energies.
For most integrable 
models it has been shown that, remarkably,  this procedure is actually exact, 
or at
least can be made exact with small modifications \cite{suth85}. 
From eq. (\ref{ABAdef}) we know that the ISE model is in this category,
as are the NNE model and the continuum Sutherland model (\ref{SUTHdef}).
For the
CSM, integrability has been confirmed \cite{BGHP93}, so one might expect
that in that case the asymptotic Bethe Ansatz should be exact too (see
\cite{Kaw93} for evidence thereof). We will assume that it is true.

As is well known \cite{KR86} the roots of $\hd(u,\hbar)$---i.e.\ the
poles of
the transfer matrix $\htrans(u,\hbar)$, see eq.\
(\ref{trmexpand})---are given
by the solutions of the Asymptotic Bethe Ansatz equations, which only
depend on the
{\em two}-particle phase-shift.  In the case of the CSM it is
$h\pi {\rm
sgn}(k_1 -k_2)$.  Notice that this phase-shift only depends on the
ordering of
the momenta $k_1$ and $k_2$.  This is why these models are
interpreted to
describe an ideal gas of particles with statistics that interpolates
between
bosons ($h =0$) and fermions ($h = 1$).  In the
dynamical model (\ref{dynmodeldef}) the particles have charge {\em
and} spin.
Therefore we get two coupled sets of `Nested' Bethe Ansatz
equations---for the general case $p\neq 2$ there are $p$ equations.
They
have been presented in \cite{Kaw93}:

\begin{eqnarray}
k_i L& =&h \left\{\sum_{j(\neq i)}\pi{\rm sgn}(k_i -k_j) +
\frac{1}{h}\left[ 2\pi I_i - \pi\sum_{\alpha =1}^{M}
{\rm sgn}(k_i -\Lambda_{\alpha})\right]\right\}\nonumber\\
&\equiv & \left( k_i^0 +\frac{1}{h}\delta k_i\right) L
\label{kdef}\\
&&\hspace{-.75truein}
\pi\sum_{\beta(\neq\alpha)}{\rm
sgn}(\Lambda_{\alpha}-\Lambda_{\beta}) +
2\pi J_{\alpha} = \pi\sum_{i=1}^{N}{\rm sgn}(\Lambda_{\alpha}-k_i).
\label{lamdef}
\end{eqnarray}
We have reinstated $L$, the circumference of the circle to get the
dimensions
correctly.  Notice how $\hbar$ drops out of these equations due to
the fact
that the full Hamiltonian is scale invariant (a peculiarity of the
$\frac{1}{r^2}$-type potentials).  So rather than sending
$\hbar\rightarrow 0$
we should let $h\rightarrow \infty$.  There are $N$ equations
defining
the $\{ k_i\}$ (one for every particle) with {\em charge} quantum numbers
$\{ I_i\}$.
Furthermore we have $M$ equations defining the auxiliary momenta $\{
\Lambda_{\alpha}\}$.  $M$ is the number of particles with a spin
$\downarrow$.
The $\{J_{\alpha}\}$ are their {\em spin} quantum numbers.  The $I$'s
and $J$'s are
distinct integers or half-odd integers depending on the parity of $N$
and
$M-N$.

Furthermore we should restrict the Hilbert space to states that only
carry spin
excitations, and no charge excitations (these elastic modes
unfortunately don't
acquire a gap as $\hbar\rightarrow 0$)\footnote{This is why we don't
use eq.\
(\protect\ref{deltaev}) which gives a much more direct expression for
the
eigenvalues of
$\hd(u,\hbar)$; however, one doesn't know a priori whether an
eigenvalue
belongs to a pure
spin- or charge excitation, or a mixture of both. This is something which 
only a physically motivated method can tell us.}.  As in for
instance the
1D-Hubbard model we accomplish this by leaving the charge quantum
numbers in
their groundstate configuration and only exciting the $\{
J_{\alpha}\}$.  Let
us therefore first analyze the absolute groundstate which has
$M=\frac{N}{2}$
(for $N$ even), so there are twice as many $k$'s as $\Lambda$'s.  The
$I$'s and
$J$'s are consecutive and spaced by one unit.  Then eq.\
(\ref{lamdef}) tells
us that between every two neighboring
 $\Lambda$'s there must be two $k$'s.  In a
spin
excited state we have $M<\frac{N}{2}$ and by leaving openings in the
$J$-distribution, we can have more than two $k$'s sit between every
pair of
$\Lambda$'s.  Notice that this equation doesn't fix the value of the
$\Lambda$'s, just their positions with respect to the $k_i$'s.  From
eq.\
(\ref{kdef}) we learn that when we order the $\{ k_i\}$ such that
$k_i < k_j$
for $i<j$:  $k_i \approx h \left( i-\frac{N+1}{2}\right)\equiv
k_i^0$.
There is however an ${\cal O}(\frac{1}{h})={\cal O}(\hbar)$
correction
through the $\Lambda$'s.  Whenever a $\Lambda$ sits between two $k$'s
they will
be drawn together by $\frac{1}{h}$.  This information is
contained in
$\delta k_i$.  Now, in the same way that the constants of motion are
contained
in the ${\cal O}(\hbar)$-term in $\hd(u,\hbar)$, their eigenvalues
are
determined by the ${\cal O}(\frac{1}{h})$ corrections to the
$k$'s.  As
$\hd(u,\hbar)$ has eigenvalues $\prod_{i=1}^{N}(u-k_i)$,
$\Gamma(u,\hbar)=\frac{d}{du} \ln\hd(u)\equiv\Gamma_0
+\hbar\Gamma_1+\ldots$
must
have eigenvalues $\sum_{i} \frac{1}{u-k_i}=\sum_i \frac{1}{u-k_i^o -
\frac{1}{h}\delta k_i}$.  So for $\Gamma_1(u)$ acting on a
state
characterized by a set $\{ \Lambda_{\alpha}\}$ we find its
eigenvalue:

\begin{eqnarray}
\Gamma_1(u) &=& \sum_{n=0}^{\infty}
\frac{1}{u^{n+1}} n \sum_{i=1}^{N}(k_i^0)^{n-1} \delta
k_i\{\Lambda_{\alpha}\}
\nonumber\\
&\equiv& \sum_{n=0}^{\infty} \frac{1}{u^{n+1}}
h_n\{\Lambda_{\alpha}\}.
\end{eqnarray}
We will now label the $\Lambda_{\alpha}$'s by $m_\alpha$, their
positions
relative to the $k$'s, i.e.\ if $\Lambda_\alpha$ sits between $k_r$
and
$k_{r+1}$, then $m_\alpha = r$. We see  that the $m_\alpha$ have to
be at least
two units apart since there are {\em at least} two $k$'s between
$\Lambda$'s
according to eq.\ (\ref{lamdef}). Now writing $\frac{1}{2}
{\rm sgn}(\Lambda_\alpha -k_i) = \theta(\Lambda_\alpha -k_i) -\half$
($\theta$
is the step function), we have:

\begin{eqnarray}
&& n\sum_{i=1}^{N}(k_i^0)^{n-1}\delta k_i\{\Lambda_\alpha\} =
n\sum_{i=1}^{N}
\left( i-\frac{N+1}{2}\right)^{n-1}\left\{ I_i-\frac{M}{2} +
\sum_{\alpha =1}^{M} \theta (\Lambda_\alpha -k_i)\right\}\nonumber\\
&&={\rm const} + \sum_{\alpha =1}^{M} \left[ n
\sum_{i=1}^{m_{\alpha}}
\left( i-\frac{N+1}{2}\right)^{n-1} \right] \nonumber\\
&&={\rm const}+\sum_{\alpha =1}^{M} \epsilon_n (m_\alpha)
\end{eqnarray}

For small $n$ we can evaluate $\epsilon_n(m_\alpha)$ exactly:

\begin{eqnarray}
\epsilon_1(m_\alpha) &=& m_\alpha\nonumber\\
\epsilon_2(m_\alpha) &=& m_\alpha (m_\alpha -N)\nonumber\\
\epsilon_3(m_\alpha) &=& \half m_\alpha (m_\alpha -N)(2m_\alpha -N) +
\frac{N^2-1}{4} m_\alpha\nonumber\\
\epsilon_4(m_\alpha) &=& \epsilon_2(m_\alpha) \left(
\epsilon_2(m_\alpha) +
\frac{N^2-1}{2}\right)
\end{eqnarray}

These results coincide nicely with the numerical values of
\cite{HHTBP92}, when
we interpret the  $m_\alpha$ as the rapidities of the ISE model! We notice
that it is consistent to interpret the momentum term $P$ in (\ref{Hexpl})
in $H_1$
and $H_3$ as  $\sum_{\alpha} m_\alpha$, i.e.\ the degree of a polynomial 
YHWS wavefunction. Non-YHWS, non polynomial wavefunctions in a Yangian
multiplet have the same value of $P$ since $P$ commutes with the Yangian
Algebra.

In conclusion we have outlined a method to obtain the constants of
motion of
the ISE model as a strong coupling limit of the CSM with particles with
internal
degrees of freedom.  Although the task to actually obtain the
invariants is
quite cumbersome, it can be done in principle.  Given the relatively
simple
structure of the invariants we expect there to be some technique that
could
simplify the computation considerably.  The construction of integrals
of motion
presented in this letter, provides us with extensive operators that
commute
with each other and the Yangian symmetry algebra.  By computing
eigenvalues of
the invariants through the Nested Bethe Ansatz, and comparing them
with
previous numerical results \cite{HHTBP92} we provided evidence for
the validity
of the approach.  It would be interesting to analyze the cause of the
miraculous absence of terms containing mixtures of permutations and
derivatives.
Recently a generalization of this method of obtaining the constants of the
motion has been generalized to spin models with a {\em trigonometric} rather
than a rational $R$-matrix in the FCR eq.\ (\ref{FCRdef})
for their  transfer matrix
\cite{uglov95}.  These trigonometric analogs of $H_{\rm ISE}$ 
do not have the Yangian $Y(sl_2)$ but rather
the quantum group $U_q(\widehat{sl}_2)$ as their symmetry algebra.

\part{Applications of the $\frac{1}{r^2}$-Spin Model.}
\label{part2}
\chapter{Dynamical $T=0$ correlations of the ISE-model in a
magnetic field}
\label{dsf}

\section{Introduction}
\label{intro3}

In this chapter we will devote ourselves to computing an observable in
the ISE-model: the dynamical spin-spin correlation function or Dynamical
Structure Functions (DSF) in a non-zero external
magnetic field. If the ISE-model were to be realized directly in
the nature, the DSF would be of foremost interest since it measures
effectively the intensity for inelastic neutron scattering off the
spin-chain. A more important reason to be concerned with it is that, 
because of the simplicity of the model, we
can actually obtain exact expressions for the DSF.
As a matter of fact, the zero field case for this model was the first
strongly interacting system
for which such correlators
could be obtained at all \cite{A93,H93}. The {\em static} structure
factor had already been obtained in 1988 \cite{H88,S88}, based on work
by Sutherland from the 70's \cite{S72}.
It turns out that both have all the qualitative features of the complicated
NNE mode, but are quantitatively much cleaner.
The underlying reason for this cleaner structure is that, where normally
a neutron, colliding inelastically with the ring of spins, tends to
excite {\em many} quasiparticles (although high numbers of them are less
likely), in the ISE-model there can be no more than {\em four}, and in
zero field just {\em two}. This 
among other things, enables the aforementioned exact approach. In this
chapter we want to generalize this to non-zero magnetic fields. The
rationale is that at $T=0$ and $B=0$ the groundstate contains no 
elementary excitations, the spinons; by tuning the magnetic field, which
acts as a chemical potential for the up-spinons, we can select any
non-zero number of them in the groundstate. This allows for a richer set
of excitations and also enables us to make contact with the excitations
over the ferromagnetic state (in high fields) which should be {\em
magnons}---with spin $\half$ rather than 1 \cite{Ino90}.

In the next section we will define the dynamical structure function and
its properties in the ISE-model. In section \ref{select} we will
discuss the various contributions to the DSFs. Finally, section
\ref{ISEnNNE} compares our results to those of M\"{u}ller et al.
\cite{M81}. They obtained approximate analytical expressions and
numerical results for dynamical correlators in the NNE chain.

\section{Dynamical Structure Functions}
\label{dsfise}

The DSF for a spin model at zero temperature is defined as:
\begin{equation}
\label{structdef}
\langle {\rm GS}|S_{m}^{i}(t)S_{n}^{j}(t')|{\rm GS}\rangle =
S^{ij}(m,n,t-t').
\end{equation}
$| {\rm GS}\rangle$ represents the groundstate of the system, and
$\svec_m(t)$ is the local spin operator, acting at site $m$ and time
$t$. Its
time-dependence is implemented via the Heisenberg representation:
$\svec_m(t)=e^{iHt}\svec_m e^{-iHt}$. The DSF measures the response of
the system to flipping/imposing a certain spin on a site $n$ in the
groundstate at time $t'$ and measuring the effect on a site $m$ at time
$t$. Since $H$ conserves $S^z$ we cannot create an up-spin somewhere and
have it evolve into a down-spin. In other words, there are restrictions
on the possible choices of $i$ and $j$ in (\ref{structdef}). We find that
only $S^{-+}, S^{+-}$ and $S^{zz}$ are non-zero
($S^{xx}=S^{yy}=\frac{1}{4}
\left[S^{+-}+S^{-+}\right]$, $S^{xy}=\frac{1}{4i}\left[S^{-+}-
S^{+-}\right]$).

The DSF (\ref{structdef}) tells us what kinds of excitations a local spin
operator can 
generate in this system. This is obvious, once we insert a complete
set of eigenstates $\{ |\nu\rangle \}$ between the two operators:
\begin{equation}
S^{ab}(m,n;t-t')=\sum_\nu \langle {\rm GS}|S^i_m |\nu\rangle
\langle\nu | S^j_n | \GS \rangle e^{-\frac{i}{\hbar}(t'-t)(E_\nu - E_0)}.
\label{interm}
\end{equation}
This can be simplified further in the case that the model is
translationally invariant (such is the case for both NNE and ISE models),
i.e.\ both $|\GS \rangle$ and $|\nu\rangle$ are momentum eigenstates. In
that case $S^{ab}$ will only depend on $m-n$ and we can consider its time
and space Fourier transform:
\begin{eqnarray}
S(Q,E) &=&\sum_\nu M^a_\nu \delta(E-(E_\nu - E_0)) \delta(Q- (p_\nu-p_0))
\nonumber\\
M^a_\nu &=& 2\pi \left| \langle \nu |S^a(Q) | \GS \rangle \right|^2,\;\;
a=\pm, z,
\label{sqedef}
\end{eqnarray}
and $\svec(Q)=\frac{1}{\sqrt{N}}\sum_{n=1}^N e^{-inQ}\svec_n$, and
$(p_\nu,E_\nu)$ and $(p_0,E_0)$ are the momentum and energy of the state
$|\nu\rangle$ and $|\GS\rangle$ respectively. Notice that the support
$S(Q,E)$ is zero except when $(Q,E)$ corresponds to the excitation
energy and momentum of a state contained in $\svec(Q)|\GS\rangle$.

For a general model, this set is constrained by global symmetries, such
as momentum conservation, and the Wigner-Eckart theorem---from the vector
character of $\svec(Q)$. For the ISE-model, however, the number of
excited states contributing to the sum (\ref{sqedef}) is very limited,
which is what we alluded to in the the introduction, when we mentioned
the simplicity of this model. We will find that the relevant excited
states differ from the groundstate only in the following way: in low
fields the number of spinons can be the same as that in the
groundstate, or it can increase by two; in high magnetic fields, there are two
additional excitations: a right or left moving {\em magnon}, or the
number of spinons can {\em decrease} by two.  

The reason for the different low and high field behavior is that the
character of the groundstate changes with magnetic field. This is caused
by the additional Zeeman term in the Hamiltonian:
\begin{equation}
H_{\rm Zeeman} = -h\sum_i S^z_i
\label{zeeman}
\end{equation}
At $h=0$ the groundstate lies in the $M=\frac{N}{2}$ sector ($M$ is the
number of down-spins). As we increase the field, there comes a point
where $h$ is equal to the $h=0$ energy difference between the (absolute) 
groundstate in
the $M=\frac{N}{2}$ sector and the one in the $M=\frac{N}{2}-1$ sector, and
the energies of these states cross.
We find that for any value of $h$ the groundstate is always YHWS.\\
\vspace{-.3truein}
\begin{wrapfigure}{r}{2.5in}
\fbox{\psfig{file=dispersion.epsi,width=2.3in}}
\end{wrapfigure}
This is
easy to see, if we consider the expression for the energy of a Yangian
multiplet in terms of the pseudomomenta (\ref{HISEdunkl}), $E\propto
\sum_i \epsilon(m_i) = \half\sum_i m_i ( m_i-N)$. 
We see that adding pseudomomenta {\em lowers}
the energy. 

I.e., the Yangian multiplet with one less pseudomomentum, and $S^z$ higher 
by +1, has 
higher energy. 
So in a given sector of $S^z$ the groundstate is a member of the subset
of YHWS (with that value of $S^z$).
From the figure it is clear that in the groundstate the pseudomomenta will 
arrange
themselves symmetrically around $m=\frac{N}{2}$. The expression for this
state is actually known analytically \cite{H88}:

\begin{equation}
\psi^M_{\GS}(n_1,\ldots,n_M) = \prod_{i<j}^M \left( z_{n_i}-z_{n_j}\right)^2
\prod_{i=1}^M
z_{n_i}^{\frac{N}{2}-M+1}
\label{GSdef}
\end{equation}
Although the groundstate magnetization $S^z=\half(N-2M)$ is a
discontinuous function of $h$, it becomes smooth in the thermodynamic
limit. With the set of pseudomomenta $\{ m_i =
\frac{N}{2}-M+2i-1\}_{i=1}^M$ ,
the groundstate energy is
$E_0(M)=({\scriptstyle \frac{2\pi}{N}})^2\frac{v_s}{\pi}
\frac{M(4M^2-3N^2-4)}{24}-\frac{h}{2}(N-2M)$. Minimizing with
respect to $M$ gives:
\begin{equation}
2\frac{S^z}{N}=2\sigma = 1- \sqrt{1-\frac{h}{h_c}}\;\;\;\;\ h_c =
\frac{\pi v_s}{2}.
\label{hcdef}
\end{equation}
As expected, with the magnetic field, the number of fully polarized
spinons $N_{sp}=2S^z$ in the groundstate grows. For $h>h_c$ the groundstate
is completely ferromagnetic.
 
For the sequel it is useful to have expression of energy and momentum of
an eigenstate of $H_{\rm ISE}$ in terms of ``spinon momenta'' as well
\cite{H91}.
If we
label
the $M+1$ orbitals from {\em right to left} by spinon momenta $-k_0\leq k\leq
k_0 =
\frac{2\pi}{N} \frac{M}{2}$ spaced by $\frac{2\pi}{N}$ we get:

\begin{eqnarray}
P & = & \sum_{k}k\,  n_{k\sigma} + Nk_0 \;{\rm mod}\; 2\pi ,\nonumber \\
\label{spinondisprel}
E & =&\sum_{k\sigma} \epbar(k)n_{k\sigma} +
\frac{1}{N}\sum_{kk',\sigma\sigma'} V(k-k') +\; E(M,N),
\end{eqnarray}
where $\epbar(k) = \frac{v_s}{\pi}(k_{0}^{2}-k^2)$, $V(k)=v_s(k_0-|k|)$
and
$n_{k\sigma}$ is the number of spinons in the orbital with momentum $k$
and
spin $\sigma$ \cite{H91}.  $E(M,N)$ only depends on the total number of sites and
spinons.
We also recognize $v_s$ as the spinon velocity $\frac{ d\epbar(k)}{dk}$
at
the z\^{o}ne boundary, $k_0$, in the groundstate. Notice from the spinon 
and magnon dispersion relation, we see that the ``mass'' of the magnon is (-2)
times that of the spinon.
The system is gapless, so this mass doesn't refer to the energy that is
required to create a pair of them,  but to the curvature in their band.

Before we turn our attention to the precise selection rules that restrict
the set of accessible intermediate states, let us first discuss an
intuitively more attractive (overcomplete) basis of the Hilbert space:
the set of {\em localized} spinon wavefunctions \cite{H94}:
\begin{equation}
\Psi(n_1,\ldots,n_M|\alpha _1,\ldots,\alpha _{N_{sp}}) =
 \prod_{i<j}\left(z_{n_i}-z_{n_j}\right)^2 \prod_{i=1}^{M}z_{n_i}
\prod_{i=1}^{M} \prod_{j=1}^{N_{sp}}\left(z_{n_i}-z_{\alpha_j}\right).
\label{locspinwf}
\end{equation}
The $\{\alpha_i\}$ are the locations of localized spinons that can point
up or down; the $\{n_i\}$ are the positions of down-spins
(other than those of possible localized spinons pointing down).  Notice that
the wavefunction prevents the $\{n_i\}$ from coinciding with the spinon sites.
We call the complement of the set of spinon sites the {\em condensate}.  It is
a singlet under the action of total spin.  Furthermore for $N_{sp}=0$ eq.\
(\ref{locspinwf}) represents the exact groundstate wavefunction for $h=0$.  The
usefulness of these states is limited by the fact that they are not mutually
orthogonal and worse, overcomplete.  However, based on numerical evidence for
up to 12 spinons, it is clear that the space spanned by states
(\ref{locspinwf}) with a fixed number $N_{sp}$ of {\em localized} spinons
contains only eigenstates of the Hamiltonian belonging to Yangian multiplets
with $N_{sp}$ or less spinons.

The subspace of states that have a fixed number of $N_{sp}$ localized spinons,
all polarized, has the pleasant property that it only contains eigenstates of
$H$ with {\em precisely} $N_{sp}$ spinons.  This is clear from the fact that
these localized spinon wavefunctions---although not eigenstates of the
Hamiltonian---are annihilated by both $J^{+}_{0}$ and $J^{+}_{1}$ 
(see Appendix A).  This is consistent with the fact that fully 
polarized spinon eigenstates
are supposed to be of a polynomial form with degree $<N$ in the
$\{z_{n_i}\}$ \cite{H91}, just like wavefunctions (\ref{locspinwf}).  In that
same reference it is also shown that these fully polarized localized spinon
states are {\em complete} as well and span {\em all} YHWS.  \\

\section{Selection Rules for Spin operators.}
\label{select}

We  continue discussing the various Yangian multiplets in terms of
their pseudo\-momenta---the zeroes of $f_2(u)$---rather than the
Drinfel'd zeroes. It is customary to denote the set of pseudomomenta by
a sequence of $N-1$ 0's and 1's, where a 1 indicates a pseudomomentum
(the locations of the big crosses in fig.\ \ref{figmotif}), and 0's (the
small crosses and circles) pad the
1's so that all integers in the range are covered. Such a binary sequence
is also called an {\em occupation pattern}.

We will now discuss the relevant selection rules for the matrix elements
$M^a_n$ in eq.\ (\ref{sqedef}).
From numerical evidence up to $N=16$ sites it has become clear that there are
only an unexpectedly small number of those matrix elements that are
non-zero.  To
resolve parity and other accidental degeneracies between the Yangian 
multiplets
we split these degeneracies\footnote{
Due to the fact that the expression of the energy of a state
(\protect\ref{HISEdunkl}) is sum of
squares of integers. } by actually diagonalizing $H + \lambda H_3$, where
$H_3$ is the second integral of the motion for this model as presented in
eq.\ (\ref{H3def}).  The eigenvalues of this operator allowed the Yangian
occupations sequences to be unambiguously identified.  States in the multiplets
are partially resolved by fixing $ S^{z}$ and $S^{tot}$ (a unique resolution of
states would be obtained by adding another term $\mu {\bf J}_0\cdot{\bf
J}_1 \propto J_0^2$
to the Hamiltonian\footnote{
This term splits the spin multiplets within
a Yangian occupation pattern, with identical $S_{\rm Tot}$ but different
rapidities $\{ \lambda_i \}$ in eq.\ (\protect\ref{ABEdef}).}.  
This would correspond to a basis of states within the
Yangian multiplet obtained through the Algebraic Bethe Ansatz \cite{F82}).

Now let us denote the eigenstates of this model as $\left|\Gamma ,\mu\right.
\rangle$ where $\Gamma$ labels a Yangian multiplet through an occupation
sequence and $\mu$ labels the state within the multiplet.  Furthermore define
$M^{\Gamma\Gamma '}(S^{a}_{i})_{\mu\mu '}\equiv\langle\left.  \Gamma\mu\right|
S^{a}_{i}\left|\Gamma\mu '\right.  \rangle$.  Then the observation made above
implies that the matrix $\underline{\underline{\bf M}}^{\Gamma\Gamma
'}(S^{a}_{i})$ vanishes if the occupation sequences $\Gamma$ and $\Gamma '$
differ ``too much'' in a sense made precise below.  This situation is analogous
to an {\em ideal} gas, where if $\hat{O}$ is a one body operator:
$\langle\left.  \alpha\right| \hat{O}\left|\beta\right.  \rangle = 0$ if the
occupation number configurations of $\left|\alpha\right.  \rangle$ and
$\left|\beta\right.  \rangle$ differ on more than one orbital.  E.g.\
$\hat{O}=\rho(x)=\sum_{kk'}e^{i(k-k')x}c_{k}^{\dagger}c_{k'}$ can add or take
out a single particle from an orbital in an ideal gas, but in an {\em
interacting} gas it could add unlimited numbers of particle/hole pairs.

We have formulated a set of empirical rules, summarizing numerical studies
on small systems with $N<16$. We expect that these will be
derived when a complete understanding of the model is developed.
According to these rules, if two multiplets $\Gamma$ and $\Gamma '$ 
do {\em not} differ too much, 
there will always be a pair of states $\mu$ and $\mu '$ in either
multiplet for which the spin matrix element is nonzero.
\begin{rule1}
If $\pi (\Gamma,m,n)$ is the total number of ones in positions $m$ through $n$
in Yangian occupation sequence $\Gamma$ then:  $\underline{\underline{\bf
M}}^{\Gamma\Gamma '}(S^{a}_{i})\neq\underline{\underline{\bf 0}}$ iff.
$|\pi(\Gamma ,m,n)-\pi(\Gamma ',m,n)|\leq 1$ for any $1\leq m<n\leq N-1$.
\label{rule11}
\end{rule1}
The rule is illustrated in Fig.\ \ref{selrulepic}.  
\begin{figure}[htb]
\begin{eqnarray}
\langle\Gamma = 100010010000,\mu | &S_i^a &|\Gamma '
=100010001000,\mu'  \rangle \neq 0\nonumber\\ \langle\Gamma =
100\underbrace{01001}0000,\mu | &S_i^a &|\Gamma ''
=001\underbrace{00000}1000,\mu''\rangle   = 0\nonumber
\end{eqnarray}
\caption{According to the selection rule, there will be states $\mu$ in
multiplet $\Gamma$ that are connected to others $\mu'$ in $\Gamma '$ through a
local spin operator.  In $\Gamma$ and $\Gamma ''$ there are none, e.g.\ since
\mbox{$|\pi (\Gamma ,4,8)-\pi(\Gamma' ,4,8)|$} \mbox{$=2-0=$} \mbox{$2>1$}.}
\label{selrulepic}
\end{figure}
A general consequence of
this rule is that when we choose $m=1$ and $n=N-1$, it follows that the total
number of ones in a sequence can't change by more than one, i.e.\ the total
number of spinons can only change by +2,0, or -2.  It is remarkable that
according to the rule this also holds on any corresponding subsequences of the
occupation number sequences.

The {\rm zero} magnetic field DSF has been computed in \cite{H93}.  The
particular structure function computed happened to be $S^{-+}(Q,E)$ (the others
are identical because of rotational invariance).  This function is governed by
excitations present in $S^{+}_{i}\left|{\rm GS}\right.  \rangle$.  Since the
zero field groundstate contains no spinons the rule tells us that we can only
expect excitations with 0 or 2 spinons.  As $ S^{+}_{i}\left|{\rm GS}\right.
\rangle$ has $ S^{z} = +1$ the former is ruled out and in the states in the
multiplets with 2 spinons, both must be polarized.  This was to be expected
since we can expand $S^{+}_{i}\left|{\rm GS}\right.  \rangle$ in a set of
localized spinon wavefunctions (\ref{locspinwf}) containing {\em two} polarized
spinons:

\begin{equation} 
\langle\left.  n_{1},\ldots,n_{M-1} \right|S^{+}_{i}\left|{\rm
GS}\right.  \rangle = \sum_{m=1}^{N-1}
\frac{2}{N}\frac{1-(-)^m}{1-z^{-m}} \Psi(n_1,\ldots,n_{M-1}|i,i+m).
\label{hnullspexp}
\end{equation}
So one of the spinons seems to be sitting on the site on which $S^{+}_{i}$
acted and the other is an even number of sites removed from it.  

\begin{figure}[htb]
\framebox[\hsize]{
\centerline{\psfig{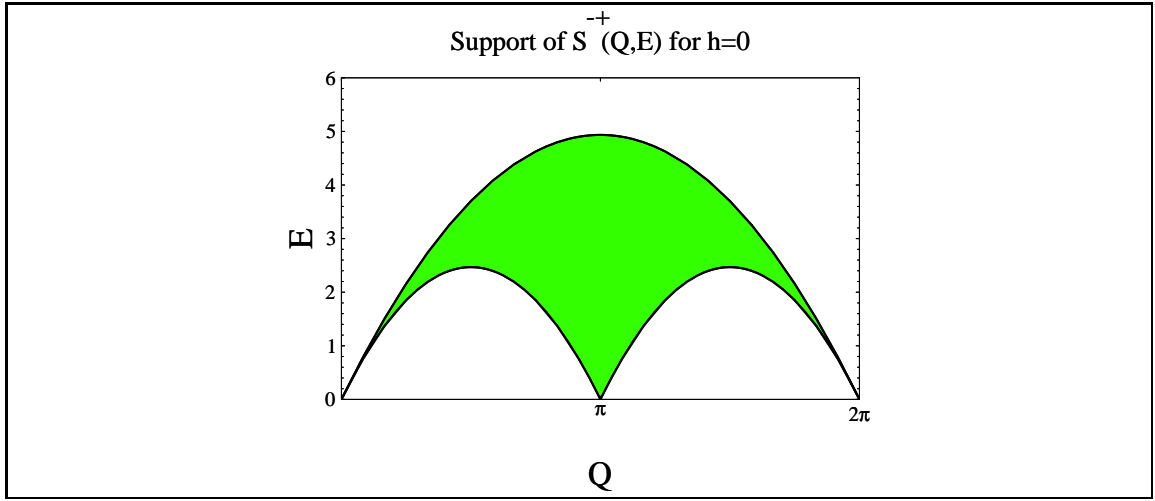}}
}
\caption{The shaded region show where $S^{-+}(Q,E)$ is nonzero for $h=0$.  The
top boundary corresponds to excitations with two spinons that have identical
momentum; on the bottom boundary one of the spinons has fixed momentum $\pm
\pi$. $E$ is given in units of $({\scriptstyle \frac{2\pi}{N}})^2 
\frac{v_s}{\pi}$.}
\label{smphne0}
\end{figure}

Since we know the spinon dispersion relation, we can demarcate the support of
$S^{-+}(Q,E)$ in the $(Q,E)$-plane in Fig.\ \ref{smphne0}.
The main steps of
the computation involving the matrix elements $M_{\nu}^{+}$, i.e.\ the weight of the
DSF at a point $(Q,E)$ on the plot, are the following:  since
$S^{+}_{i}\left|{\rm GS}\right.  \rangle$ only contains states with two fully
polarized spinons, it must be built out of YHWS.  These wavefunctions are
functionally identical to eigenfunctions of the Calogero-Sutherland model at
coupling $\lambda =2$ of particles moving on a ring.  Since both wavefunctions
are of a polynomial form with degree $<N$ the computation involving 
a sum over sites
is identical to taking an integral over the ring in the continuum model:
$\int_0^N dn\, e^{ikn}=\sum_{n=0}^{N-1} e^{ikn}$ if $k\in
\{-(N-1),\ldots,N-1\}$. The
action of $S^{+}_{i}$ in the spin chain is translated into a particle
destruction operator $\Psi(x,t)$ ($ S^{+}_{i}$ removes a down spin).  So the
$S^{-+}(Q,E)$ DSF reduces to the Greens function $\langle\left.  {\rm
GS}\right| \Psi^{\dagger}(x,t)\Psi(0,0)\left|{\rm GS}\right.  \rangle$ in the
Calogero-Sutherland model.  It can be computed in the thermodynamic limit, in
which case it can be mapped unto a Gaussian hermitian matrix model correlator.
The result is:

\begin{equation}
\label{psidagpsi}
S^{-+}(Q,E) = \frac{1}{16\pi}\left(\frac{(v_1-v_2)^2}{( v_s ^2 -v_{1}^{2})( v_s
^2-v_{2}^{2}) }\right)^{\frac{1}{2}},
\end{equation}
with $ Q= -\frac{\pi}{2 v_s} (v_1+v_2)$ and $E=\epbar(v_1)+\epbar(v_2)$.
The DSF matrix element is parametrized by $v_1,v_2$, which are quickly
identified with the velocities of the two spinons in the excited state.  The
$\langle\Psi^\dagger\Psi\rangle$ Greens function has been obtained recently at
finite $N$ as well: both the groundstate, the YHWS $|\nu
\rangle$ it connects
with, and the action of $S^+_i$ are expressible as symmetric Jack-polynomials,
and the matrix element is essentially the Clebsch-Gordan coefficient of
the three \cite{Ha94}.

For $h\neq 0$ the three different structure functions $S^{-+}(Q,E)$,
$S^{zz}(Q,E)$ and $S^{+-}(Q,E)$ will not be equal, since $\left|{\rm GS}\right.
\rangle$ is no longer a singlet.  In fact it has $ S^{z} =
S^{tot}=\frac{N_{sp}}{2}\equiv S_0$, where $N_{sp} = N_{sp} (h)$ is given by
(\ref{hcdef}).  This difference between the three correlation function is also
expressed in two additional global $SU(2)$ selection rules \cite{M81}, which
rule out certain matrix elements based on the total spin $S^{tot}$ and $ S^{z}$
of the final state $\mu$ {\em inside} the Yangian multiplet $\Gamma$.

In the first place there is the Wigner-Eckart theorem for vector operators such
as $S^{a}_{i}$, which tells us that in order for $\langle\left.  \Gamma
,\mu\right| S^{a}_{i}\left|{\rm GS}\right.  \rangle$ to be nonzero the total
spin $S$ of $\left|\Gamma,\mu\right.  \rangle$ must satisfy $S_0 -1 \leq S\leq
S_0 +1$.  Secondly, for any state in a multiplet with $N_{sp}$ spinons, we have
$\frac{N_{sp}}{2}\geq S\geq S^{z}$---where equality only holds for the YHWS,
which has all its spinons up.  Now put this together with the fact that
$ S^{+}_{i}$ raises $ S^{z}$ by +1, $ S^{-}_{i}$ lowers it by 1, and $
S^{z}_{i}$ leaves it the same.  Classifying states according to their $ S^{z}$
and $S^{tot}$ as types $(i)$ - $(vi)$, following M\"{u}ller et al.\ \cite{M81}
we find the following contributions:
\begin{description}
\item[\underline{$ S^{+}_{i}\left|{\rm GS}\right.  \rangle$}] contains states
with $S=S_0+1$ and $\Delta N_{sp} =+2$ (type $(iii)$ ).
\item[\underline{$ S^{z}_{i}\left|{\rm GS}\right.  \rangle$}] contains states
with $S=S_0+1$ and $\Delta N_{sp} =+2$ (type $(i)$ ) or $S=S_0$ and $\Delta
N_{sp} =+2,0$ (type $(ii)^{a,b}$ ).
\item[\underline{$ S^{-}_{i}\left|{\rm GS}\right.  \rangle$}] contains states
with $S=S_0+1$ and $\Delta N_{sp} =+2$ (type $(iv)$ ) or $S=S_0$ and $\Delta
N_{sp} =+2,0$ (type $(v)^{a,b}$ ), or $S=S_0-1$ and $\Delta N_{sp} =+2,0,-2$
(type $(vi)^{a,b,c}$ ).\\
\end{description}
Since we have an additional quantum number to label states:  $N_{sp}$, we added
Latin superscripts $a,b,c$ to the Roman numerals.  All 10 contributions are
summarized in table (\ref{sumtable}).

\begin{table}
\caption{List of matrix elements contributing to the DSF.  The correlation
functions in the last column refer to those in the Calogero-Sutherland model as
they appear  in eqs.\ (\protect\ref{psidagpsi},
\protect\ref{psidagpsisz}, \protect\ref{rhorho}, \protect\ref{psidagpsismin},
\protect\ref{rhorhosmin}, \protect\ref{psipsidag}).  The entries marked with a
$(\dag )$ do not survive in the thermodynamic limit.  For the example
occupation sequences we act on the specific groundstate:
$000\protect\rule[-.25em]{.3ex}{1.1em}
101010101\protect\rule[-.25em]{.3ex}{1.1em} 000$, with the symbols
$\protect\rule[-.25em]{.3ex}{1.1em}$ delimiting the center region.  In the
column under ``Excitation'' ${\cal S}$ denotes a spinon and ${\cal M}$ denotes
a magnon.  }
\vspace{1em}
\psfig{file=table2.epsi,width=\hsize}
\samepage
\comment{
\begin{tabular}{ccccccc}
\hline
\hline
DSF & Type & $S^{tot}$ & $\Delta N_{sp} $ &Typical contributing & Excitation &
Matrix element \\ 
    &      &           &            & Yangian multiplet   &            &    \\
\hline
$S^{-+}(Q,E)$ & $(iii)$ &$S=S_0 +1$& $\Delta N_{sp} =+2$ &
$000\rule[-.25em]{.3ex}{1.1em} 100100101\rule[-.25em]{.3ex}{1.1em} 000$ &
$2{\cal{S}}$ &$\langle\Psi^{\dag}\Psi\rangle$ \\
\nopagebreak
\rule{2em}{0em}${\scriptstyle ( S^{z} = S_0 +1)}$& & & & & &\\
\nopagebreak
& & & & & &\\
\nopagebreak
$S^{zz}(Q,E)$ &$(i)^{\dag}$ & $S=S_0 +1$ & $\Delta N_{sp} =+2$
&$000\rule[-.25em]{.3ex}{1.1em} 100101001\rule[-.25em]{.3ex}{1.1em} 000$ &
$2{\cal{S}}$ & $\frac{1}{2( S_{0} +1)}\langle\Psi^{\dag}\Psi\rangle$ \\
\nopagebreak
\rule{2em}{0em}${\scriptstyle ( S^{z} = S_0)}$ & $(ii)^{a}$& $S=S_0 $ &
$\Delta N_{sp} =+2$ &$000\rule[-.25em]{.3ex}{1.1em}
100101001\rule[-.25em]{.3ex}{1.1em} 000$ & $2{\cal{S}}$ & \\
\nopagebreak
&$(ii)^b$ & $S=S_0$ & $\Delta N_{sp} =0$ & $100\rule[-.25em]{.3ex}{1.1em}
101001001\rule[-.25em]{.3ex}{1.1em} 000$ & $2{\cal{S}} + {\cal{M}}$ &
$\langle\rho\rho\rangle$ \\
\nopagebreak
& & & & & &\\
\nopagebreak
$S^{+-}(Q,E) $ & $(iv)^{\dag}$& $S=S_0 +1$ & $\Delta N_{sp} =+2$
&$000\rule[-.25em]{.3ex}{1.1em} 100100101\rule[-.25em]{.3ex}{1.1em} 000$ & $
2{\cal{S}}$ & $\frac{1}{( S_{0} +1)(2 S_{0} + 1)}\langle\Psi^{\dag}\Psi\rangle
$ \\
\nopagebreak
\rule{2em}{0em}${\scriptstyle ( S^{z} = S_0 -1)}$ & $(v)^{a\dag}$& $S=S_0 $ &
$\Delta N_{sp} =+2$ & $000\rule[-.25em]{.3ex}{1.1em}
010010101\rule[-.25em]{.3ex}{1.1em} 000$ & $2{\cal{S}}$ & \\
\nopagebreak
& $(vi)^a$& $S=S_0 -1$ & $\Delta N_{sp} =+2$ &$000\rule[-.25em]{.3ex}{1.1em}
010010101\rule[-.25em]{.3ex}{1.1em} 000$ & $2{\cal{S}}$ & \\
\nopagebreak
& $(v)^{b\dag}$& $S=S_0 $ & $\Delta N_{sp} =0$ & $000\rule[-.25em]{.3ex}{1.1em}
100100101\rule[-.25em]{.3ex}{1.1em} 010$ & $2{\cal{S}} + {\cal{M}}$ &
$\frac{2}{ S_{0}}\langle\rho\rho\rangle $ \\
\nopagebreak
& $(vi)^b$& $S=S_0 -1$ & $\Delta N_{sp} =0$ & $100\rule[-.25em]{.3ex}{1.1em}
100101001\rule[-.25em]{.3ex}{1.1em} 000$& $2{\cal{S}} + {\cal{M}}$ & \\
\nopagebreak
& $(vi)^c$& $S=S_0 -1$ & $\Delta N_{sp} =-2$ & $100\rule[-.25em]{.3ex}{1.1em}
100100101\rule[-.25em]{.3ex}{1.1em} 010$ & $2{\cal{S}} +
2{\cal{M}}$, ${\cal{M}}$ & $\langle\Psi\Psi^{\dag}\rangle$\\
\hline\hline
\end{tabular}
}
\label{sumtable}
\end{table}

We will now investigate all three structure functions individually following
these selection rules.

\subsection{${\bf S^{-+}(Q,E)}$}
\label{smpsec}

\vspace{2.5ex}
\underline{Type $(iii)$: $\Delta N_{sp} =+2$, $\Delta S=+1$.} 

For the occupation sequence of the groundstate in a given magnetic field (e.g.\ \\
000010101010000) let us label the zeroes in the leftmost orbital as the left
spinon condensate and the ones in the rightmost orbital as the right spinon
condensate.  From table \ref{sumtable} we learn that action of $ S^{+}_{i}$ on
the groundstate only produces states with two more spinons, i.e.\ one less 1.
This 1 has to come out of the center $\cdots 10101\cdots$ region.  We can't
take more than a single 1 out of the center region---and place it into the left
or right spinon condensate---since this would imply a violation of Rule
\ref{rule11} applied to the center region.  Taking out a 1 in the center region
is equivalent to inserting 2 spinons there.  A typical nonzero matrix element
would be $\langle\left.  0001001001000,\mu\right| S^{+}_{i} \left|
0001010101000\right.  \rangle$, with the two spinons residing in orbital two
and three.

All this means is that we get a simple two spinon spectrum, as in the zero
magnetic field case.  The only difference is that now the momenta of the
spinons can only vary from $-k_0$ to $k_0$, where
$k_0=\frac{\pi}{4N}(N-N_{sp})$ decreases with increasing magnetic field as
$N_{sp}=N_{sp}(h)$ according to eq.\ (\ref{hcdef}).  The support of
$S^{-+}(Q,E)$ is essentially a squeezed version of Fig.\ \ref{smphne0}.

As for the weight associated with 2-spinon excitations:  the calculation for
the zero magnetic field case (\ref{psidagpsi}) carries over without problems.
The reason for this is twofold.  In the first place, the nonzero magnetic
field groundstate wavefunction (\ref{GSdef}) is of the same Jastrow-form as the
zero field one, with just an extra phase factor
$\prod_{i}z_{n_i}^{\frac{N}{2}-M}$ appended.  When we take the matrix element,
the phase factors from ket and bra part cancel each other.  Secondly, the
excited states are again of the YHWS type $(S=\frac{N_{sp}}{2})$ and have to be
polynomials.  The mapping onto a Calogero-Sutherland model matrix element
remains therefore legitimate.

The contribution of just two-spinon YHWS excitations was to be expected since
we can expand any fully polarized localized spinon wavefunction with $N_{sp}$
spinons acted upon with $ S^{+}_{i}$ in terms of a set containing $N_{sp}+2$
spinons:

\begin{eqnarray}
\lefteqn{
( S^{+}_{i}\Psi_{\alpha_i})( n_{1},\ldots,n_{M-1} ) =}\nonumber\\
&& \sum_{p\in V}\prod_{r\in V}\frac{(z_r-z_i)}{(z_r-z_p)} \Psi(
n_{1},\ldots,n_{M-1} |\alpha_1,\ldots,\alpha_{N_{sp}},i,p).
\label{spin2exp}
\end{eqnarray}
Here $V$ is a set of $M$ random sites on the circle excluding the spinon sites
$\{\alpha_i\}$ (see Appendix B).  Eq.  (\ref{hnullspexp}) 
is a special case of
this expansion with $V$ equal to the sites that are an even number of steps
removed from the site on which the local spin operator acts.  We could also
have realized that the number of spinons can't go up by more than 2 when we
consider that $ S^+$ and $J^{+}_{1}$ annihilate $ S^{+}_{i}\left|{\rm
GS}\right.  \rangle$, indicating that the latter must consist of purely YHWS
{\em with} $ S^{z} = S_{0} +1$!

\subsection{${\bf S^{zz}(Q,E)}$}
\label{szzsec}

For the DSF $S^{zz}(Q,E)$ we find similar simple excitations that contribute,
although at present not all resulting matrix elements can be computed.  From
the combined selection rules we find three types of excitations:  $(i)$,
$(ii)^a$ and $(ii)^b$, in table \ref{sumtable}.  They all have in common that
$\Delta N_{sp} =0$ or +2.  This isn't surprising:  let us consider the state
$J^{+}_{1}\left( S^{z}_{i}\left|{\rm GS}\right.  \rangle\right)$.  This state
is annihilated by $ J^+_1$ and $ S^+$, so it must be YHWS (see Appendix
C).
Therefore, since the first action of $ J^+_1$ doesn't change the number of
spinons, $ S^{z}_{i}\left|{\rm GS}\right.  \rangle$ must be a mixture of states
that contain not more than $2S_0 +2$ spinons, where $2S_0$ is the number of
spinons in the groundstate.

We will now discuss the individual types and where possible compute the values
of the matrix elements.

\vspace{2.5ex}
\underline{Type $(i)$: $\Delta N_{sp} =+2$, $\Delta S=+1$.} 

Having identical selection rules, these states sit in the exact same Yangian-
and spin multiplets as the type $(iii)$ states, and therefore they also contain
two additional spinons.  However their $ S^{z} =S^{tot} -1$, so they are no
longer YHWS like the type $(iii)$ states.  Nevertheless they are related by a
simple application of $ S^{-}$:

\begin{equation}
|{\scriptstyle \Gamma,\Delta N_{sp} =+2},\!
		\renewcommand{\arraystretch}{.5}
		\begin{array}{c}
		{\scriptscriptstyle S^{tot}= S_0 +1} \\
		{\scriptscriptstyle  S^{z} =S_0}
	        \end{array} \rangle
= {\scriptstyle \frac{1}{\sqrt{2(S_0 +1)} }}  S^{-}\left|\Gamma\right. \rangle,
\end{equation}
where $\left|\Gamma\right.  \rangle$ denotes the YHWS of the multiplet with
occupation sequence $\Gamma$ of type $(iii)$.  This allows us to reduce a type
$(i)$ matrix element to one that is a multiple of a type $(iii)$ given in eq.\
(\ref{psidagpsi}):
\begin{eqnarray}
\left|\langle{\scriptstyle \Gamma,\Delta N_{sp} =+2} ,\!
		\renewcommand{\arraystretch}{.5}
		\begin{array}{c}
		{\scriptscriptstyle S^{tot}= S_0 +1} \\
		{\scriptscriptstyle  S^{z} =S_0}
	        \end{array} |  
S^{z}_{i}\left|{\rm GS}\right. \rangle\right|^2
&=& {\scriptstyle \frac{1}{2(S_0 +1)}}\left|\langle\left.
\Gamma\right | [ S^+, S^{z}_{i} ]\left|{\rm GS}\right.  \rangle\right|^2
\nonumber\\
&=&{\scriptstyle \frac{1}{2(S_0 +1)}}\left|\langle\left.
\Gamma\right| S^{+}_{i}\left|{\rm GS}\right.  \rangle\right|^2,
\label{psidagpsisz}
\end{eqnarray}
where we used that $ S^+$ annihilates the groundstate.

\vspace{2.5ex}
\underline{Type $(ii)^a$: $\Delta N_{sp} =+2$, $\Delta S=0$.}

This type of state is a member of the same kind of {\em Yangian} multiplet as
type $(iii)$ and $(i)$---i.e.\ with two extra spinons---but it sits in a {\em
spin} multiplet that does not contain the YHWS.  
\begin{figure}[htb]
\framebox[6in]{\centerline{\psfig{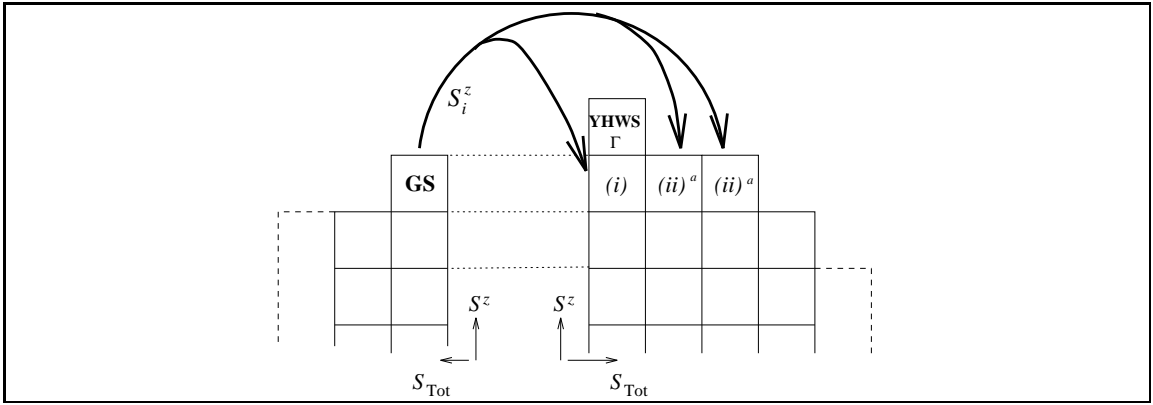}}}
\caption{Graphical illustration of the spin multiplets accessed in
$S^z_i$-excitations of types $(i)$ and $(ii)^a$.}
\label{yhwsfig}
\end{figure}
States of this kind are supplied by the Algebraic Bethe Ansatz, eq.\
(\ref{ABEdef}): $T^-(\lambda_0)|\Gamma\rangle$. $\lambda_0$ is given
by eq.\ (\ref{ABAeq}) for $n=1$: $\frac{P(\lambda_0+1)}{P(\lambda_0)}=1$,
where $P(u)$ is the Drinfel'd polynomial of occupation sequence
$\Gamma$. This equation has a number of {\em trivial} solutions, when
$\lambda_0$ corresponds to two adjacent Drinfel'd roots. These trivial
solutions are to be discarded (they correspond to an orbital with more
than one spinon in it; these spinons  
can {\em only be in a symmetric state}, and there is a reduction in the
naively expected number $2(S_0+1)-1$ of $S=S_0$ states). 
The problem is, how to commute the $T^-(\lambda_0)$ through $S^z_i$ in the
matrix element $\langle{\rm GS}|S^z_i T^-(\lambda_0)|\Gamma\rangle$.

\vspace{2.5ex}
\underline{Type $(ii)^b$: $\Delta N_{sp} =0$, $\Delta S=0$.}

Since this type of state has $\Delta N_{sp} =0$ the number of ones in its
occupation sequence must be identical to that in the groundstate.  As with the
previous three types we can delete just a single 1 from the center, leaving
behind two spinons pointing up.  This 1 then must go into either the left or
the right spinon condensate.  Therefore a typical nonzero matrix element would
be:  $\langle\left.00100\rule[-.25em]{.3ex}{1.1em}
1001001\rule[-.25em]{.3ex}{1.1em} 00000\right| S^{z}_{i}\left|
00000\rule[-.25em]{.3ex}{1.1em} 1010101\rule[-.25em]{.3ex}{1.1em} 00000\right.
\rangle$ where the $\rule[-.25em]{.3ex}{1.1em}$ just helps to draw attention to
the center region.  Rule \ref{rule11} rules out any additional ones leaving the
center region.  The additional 1 on the left or right can be interpreted as a
{\em magnon} with $ S^{z} = -1$.  The limiting case where the magnon ``fuses''
with two spinons at a boundary between a condensate and the center region gives
us the groundstate.

These states must be YHWS since they have $\frac{N_{sp}}{2} =S= S^{z}$, like
$\left|{\rm GS}\right.  \rangle$.  This fact allows us to calculate the
corresponding matrix elements.  Since both groundstate and excited state are
YHWS, a mapping onto the Calogero-Sutherland model is valid.  In this case we
need a groundstate density-density correlator $\langle\left.  {\rm GS}\right|
\rho(x,t)\rho(x',t')\left|{\rm GS}\right.  \rangle$ since $ S^{z}_{i}$
measures the presence or absence of a down-spin (i.e.\ a particle in the
CS-model).  This calculation has been done by Altshuler et al.\ \cite{A93}, in
the thermodynamic limit, by studying the repulsion of energy levels in a
random matrix model under a varying perturbation.  The energy levels are
identified with the positions of the particles and the strength of the
perturbation corresponds to imaginary time.  Their original expression depends
on three parameters (called $\lambda,\lambda_1$ and $\lambda_2$), the latter
two of which are compact, and the first one is unbounded.  This is precisely
what we expect from our selection rule:  2 spinons restricted to the center
region with momenta in the range $-k_0\ldots +k_0$ and a magnon that can go
off all the way to the right or left (i.e.\ $\pm \infty$ in the thermodynamic
limit).  In terms of the velocity, $v$, of the magnon---with dispersion
relation (\ref{HISEdunkl})---and spinon velocities $v_1,v_2$ 
their result is as
follows:

\begin{eqnarray} \lefteqn{\left|\langle\left.  v,v_1,v_2\right|
\rho(Q)\left|{\rm GS}\right.  \rangle\right|^2 \propto}\nonumber\\
&&\frac{(v- v_s )(v+ v_s )|v_1-v_2|}{\sqrt{(v_s^2-v_1^2)(v_s^2-v_2^2)}
(v-v_1)^2(v-v_2)^2} \left[
(v-v_1)+(v-v_2)\right]^2 .
\label{rhorho}
\end{eqnarray}
$E=\epbar(v_1)+\epbar(v_2)+\epsilon(v)$ and
$Q=\frac{\pi}{v_s}(v-\half(v_1+v_2))$.
For completeness we give the relation between the $v$'s and the
$\lambda$'s :

\begin{eqnarray}
\lambda & =& v/ v_s \nonumber\\
\lambda_1\lambda_2 & = &\mbox{$\frac{(v_1 +v_2)}{2 v_s }$}\nonumber\\
(1-\lambda_{1}^{2})(1-\lambda_{2}^{2}) & = & \mbox{$\left(\frac{v_1 -v_2}{2
v_s }\right)^2 $}.
\end{eqnarray}
\begin{figure}[htb] 
\framebox[\hsize]{\centerline{\psfig{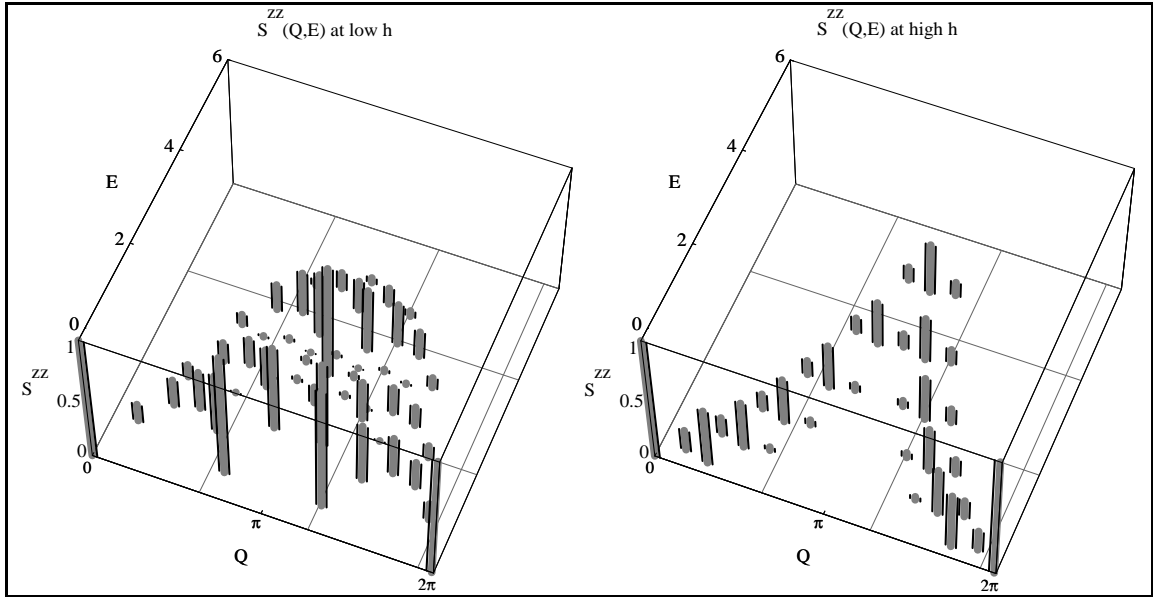}}}
\caption{ $S^{zz}(Q,E)$ for small $h$ $(\sigma = \frac{S_z}{N}=.05)$ and large
$h$ $(\sigma=.4)$ on $N=14$ sites. $E$ is in units of
$({\scriptstyle\frac{2\pi}{N}})^2 \frac{v_s}{\pi}$.}
\label{szzdata} 
\end{figure}
Numerical data for $S^{zz}(Q,E)$ can be found in fig.\ (\ref{szzdata}) 
at values
of $h$ close to 0 and $h_c$.  Fig.\ (\ref{szzsup}) shows the corresponding
support of $S^{zz}(Q,E)$ in the $(Q,E)$ plane, as predicted from the selection
rule, in the thermodynamic limit.  We notice that for finite-size systems some
of the features near the lower boundary are absent.
\begin{figure}[htb]
\framebox[\hsize]{\centerline{\psfig{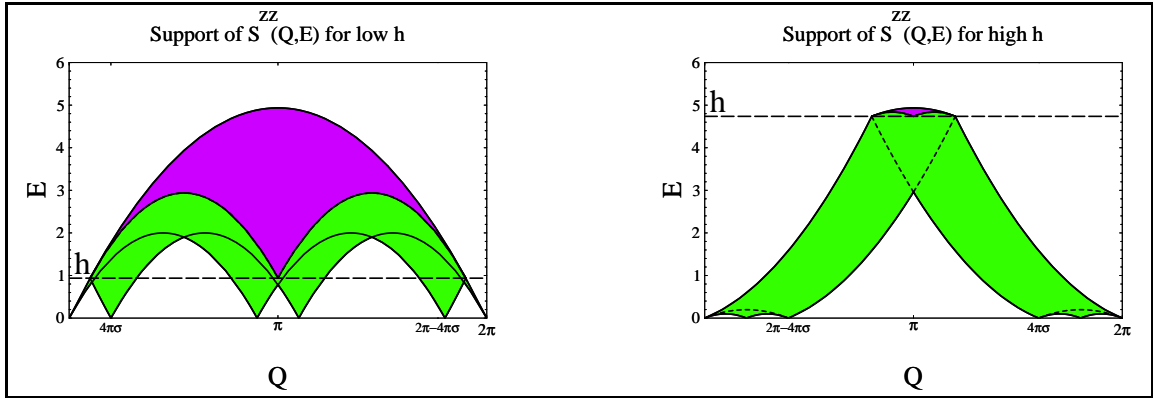}}}
\caption{The regions where $S^{zz}(Q,E)$ is nonzero for low and high $h$.  The
area shaded dark contains the contributions from the excitations with $\Delta
N_{sp} = +2$ (types ($(i)$ and $(ii)^a$).  The excitations with $\Delta
N_{sp}=0$---type $(ii)^b$---can also occupy the lightly shaded area.}
\label{szzsup} 
\end{figure}

\subsection{${\bf S^{+-}(Q,E)}$}
\label{spmsec}

Finally we discuss the $S^{+-}(Q,E)$ DSF which is governed by the excitations
of types $(iv)$ - $(vi)^c$ that are present in $ S^{-}_{i}\left|{\rm GS}\right.
\rangle$.

\vspace{2.5ex}
\underline{Type $(iv)$: $\Delta N_{sp} =+2$, $\Delta S=+1$.}

This type of state is very similar to types $(iii)$ and $(i)$, as a matter of
fact they all reside in identical Yangian- and spin multiplets.  Therefore type
$(iv)$ states differ from the groundstate only by two extra spinons in the
center region.  They are related to their YHWS $\left|\Gamma\right.  \rangle$
(of type $(iii)$) through:

\begin{equation} 
| {\scriptstyle \Gamma,\Delta N_{sp} =+2},\!  
\renewcommand{\arraystretch}{.5}
\begin{array}{c}
{\scriptscriptstyle S^{tot}= S_0 +1} \\ 
{\scriptscriptstyle S^{z} =S_0 -1} 
\end{array} 
\rangle ={\scriptstyle \frac{1}{\sqrt{(4S_0 +2)(2S_0 +2)}}}
(S^{-})^2\left|\Gamma\right.  \rangle.  
\end{equation} 
Analogous to the calculation for type $(i)$ states, a matrix element of type
$(iv)$ can now easily be reduced to one involving type $(iii)$:

\begin{eqnarray}
\left|\langle{\scriptstyle \Gamma,\Delta N_{sp} =+2},\!
		\renewcommand{\arraystretch}{.5}
		\begin{array}{c}
		{\scriptscriptstyle S^{tot} = S_0 +1} \\
		{\scriptscriptstyle  S^{z} =S_0 -1}
	        \end{array} | 
S^{z}_{i} \left|{\rm GS}\right. \rangle\right|^2 
&=&{\scriptstyle \frac{1}{(4S_0 +2)(2S_0 +2)}}\left|
\langle\left.  \Gamma\right| ( S^+)^2 S^{-}_{i}\left|{\rm GS}\right.
\rangle\right|^2 \nonumber \\
&=& {\scriptstyle \frac{1}{(2S_0 +1)(S_0
+1)}}\left|\langle\left.  \Gamma\right| S^{+}_{i}\left|{\rm GS}\right.
\rangle\right|^2.
\label{psidagpsismin}
\end{eqnarray}
The last matrix element in this equation has already been computed for the
$S^{-+}(Q,E)$ DSF.  However the energy of the corresponding excited state here
is shifted by $2h$ in comparison, because of the Zeeman term in the
Hamiltonian.

\vspace{2.5ex}
\underline{Types $(v)^a$ and $(vi)^a$:$\Delta N_{sp} =+2$, $\Delta S=0, -1$.}

States of types $(v)^a$ and$(vi)^a$ contain a two-spinon excitation as
well---like types $(iii)$, $(i)$ and $(iv)$.  However, since they don't reside
in the spin multiplet of their YHWS ($\frac{N_{sp}}{2} >S$), the associated
matrix elements are unknown. Type $(v)$ can be related to those of type
$(ii)^a$, though.
\begin{equation}
\left|\langle{\scriptstyle \Gamma,\Delta N_{sp}=0},\!
		\renewcommand{\arraystretch}{.5}
		\begin{array}{c}
		{\scriptscriptstyle S^{tot}= S_0} \\
		{\scriptscriptstyle  S^{z} =S_0 -1}
	        \end{array} |S^-_i |\GS\rangle\right|^2 =
{\scriptstyle \frac{1}{2S_0}}
\left|\langle{\scriptstyle \Gamma,\Delta N_{sp}=0},\!
		\renewcommand{\arraystretch}{.5}
		\begin{array}{c}
		{\scriptscriptstyle S^{tot}= S_0} \\
		{\scriptscriptstyle  S^{z} =S_0 }
	        \end{array} |S^z_i |\GS\rangle\right|^2 
\end{equation}
Type $(vi)^b$ is even worse, since it involves the application of two
$T^-(\lambda)$ operators.

\vspace{2.5ex}
\underline{Types $(v)^b$: $\Delta N_{sp} =0$, $\Delta S=0$.}

Type $(v)^b$ states are very similar to those of type $(ii)^b$, they only
differ in $ S^{z}$ by -1.  Therefore both contain two excited spinons and a
single left- or right moving magnon.  They are related to each other by:

\begin{equation}
| {\scriptstyle \Gamma,\Delta N_{sp} =0},\!
		\renewcommand{\arraystretch}{.5}
		\begin{array}{c}
		{\scriptscriptstyle S^{tot}= S_0} \\
		{\scriptscriptstyle  S^{z} =S_0 -1}
	        \end{array} \rangle
={\scriptstyle \frac{1}{\sqrt{2S_0}}} S^{-}\left|\Gamma\right. \rangle,
\end{equation}
and $\left|\Gamma\right.  \rangle$ is the type $(ii)^b$ YHWS of the multiplet
with occupation sequence $\Gamma$.  We can now trivially relate the matrix
elements of type $(ii)^b$ and $(v)^b$:

\begin{eqnarray}
\left|\langle{\scriptstyle \Gamma,\Delta N_{sp} =0},\!
		\renewcommand{\arraystretch}{.5}
		\begin{array}{c}
		{\scriptscriptstyle S^{tot}= S_0} \\
		{\scriptscriptstyle  S^{z} =S_0 -1}
	        \end{array} |  
S^{-}_{i}\left|{\rm GS}\right. \rangle\right|^2 
&=&{\scriptstyle \frac{1}{2S_0}}\left|\langle\left.
\Gamma\right|[ S^+, S^{-}_{i}]\left|{\rm GS}\right.  \rangle\right|^2
\nonumber\\
&=&{\scriptstyle \frac{2}{S_0}}\left|\langle\left.\Gamma\right|
S^{z}_{i}\left|{\rm GS}\right.  \rangle\right|^2 .
\label{rhorhosmin}
\end{eqnarray}
The last matrix element is listed in eq.\ (\ref{rhorho}).

\vspace{2.5ex}
\underline{Type $(vi)^b$: $\Delta N_{sp} =0$, $\Delta S=-1$.}

These states reside in the same Yangian multiplets as the previous type however
they are not in the spin multiplets of the YHWS:  $\frac{N_{sp}}{2}>S=S_0 -1$.
The problem is very similar to that for case $(ii)^a$.

\vspace{2.5ex}
\underline{Type $(vi)^c$: $\Delta N_{sp} =-2$, $\Delta S=-1$.}

This last class of states has $\Delta N_{sp} =-2$ and therefore the number of
ones in its occupation sequence goes up by one compared to the groundstate.
The selection Rule \ref{rule11} only allows the extra 1 to go into the left or
right spinon condensate.  As before we are also allowed to take a 1 out of the
center region and bring it into the left or right spinon condensate.  Notice
however that the rule forbids both of the ones to go into the {\em same}
condensate:  one has to be left moving and the other must be right moving.  The
result is an excited state with two magnons and two spinons.  A typical nonzero
matrix element would be $\langle\left.  0100\rule[-.25em]{.3ex}{1.1em}
100101001\rule[-.25em]{.3ex}{1.1em} 0001\right| S^{-}_{i}\left|
0000\rule[-.25em]{.3ex}{1.1em} 101010101\rule[-.25em]{.3ex}{1.1em} 0000\right.
\rangle$.  The $\rule[-.25em]{.3ex}{1.1em}$ just helps to guide the eye.  Also
present are YHWS of multiplets from the limiting cases where one of the magnons
fuses with the two spinons in the center; this leaves a multiplet with nothing
but one magnon; example:  $\langle\left.  0000\rule[-.25em]{.3ex}{1.1em}
101010101\rule[-.25em]{.3ex}{1.1em} 0010\right| S^{-}_{i}\left|
0000\rule[-.25em]{.3ex}{1.1em} 101010101\rule[-.25em]{.3ex}{1.1em} 0000\right.
\rangle$.  This single magnon excitation is familiar from the strong field
regime.

Since type $(vi)^c$ states are YHWS ($\frac{N_{sp}}{2}=S$), as is the
groundstate, we can repeat the calculation of the matrix elements by a mapping
onto the Calogero-Sutherland model.  Because $ S^{-}_{i}$ creates a down spin,
it corresponds to a particle creation operator in the CS-model.  The relevant
correlation function is therefore $\langle\left.  {\rm
GS}\right|\Psi(x,t)\Psi^{\dag}(x',t')\left|{\rm GS}\right.  \rangle$.  As in
the $\langle\Psi^{\dag}\Psi\rangle$ case for type $(iii)$ states, a further
mapping onto a Gaussian Hermitean matrix model allows one to calculate the
Fourier transform of this correlation function \cite{Z93}.  The result
consists of two parts $G^{(1)}$ and $G^{(2)}$. $G^{(2)}$ is 
parametrized by four variables, two of which are compact:  $|v_1|,|v_2|<v_s$,
 and two are noncompact:$v<-v_s,\,v'>v_s$:
\begin{eqnarray}
&&\hspace{-0.35truein}
\left|\langle\left.  v,v',v_1,v_2\right| S^{-}(Q)\left|{\rm
GS}\right.  \rangle\right|^2 =\hspace{0.2truein}
\left\{ \frac{(v-v_1)(v'-v_2)+(v'-v_1)(v-v_2)}{(v-v')}\right\}^2 \nonumber\\
&& \hspace{0.2truein}\times
\frac{1}{\sqrt{8\pi}}
\frac{(v^2-v_s^2)(v'^2-v_s^2)|v_1-v_2|}{
\sqrt{(v_s^2-v_1^2)(v_s^2-v_2^2)}\left[
(v-v_1)(v-v_2)(v'-v_1)(v'-v_2)\right]^2} 
\label{psipsidag}
\end{eqnarray}
Energy and momentum in terms of the $v$'s are given by
$Q=\frac{\pi}{v_s}(v+v'-\frac{1}{2} (v_1 + v_2))$ and $E=\frac{\pi}{2v_s} (v^2
+ v'^2 - \frac{1}{2}(v_{1}^{2}+v_{2}^{2}) -v_{s}^{2})$.  It is obvious that 
the
compact parameters are to be identified with the spinon velocities and the
non-compact ones with the magnon velocities.
$G^{(1)}$ is parametrized by just one variable $|v|>v_s$, which is noncompact
\begin{equation}
\left| \langle\left. v \right| S^-(Q)\left|\GS\right. \rangle\right|^2 = 
\frac{1}{2\pi}\frac{v-v_s}{v+v_s}.
\end{equation}
This contribution corresponds to the case where one of the magnons has
$|v|=v_s$, and 
``fuses'' with the two spinons, thus leaving but one magnon.
\begin{figure}[htb]
\framebox[\hsize]{\centerline{\psfig{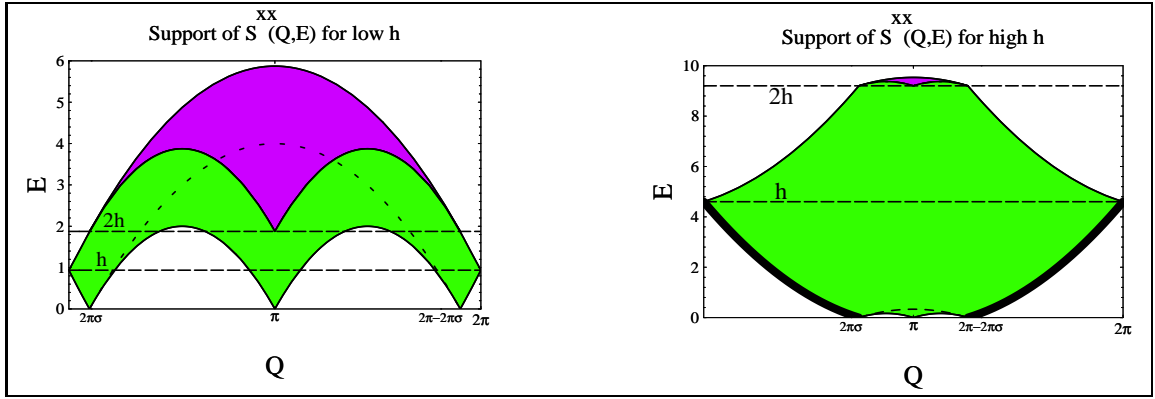}}}
\caption{Non-zero region of $4S^{xx}(Q,E)=S^{+-}(Q,E)+S^{-+}(Q,E)$ in the  
$(Q,E)$-plane for low and high
$h$.  The contributions of $\Delta N_{sp}=+2$ (types $(iv),(v)^a$ and $(vi)^a$)
live in the dark shaded region.  These excitations will survive in the limit
$h\rightarrow 0$.  The $\Delta N_{sp}=0,-2$ can also occupy the area shaded
light.  For $h\rightarrow h_c$ only the 1 magnon contributions survive, as can
be seen in Fig.\ \protect\ref{smindata}.  These high field magnon excitations
are indicated by the thick lines.}
\label{sminsup} 
\end{figure}
Since we now know all possible excitations contributing to $S^{+-}(Q,E)$ we can
draw its support in Fig.\ \ref{sminsup} for low and high values of $h$.  Fig.\
\ref{smindata} \footnote{ I would like to thank Dr.\ K.\ Lefmann and Dr.\
C.\ Rischel for pointing out an error in this figure, in the original
publication.} show numerical data on $S^{+-}(Q,E)$ for those values of $h$.
Table \ref{sumtable} summarizes the selection rules and available information
on matrix elements.
\begin{figure}[htb]
\framebox[\hsize]{\centerline{\psfig{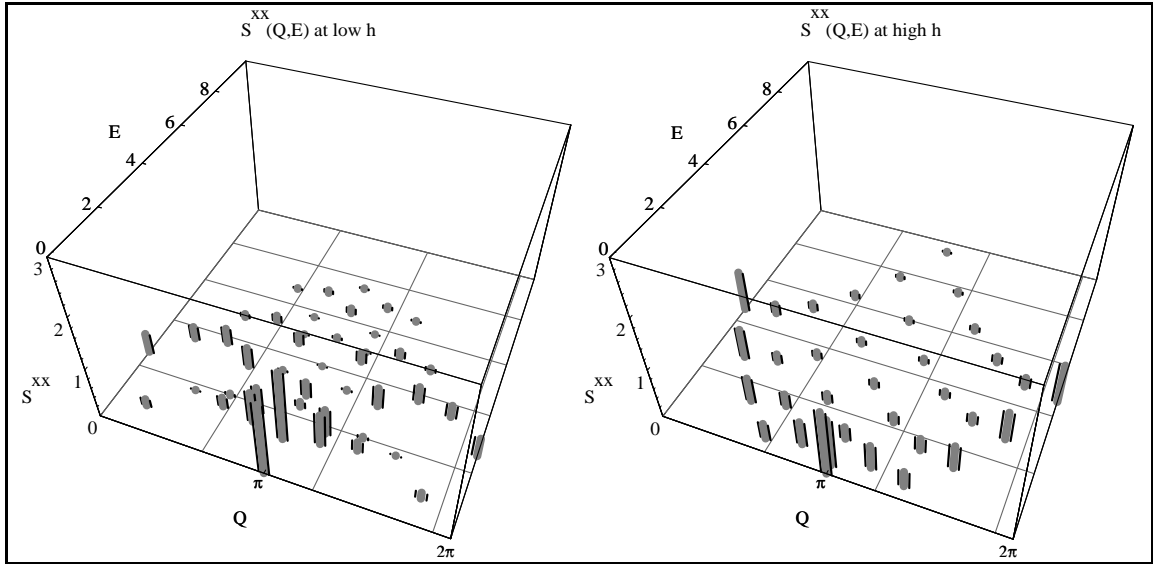}}}
\caption{$S^{xx}(Q,E)$ for low and high $h$, from numerical diagonalizations 
on a 10-site chain. $E$ is in units of $({\scriptstyle \frac{2\pi}{N}})^2
\frac{v_s}{\pi}$. }
\label{smindata}
\end{figure}

\section{Comparison to the Bethe Ansatz Model}
\label{ISEnNNE}

\begin{figure}[htb]
\framebox[\hsize]{\centerline{\psfig{file=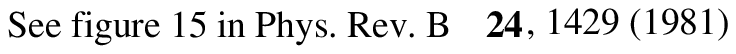,width=5.0in,angle=-90}}}
\caption{The $S^{zz}$ structure functions according to M\"{u}ller et
al., Phys.\ Rev.\ B {\bf 24}, 1429 (1981),
for $N=10$ in low ($S^z_{\rm GS}=2$) and high 
($S^z_{\rm GS}=4$) magnetic
fields. Notice the good match with its ISE counterpart fig.\
\protect\ref{szzdata}, except for the low weight NNE states outside the
2-spinon/1-magnon region. The solid lines are boundaries of regions that
receive contributions from just 2-spinon excitations. Their dispersion is
trigonometric, rather than quadratic, but qualitatively similar.}
\label{muellerfig}
\end{figure}

In 1980 M\"{u}ller et al.\ \cite{M81} did a similar calculation of DSFs for
the nearest neighbor Heisenberg chain.  They identified certain types of
states called {\em spin wave continuum} states (SWC) as carrying the dominant
contribution to the DSFs.  These SWC states can be described in a Bethe Ansatz
rapidity language by occupation sequences, just like the Yangian multiplets in
the ISE model.  As it turns out, these SWC states correspond to {\em exactly
the same } rapidity sequences that are favored by our selection rules in the
ISE-model!  Although the dispersion relations for the NNE Bethe Ansatz
rapidities are
different from those in the ISE-model, the authors find the support of the
DSFs in the nearest neighbor model to have essentially the same shape as we do
in the the inverse exchange case, see fig.\ \ref{muellerfig}.

However, in the nearest-neighbor Heisenberg chain there are some ``anomalous''
states, characterized by a change of more than $\pm 2$ spinons, which
contribute to a lesser degree to the structure functions and {\em don't} lie
within the bounds found by the authors.  In the ISE-model these contributions are
completely absent and once again we find this model to have a surprisingly
clean structure.  So in this sense the ISE-model is an ideal spinon
gas, whereas in the NNE Heisenberg chain the spinons interact.

\begin{figure}[htb]
\framebox[\hsize]{\centerline{\psfig{file=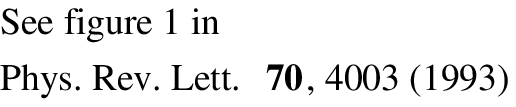,height=4.0in,angle=-90}}}
\caption{Neutron scattering data from a quasi one dimensional spin chain
${\rm KCuF}_3$ according to Tennant et al.\, Phys.\ Rev.\ Lett.\ {\bf
70}, 4003 (1993). 
Response is large in regions where the incoming neutrons can
excite just two spinons.}
\label{tenlow}
\end{figure}

The characteristic shape of the region in the $Q,E$ plane that carries all
(ISE) or most (NNE-model) weight has recently been seen experimentally too
\cite{T93}.
Tennant et al.\ scanned along a line in the energy-momentum region with the
help of neutrons scattering off ${\rm KCuF}_3$ at low (but not too low)
temperatures, see fig.\ \ref{tenlow}. 
The intensity obviously peaks, when
traversing the ``allowed region''. In the forbidden regions the intensity
drops, but not all the way to zero. This might happen
because ${\rm KCuF}_3$ is better described by a NNE model than by the
ISE-model, but the fact that $T\neq 0$ can cause a more smeared 
distribution too. Finally,
there may be non-magnetic scattering centers that contribute to the
structure function as well.

M\"{u}ller et al.\ also gave general rules, based on comparing Clebsch-Gordon
coefficients, determining which matrix elements will survive in the
thermodynamic limit.  Their conclusion is that the only surviving ones have
$S_{\rm Tot}= S^{z}$.  This means only excitations $(iii)$ for $S^{-+}(Q,E)$ ,
$(ii)^a$ and $(ii)^b$ for $S^{zz}(Q,E)$ and $(vi)^a$, $(vi)^b$ and $(vi)^c$ for
$S^{+-}(Q,E)$ remain relevant.  (Exceptions are single excitations with $Q=0$,
since these correspond to $ S^+$, $ S^{z}$ and $ S^{-}$ which give macroscopic
contributions, as we can see in the figures).  These contributions are trivial
to compute.

\section{Conclusion}

We found a remarkably simple selection rule for nonzero matrix elements of
local spin operators between eigenstates of the ISE-model, which is
reminiscent of the ideal gas single-particle selection rules.  One of the
consequences of this rule is that the {\em total} number of spinons can only
change by $0,\pm 2$.  Within the occupation sequences this holds {\em locally}
as well.

In the particular case that one of the states in the matrix element is also the
groundstate in a magnetic field (i.e.\ fully polarized spinons, condensed into
the left- and rightmost orbitals in equal amounts), the general selection rule
only allows excitations with no more than 2 spinons (the rule applied to the
center region) and one left- and one right moving magnon (the rule applied to
the condensates on the left and right) \footnote{
The particular application of our selection rule to matrix
elements of the form $\langle\left.\Gamma\right| S^a_i\left|{\rm GS}\right.
\rangle$ where $\left|{\rm GS}\right.  \rangle$ is the groundstate in a
magnetic field, $a=+$ and $\left|\Gamma\right.  \rangle$ is a {\em YHWS}, has
now been confirmed mathematically in \protect\cite{Ha94}
using the theory of Jack-polynomials. Additional evidence for the $S^z_i$ case
can be found in those references as well.
}.  This implies that the
structure functions based on the matrix elements involving these states have a
finite support in a region dictated by convolving the dispersion relations of
these particles (Figs.\ref{szzsup} and \ref{sminsup}).  These regions have the
same shape as those for the nearest neighbor Heisenberg chain.  The latter
model carries some weight outside these regions as well.  Therefore the
ISE-model has a much cleaner spinon structure than the NNE-model.

Matrix elements that connect a number of the states in these regions to the
groundstate through the local action of a spin operator have been presented.
However, information is lacking on those types that involve states not in the
spin multiplet of the YHWS.  This is particularly bothersome for types
$(ii)^a$, $(vi)^a$ and $(vi)^b$ since these will survive in the thermodynamic
limit.  Their calculation would allow a full reconstruction of the $h\neq 0 $
DFS for the ISE-model.  A more algebraic treatment involving Yangian operators should
provide more insight.

We believe that eventually all dynamical 
correlation functions of the model will be
explicitly found, namely $\langle S(Q,E)\rangle$ at finite magnetic field
{\em and} temperature. This is a goal that we recommend to future workers 
on the model. The ability to calculate matrix elements of $\svec_i$ between
non-YHW states will be a key requirement. 
\newpage

\section{Appendix.} 
\subsection{Appendix A.}
\label{app3A}

We show that localized spinon wavefunctions with $N_{sp}$ spinons all pointing
up are necessarily linear combinations of YHWS with exactly $N_{sp}$
spinons.  This follows easily from the fact that both $ S^+$ and $J^{+}_{1}$
annihilate these states. The derivations is very similar to that of Appendix
A of chapter \ref{chapter1}. If we write $\Psi( n_1,\ldots,n_{M}
)=\psi(z_{n_1},\ldots,z_{n_M})$ where $z_{n} = \exp({\frac{2\pi i n}{N}})$ and

\begin{equation} 
\psi(w_1,\ldots,w_M)  =  \prod_{i<j}(w_i-w_j)^2\prod_{i}w_i  
\prod_{i=1}^{M}\prod_{j=1}^{N_{sp}}(w_i-z_{\alpha_j}), 
\end{equation}
then 
\begin{eqnarray}
( S^+\Psi)( n_{1},\ldots,n_{M-1} ) 
&=&\sum_{j=1}^{N}\Psi( n_{1},\ldots,n_{M-1} ,j)
 = \sum_{j=1}^{N}\psi(
z_{n_{1}},\ldots,z_{n_{M-1}} ,e^{\frac{2\pi ij}{N}})\nonumber\\
& = &\psi(z_{n_1},\ldots,z_{n_M},0) =0,
\end{eqnarray}
(where the $\sum_{j}$ was recognized as the zero-mode of a Fourier expansion).
And with $w_{jk} = \frac{z_j + z_k}{z_j - z_k}$:
\begin{equation}
(J_{1}^{+}\Psi)( n_{1},\ldots,n_{M-1} )\sim
 -\sum_{j=1}^{M}\sum_{i\neq n_j}^{N} w_{n_j,i} \Psi(
n_{1},\ldots,n_{M-1} ,i),
\end{equation}
where we used the fact that $2 S^{z}_{i} = 1-2\sum_{j=1}^{M}\delta_{i,n_j}$ and
that $\Psi$ is a symmetric function which vanishes when two of its arguments
coincide. Employing eq.\ (\ref{cotconv}), we have $J^{+}_{1}\Psi = 0$.
since $\psi$ has a double zero when two of
its arguments coincide and it vanishes at $z=0$.

\subsection{Appendix B.}
\label{app3B}

In this Appendix we want to prove eq.\ (\ref{spin2exp}).  Let us first
introduce the following identity which holds for any set of distinct complex
numbers $\{ \omega_i\}$:

\begin{equation}
\prod_{j=1}^{M-1}(Z_j-z)=\sum_{k=1}^{M}\prod_{l(\neq k)}^{M}\frac{\omega_l
-z}{\omega_l - \omega_k}\prod_{i=1}^{M}(Z_i -\omega_k).
\label{lagrid}
\end{equation}
The RHS is just the Lagrange interpolation formula applied to
the function in $z$ on the LHS!  

Say we want to write $S_{i}^{+}\Psi_{\{\alpha_i\} }$ as a linear combination of
localized spinon wavefunctions with two more spinons than $\Psi_{\{\alpha_i\}
}$, and all spinons pointing up.  We fix one of the additional two
spinons at $i$, the site on which $S^{+}_{i}$ acts, i.e.\ :

\begin{eqnarray}
( S^{+}_{i}\Psi_{\{\alpha_i\} })( n_{1},\ldots,n_{M-1} )
&\equiv& \Psi( n_{1},\ldots,n_{M-1} ,i|\alpha_1,\ldots,\alpha_{N_{sp}})
\nonumber\\
&=&\sum_{p(\neq i,\{\alpha_k\})}a_p\Psi( n_{1},\ldots,n_{M-1}
|\alpha_1,\ldots,\alpha_{N_{sp}},i,p).\nonumber\\
\end{eqnarray}
Using eq.\ (\ref{locspinwf}) we can divide out common factors of
$(z_{n_k}-z_{n_l})$ etc.\ , and we are left with:

\begin{equation}
z_i \prod_{j}(z_{n_j}-z_i) = \sum_{p\neq (i,\{\alpha_k\}
)}a_p\prod_{j}(z_{n_j}-z_p).
\end{equation}
The result follows when we apply the identity (\ref{lagrid}) to this equation
with $z=z_{i}$ and $a_p=z_{i}\prod_{k}\frac{z_k - z_i}{z_k -z_p}$ where the
$z_k$ are randomly chosen distinct sites which don't coincide with the
localized spinons.

\subsection{Appendix C.}
\label{app3C}

We show that $ S^{z}_{i}\left|{\rm GS}\right.  \rangle$ can only have 0 or 2
more spinons than $\left|{\rm GS}\right.  \rangle$ where $\left|{\rm GS}\right.
\rangle$ is a YHWS groundstate in a nonzero magnetic field with

\begin{eqnarray}
\langle\left.  n_1,\ldots,n_{M} \right|{\rm GS}\rangle &=&\Psi_{0}(
n_1,\ldots,n_{M} ) \nonumber\\
&=& \prod_{i<j}(z_{n_i}-z_{n_j})^2\prod_{i}z_{n_i}^{\frac{N}{2}-M+1}.
\end{eqnarray}
The proof hinges on the fact that in either of those cases ($\Delta N_{sp} =0$
or $\Delta N_{sp} =+2$) acting on $ S^{z}_{i}\left|{\rm GS}\right.  \rangle$
{\em twice} with $J_{1}^{+}$ will annihilate that state.  Potential $\Delta
N_{sp} =4,\ldots$ contributions should survive as they are at least 2 levels
from the top of their Yangian multiplet.  Now

\begin{equation}
(2 S^{z}_{i} \Psi_{0})\left({\scriptstyle n_1,\ldots,n_{M} }\right) =
\left(1-2\sum_{j=1}^{M} \delta_{i,n_j}\right) \Psi_{0}\left({\scriptstyle
n_1,\ldots,n_{M} }\right),
\end{equation}
and

\begin{eqnarray}
\lefteqn{\left( 2J_{1}^{+} S^{z}_{i}\Psi_{0}\right) ( n_{1},\ldots,n_{M-1}
)\sim } \nonumber\\
&&  - \sum_{k=1}^{M-1}\sum_{l\neq n_k}^{N} w_{n_k,l} \times
\left[\Psi_{0}({\scriptstyle n_{1},\ldots,n_{M-1}
,l})-2\sum_{j=1}^{M-1}\delta_{i,n_j}\Psi_{0}({\scriptstyle
n_{1},\ldots,n_{M-1},l})\right]\nonumber\\
&&-2\sum_{k=1}^{M-1}w_{n_k,i}\Psi_{0}( n_{1},\ldots,n_{M-1},i)
\end{eqnarray}
The first two terms vanish when we apply the convolution theorem with $w_{kl}$,
as in Appendix \ref{app3A}; only the last term survives.  Notice that the
$\{n_i\};i=1\ldots,M-1$ cannot be equal to $i$ anymore.  This state is
trivially annihilated by $ S^+$ as it vanishes at $z=0$.  Furthermore:

\begin{eqnarray}
\left(\left(J_{1}^{+}\right)^2 S^{z}_{i}\Psi_{0}\right)( n_{1},\ldots,n_{M-2}
)& =& 
\rule{1em}{0em} \sum_{p=1}^{M-2}\sum_{q\neq i,n_p}^{N}
w_{n_p,q}\sum_{k=1}^{M-2} w_{n_k,i}\Psi_{0}( n_{1},\ldots,n_{M-2} ,q,i)
\nonumber\\
& &+ \sum_{p=1}^{M-2}\sum_{q\neq i,n_p}^{N} w_{n_p,q}
w_{q,i}\Psi_{0}( n_{1},\ldots,n_{M-2} ,q,i).
\end{eqnarray}
With the help of the convolution theorem we can set the first term to zero (we
can stick in the extra term with $q=i$, which is missing, at no cost since
$\Psi_{0}$ vanishes when $q=i$).  For the second term we use the identity that
lies at the heart of the integrability of the Haldane-Shastry type models:
$w_{ij}w_{jk}+w_{jk}w_{ki}+w_{ki}w_{ij} = -1$.

\begin{eqnarray}
\lefteqn{\left(\left(J_{1}^{+}\right)^2 S^{z}_{i}\Psi\right)(
n_{1},\ldots,n_{M-2} ) =} \\
& &-\sum_{p=1}^{M-2}\sum_{q\neq
i,n_p}^{N}(1+w_{i,n_p}w_{n_p,q}+w_{q,i}w_{i,n_p})\Psi({\scriptstyle
n_{1},\ldots,n_{M-2} ,q,i})\nonumber.
\end{eqnarray}
The first term is zero since $\Psi$ is a polynomial that vanishes at the
origin, whereas terms two and three can also be put to zero with the help of
the convolution theorem.


\chapter{New Types of Off-Diagonal Long Range Order in Quantum Spin-Chains}
\label{chapter4}
\markright{Chapter \thechapter: New Types of ODLRO in Quantum Spin-Chains}

\section{Introduction}

In this final chapter we attempt to link properties of the ISE-model to
some other one and two dimensional correlated systems. We want to show
that there exists a new form of Off Diagonal Long Range Order (ODLRO) in
1D spin models; this ODLRO, inspired by electron pair hopping in
coupled Hubbard chains or planes, was suggested by Anderson in
1991 \cite{pwaconj}. The corresponding order parameter also 
has an interesting analogy with the Read-Girvin-MacDonald
order parameter in the 
Fractional Quantum Hall Effect (FQHE) \cite{GM87,Read89}. To be specific, consider
the following linear operator $\Del(j)$, acting between the
$2^{N-2}$-dimensional $(N-2)$-site spin Hilbert space, and the $2^{N}$
dimensional $N$-site Hilbert space.  Acting at site $j$, it consists of
slipping in $2$ sites, with two spins in a {\em singlet},
 between sites $j$ and
$j+1$, in the groundstate of a particular model. 
This operator acquires a non-zero expectation value in the
groundstate in the following sense \cite{A66}:
\begin{eqnarray}
\left| \langle 0,N-2|\Del^\dagger(j)\Del(0)|0,N-2\rangle\right|^2 & = &
\left|\sum_\nu \langle 0,N-2 | \Del^\dagger(j)|\nu,N\rangle
\langle \nu,N|\Del(0)|0,N-2\rangle\right| \nonumber\\
& = & \left| \langle 0,N |\Del(0)|0,N-2\rangle\right|^2 +\ldots\nonumber\\
&\rightarrow & \; {\rm const}\; \;\;{\rm for}\; N\gg j \gg 1.
\label{ODLRO}
\end{eqnarray}
$|0,N-2\rangle$ is the groundstate on $N-2$ sites, and $\{ 
|\nu,N\rangle\}$ is a complete set of eigenstates of the $N$ site
model, where $\nu=0$ labels the groundstate.  In the case of
translationally invariant models, the first term in the expansion
(\ref{ODLRO}) guarantees the ODLRO for
$\Del(j)$ provided it exceeds $\half$; in that case the eigenstates have a
definite momentum and $\mbox{$\langle 0,N-2|\Del^\dagger(j)|\nu,N\rangle$} =
e^{i(p_0^{N-2}-p_\nu^{N})j}
\mbox{$\langle 0,N-2|\Del^\dagger(0)|\nu,N\rangle$}$. 
The sum
(\ref{ODLRO}) is bounded from below by $2\left|\langle
0,N-2|\Del(0)|0,N\rangle\right|^2-1$. With the notation:\\
$a_\nu = e^{i(p_0^{N-2}-p_\nu^{N})j} \cdot$
$\left|\langle \nu,N|\Del(0)|0, 
N-2\rangle\right|^2 $, we find this, since 
$\left|\sum_\nu a_\nu \right|\geq$ 
 $ |a_0|-\sum_{\nu>0} |a_\nu|$ $=2|a_0|-1$ if $|a_0|\geq\half$ and
$\sum_\nu |a_\nu|=1$. The last equation is a consequence of the fact that
$\hat{\Delta}(0)|0,N-2\rangle$ has norm 1.

We tested for this form of ODLRO for three different 1D models: the
short ranged NNE and XY models, and the long ranged ISE-model. In all
three cases we find the ODLRO for $\Del$,  and (even more remarkably)
the constant in eq.\ (\ref{ODLRO}) is almost insensitive to the specific
model that we test for: $\sim (0.81)^2\approx 0.65$. $\Del(j)$ is mentioned
implicitly in \cite{PS}, but the context is that of a 1D Hubbard model. We
will turn our attention to that in section \ref{chain1d}.

One might argue that in one dimension only quasi-long range order with
algebraic decay of the order parameter 
correlation function should occur---{\em true} LRO being ruled out by the
Mermin-Wagner theorem. The way out is that true LRO is allowed if 
the order parameter either commutes with the Hamiltonian, or if it breaks
a discrete symmetry. The latter is the case here, we will find that the
singlet insertion breaks the $Z_2$ symmetry of shifting by one lattice
site.

In the next section we will discuss numerical results and analytical
calculations, indicating the presence of the new ODLRO. In section
\ref{FQHE}, we will elucidate the analogy between the singlet insertion
operator $\Del$ and the Read order parameter in the $\nu=\half$ FQHE.
Finally, section \ref{chain1d} will illustrate why the LRO for a singlet
insertion may have important repercussions for superconducting
correlations. 
\newpage

\section{Evidence of  Singlet Insertion ODLRO.}
\label{evidence}

\subsection{Spin- and Spinon-Singlet Insertion in the ISE-model}

\begin{figure}[htb]
\framebox[\hsize]{\centerline{\psfig{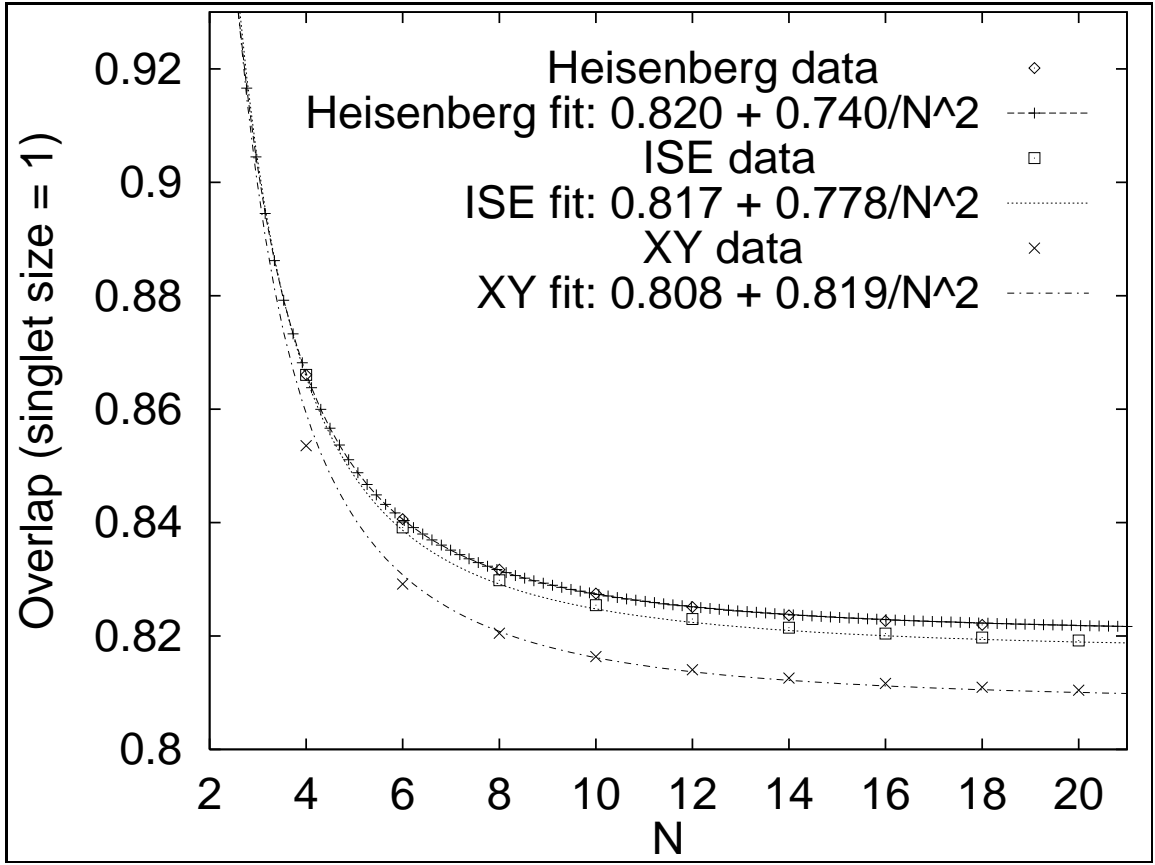}}}
\caption{Overlap between the $(N-2)$-site groundstate with a singlet appended,
with the $N$-site groundstate, as a function of N, for three spin
models.}
\label{medec}
\end{figure}

In fig.\ (\ref{medec}) 
we have plotted numerical results for this singlet insertion \linebreak
$\left|\langle 0,N|\Del(j)|0,N-2\rangle\right|$, for the groundstate of the
three models mentioned earlier, with Hamiltonian:
\begin{equation}
H=\sum_{i\neq j} J(i-j)\left\{ \half(S_i^+ S_j^- +S_i^-S_j^+ )+
\Delta(S_i^z S_j^z)\right\}.
\label{hspindef}
\end{equation}
For the XY and NNE models $J(n)=\delta_{n,1}+\delta_{n,-1}$ and
$\Delta=0,\, \Delta=1$ respectively. For the ISE-model
$J(n)=\frac{1}{(\frac{N}{\pi}\sin(\frac{n\pi}{N}))^2}$, $\Delta=1$.

$\Del^\dagger(j)\Del(0)$ can be rewritten as 
\begin{equation}
\label{shift}
\Del^\dagger(j)\Del(0) = P_{1,2}^S (P_{12\cdots j})^2
\end{equation}
where $P_{12}^S$ is the projector for the spins on sites 1,2 onto a
singlet: $\frac{1}{4}-\svec_1\cdot\svec_2$, and $P_{12\cdots j}$ is the
operator that cyclically permutes spins on sites $1\cdots j$. With the help
of eq.~(\ref{shift}) we have
verified that $\left|\langle 0,N|\Del(0)|0,N-2\rangle\right|^2$ 
is indeed the dominant
contribution to $\langle 0|\Del^\dagger(j)\Del(0)|0\rangle $
for $N\rightarrow\infty$.

These overlaps are shown as functions of $N$ in fig.\ \ref{medec} together
with fits to the results of the form $0.817 + 0.778 N^{-2}$ for the ISE model,
$0.820 + .740 N^{-2}$ for the NNE Heisenberg model, and $0.808 + 0.819 N^{-2}$
for the XY model.  The results strongly suggest that all overlaps remain
finite in the limit where the system size goes to infinity.  Note that since
the phases of the $N$ and $N-2$ site wavefunctions may be chosen independently
the phase of the overlaps is meaningless and further, since the groundstate
momenta of the $N$ and $N-2$ site groundstate wavefunctions differ by $\pi$,
the overlap is multiplied by minus one if the location of the singlet pair is
shifted by one site.  We have chosen the overlap real and positive for
convenience.  As a consequence of this alternation and the finiteness of the
overlap, the singlet insertion-singlet deletion correlation function has fixed
magnitude and a sign determined by the number of sites between the insertion
and the deletion modulo 2 in the limit of infinite separation of the insertion
and deletion; the models therefore have a hidden ODLRO which breaks a $Z_2$
symmetry.  As fig.\ (\ref{medec}) shows, the overlaps for the $N$-site
groundstate with the $N-2$ site groundstate with a nearest neighbor singlet
insertion in the ISE and Heisenberg models are not only finite in the
$N\rightarrow \infty$ limit, but also surprisingly close to each other,
despite the fact that the range of the interaction in both models is quite
different.

\begin{figure}[htb]
\framebox[\hsize]{\centerline{\psfig{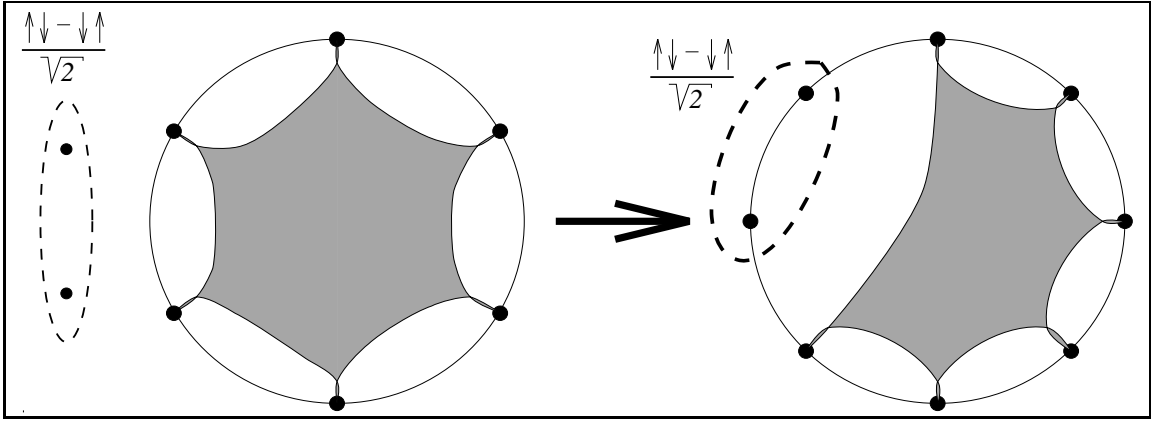}}}
\caption{Graphical illustration of the deformation of the singlet
insertion to a spinon singlet insertion. The gray area is the
groundstate wavefunction, or an adiabatic deformation thereof.}
\label{trampofig}
\end{figure}

We now present an analytical calculation in the ISE model that
gives some understanding of origin of  the finite overlap.
The unnormalized 
groundstate wavefunction on $N$ sites of the ISE model in a basis 
labeled by the positions of the down-spins
$\{n_1,\ldots,n_M\}$, $M=\frac{N}{2}$ is given by eq.\ (\ref{psi0}):
\begin{equation}
\Psi^{(0)}_N\left(n_1,\ldots,n_{\frac{N}{2}}\right) = \prod_{i=1}^{\frac{N}{2}}
(-)^{n_i}
\prod_{i<j}^{\frac{N}{2}} \sin^2\left(\frac{n_i -n_j}{N}\pi\right)
\end{equation}
In a basis of local spins
$ \{ |\sigma_1\cdots \sigma_M\rangle\}$, $\sigma_i = \pm \frac{1}{2}$ this
reduces to \cite{H91}:
\begin{equation}
\Psi^{(0)}_{M}\{\sigma_1\cdots\sigma_M\} = \prod_{i<j}^{M}(z_i-z_j)^{\delta_{\sigma_i ,\sigma_j}} 
e^{\frac{\pi i}{2}\rm{sgn}(\sigma_i - \sigma_j)}.
\end{equation}
For the $N$ site ISE groundstate $M=N$ and $\{z_i\}\equiv C_N =
\{e^{\frac{2 \pi i
n}{N}} \}_{n=1}^{N}$, while for the $N-2$ site groundstate $M=N-2$ and $\{
z_i\} \equiv C_{N-2} =\{e^{\frac{2\pi i n}{N-2}} \}_{n=1}^{N-2}$. 
To compute the ISE
overlap we add $\sigma_{N-1},\; \sigma_{N}$, sitting in a singlet, to the
$N-2$ site groundstate.  This overlap is not 
calculable since the $z_i$'s from both
sets are not commensurate with each other.  However, if we slightly deform the
set $C_{N-2}$ to be $\{e^{\frac{2\pi i n}{N}}\}_{n=1}^{N-2}$ and leave
$\sigma_{N-1},\sigma_{N}$ in a singlet then we obtain a new state
$\Psi^{(2)}_N$,
that can be recognized as a {\em localized} 
2-{\em spinon state}~\cite{H91,TH94}, eq.\ (\ref{locspinwf}), 
as in fig.\ (\ref{trampofig}).
This is an uncontrolled approximation whose justification
lies in the agreement with our numerical results.
This state is 
an admixture of eigenstates that contain only
0 or  2 ``real'' spinons.  Here the localized 
spinons sit at sites $N-1$ and $N$ in a singlet.  In general we could have
put them at sites $\alpha,\beta$ by deforming $C_{N-2}$ into $\{e^{\frac{2\pi i
n}{N}}|i=1,\ldots,N\}/\{e^{\frac{2\pi i\alpha}{N}},e^{\frac{2\pi
i\beta}{N}}\}$.

In the basis of states with
$M$ overturned spins with respect to the ferromagnetic state 
labeled by their positions along the chain: $n_1,\ldots,n_M$, 
$\Psi^{(0)}_N$ and $\Psi^{(2)}_{N}$
are given by\footnote{
The conversion from $\{ \sigma_i \}$ to $\{ n_i \}$ basis is based on the
identity $\prod_{j=1}^N (1-\frac{z}{z_j})={\rm const}$ when
$z_j=e^{\frac{2\pi i}{N} j}$ and $z$ lies on the lattice $z^N=1$.} :
\begin{equation}
\Psi^{(2)}_N\left(n_1,\ldots,n_{\frac{N}{2}-1}\right) 
= \prod_{i=1}^{\frac{N}{2}-1} (-)^{n_i}
\prod_{i<j}\sin^2\left({\scriptstyle\frac{n_i-n_j}{N}\pi}\right) 
\times \prod_{i=1}^{\frac{N}{2}-1} \sin\left({\scriptstyle 
\frac{n_i-\alpha}{N}\pi}\right)
\sin\left({\scriptstyle\frac{n_i-\beta}{N}\pi}\right).
\end{equation}
The overlap between $\Psi^{(0)}_N$ and $\Psi^{(2)}_N$ 
for arbitrary
separation $\alpha-\beta$ between the spinons is calculated in Appendix A.
We find that the normalized overlap is given by:
\begin{equation}
\frac{2}{N}\frac{\sin\left(\pi\frac{\alpha-\beta}{2}\right)}{ 
\sin(\pi\frac{\alpha-\beta}{N})}
\sqrt{\frac{\pi(\alpha-\beta)}{{\rm Si}(\pi(\alpha-\beta))}},
\label{twospinonov}
\end{equation}
where ${\rm Si}(x)$ is the Sine integral function.
Thus for a nearest neighbor the overlap is $\frac{2}{\pi}\sqrt{\frac{\pi}{
{\rm Si}(\pi)}}\simeq 0.82917$, which is within 1.5\% of the Heisenberg and ISE
groundstate singlet insertion overlaps.

\subsection{Relation to the FQHE.}
\label{FQHE}

We noticed that the action of the generalized singlet insertion operation
at sites $\alpha, \beta$, on the ISE-model groundstate was essentially
the following: it creates two ``holes'' at those sites---where the down
spins can't reside---by way of multiplying the wavefunction by a factor
$\prod_i (z_{n_i}-z_\alpha)(z_{n_i}-z_\beta)$. Then it inserts a down
spinon at site $\alpha$, which, in the down spin - particle analogy is
just a (hard-core) boson\footnote{ Actually, we have the down spinon at
$\alpha$ half the time, and $\beta$ the other half, but just fixing it at
$\alpha$ leaves the matrix element the same, up to a factor $\sqrt{2}$,
see eq. (\ref{sigmadef}).}. This all looks very similar to what we know
about the {\em bosonic} $\nu=\half$ Fractional Quantum Hall Effect. 
In this theory,
particles are constrained to a 2D plane in a very high magnetic field,
such that the cyclotron frequency dominates over the interactions between
the particles. In that case the Hilbert space is effectively just the set
of lowest Landau level wavefunctions, of (kinetic) energy $\half \hbar\omega_c
=\frac{\hbar eB}{2m}$, which are degenerate\footnote{Degeneracy
$=g=\frac{\Phi}{\Phi_0}
=\frac{B A e}{h}$; $A$ is the area of the sample.}. When we
turn on interactions, some dramatic things happen: when the filling
fraction $\nu=\frac{N}{g}$ is close to a rational number of the from
$\frac{1}{m}$, the system develops a gap, and becomes an incompressible
fluid with approximate groundstate wavefunction \cite{Laugh}:
\begin{equation}
\Psi(z_1,\ldots,z_M)=\prod_{i<j}(z_i-z_j)^m e^{-\frac{|z_i|^2}{4l^2}},
\label{lllwave}
\end{equation}
where $l^2=\left|\frac{\hbar}{eB}\right|$ and $z_i=x_i+iy_i$ is the
complex co-ordinate of particle $i$. Obviously, for $m$ even
the particles must be bosons, and fermions otherwise.
Wavefunction (\ref{lllwave}) is valid for particles confined to a circular
droplet.

Elementary excitations of this wavefunction are so-called quasiholes and
quasiparticles. We will concentrate on the quasiholes. To create one at
co-ordinate $z$, we have to introduce an extra zero in the groundstate
wavefunction at that point:
\begin{equation}
\Psi(z_1,\ldots,z_N|z)=\prod_{i=1}^N (z-z_i)\,\Psi_0(z_1,\ldots,z_N).
\end{equation}
This is mathematically identical to the up-spinon insertion in the ISE spin
wavefunction. A quasihole is a vortex in the quantum hall fluid, that
allows one extra flux quantum to pierce the fluid.  

The entry into a Quantum Hall phase---with gap---can be signalled by the
non-zero expectation value of an order parameter. The Quantum Hall order
parameter is not of simple the Bose-condensation type $\psi^\dagger({\bf r})$,
or Cooper pair type $\psi_{k\uparrow} \psi_{-k\downarrow}$. A first
version was suggested by Girvin and MacDonald \cite{GM87}. Soon
afterwards, Read introduced his version \cite{Read89}:
\begin{equation}
\hat{O}_m(z) = \exp\left(m\int d^2 z'\,
\ln(z-z')\rho(z')\right)\psi^\dagger(z),
\label{read}
\end{equation}
whose effect is to introduce a {\em neutral} complex of $m$ quasiholes,
and a particle, at co-ordinate $z$. This operator has true long range
order in a $\nu=\frac{1}{m}$ Quantum Hall state. Thus, in the limit where
the number of sites in the ISE-model becomes large, the spinon-insertion
expectation value is essentially the Read order-parameter in the bosonic
$m=2$, $\nu=\half$ FQHE, acting at the edge of the circular droplet. 

\begin{figure}[htb]
\framebox[\hsize]{\centerline{\psfig{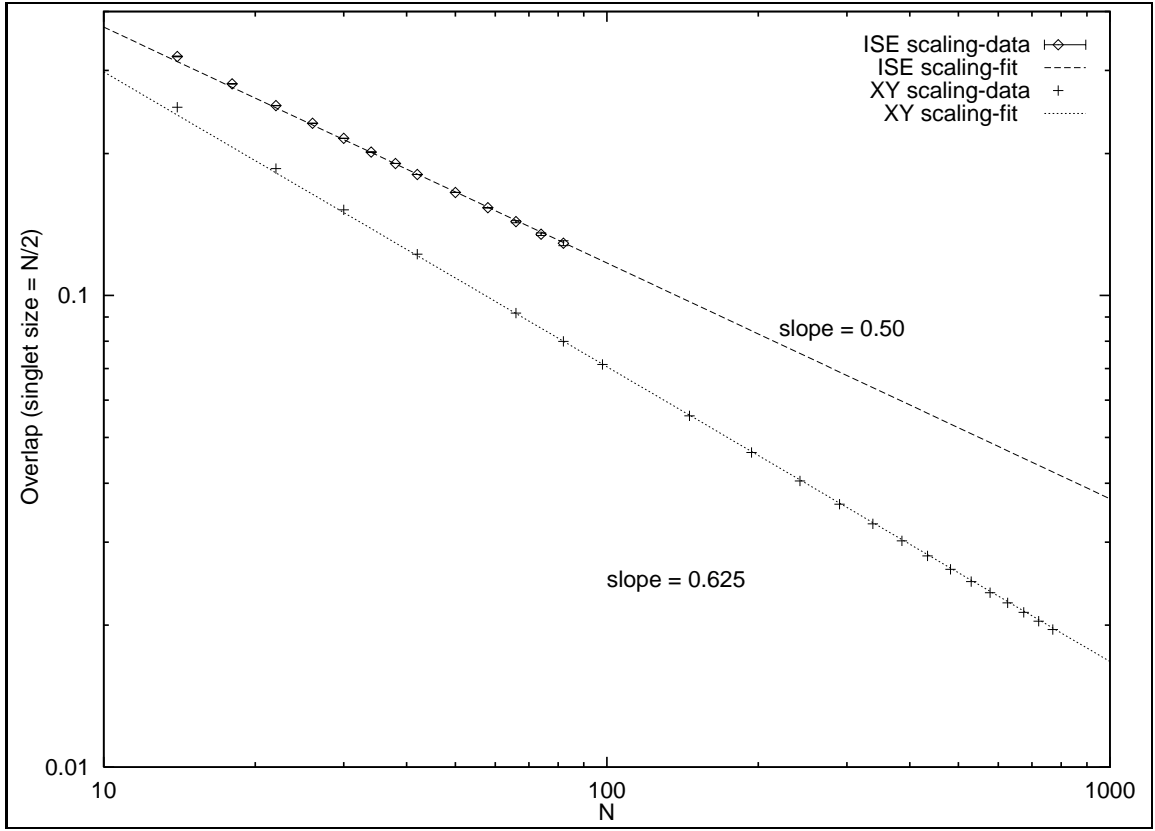}}}
\caption{The overlaps between the $N$ site groundstate and the $N-2$ site
groundstate with two spins in a singlet inserted at sites 1 and
$\frac{N}{2}$. XY model results are from exact numerical results,
whereas the ISE-model data for $N>18$ is from Monte Carlo simulations.}
\label{fig2a}
\end{figure}

The analogy can be extended further. When the
spinon sites are dragged apart, such that $1\ll |\alpha - \beta|\ll N$,
we are essentially measuring the spinon propagator. Recall that the
localized two spinon wavefunction is a linear combination of eigenstates
with two {\em real} spinons or none (the groundstate). When we overlap with
the latter, we measure the amplitude for one of the spinons to propagate
to the other and annihilate (provided that they are in a singlet
configuration, since $\svec^2$ is conserved). The decay of this propagator
$\langle 0,N|(0,\alpha);N-2\rangle$ (where the ket designates the two
spinon state) can be derived from eq. (\ref{twospinonov}). It is
proportional to $\alpha^{-\half}$; $1\ll\alpha \ll N$. We can do the same
thing for the singlet insertion, by not inserting the spins in the
singlet next to each other, but a distance $\alpha$ apart. This
correlator is plotted in fig.\ (\ref{fig2a}), and shows the same decay
$\propto \alpha^{-\half}$. 
We have evidence for the same exponent in the
NNE model\footnote{Actually, fig.\ (\protect\ref{fig2a}) shows the decay
as a function for the system-size $N$, not $\alpha$. The reason is that for
periodic boundary conditions, if
the matrix element decays algebraically, we expect it to be of the form 
\begin{equation}
\langle 0, N |(\alpha,0);N-2\rangle\propto
\left|\frac{1}{\frac{N}{\pi}\sin\left(
\frac{\pi\alpha}{N}\right)}\right|^{2\delta}.
\end{equation}
That is, we assume that the ISE-model is conformally invariant at $T=0$
and the correlators of the primary operators in this theory, including
the singlet insertion, decay algebraically on the infinite (1+1) dimensional
space time: $\langle
\phi(z)\phi(0)\rangle\rangle\sim |z|^{-2\delta}$, and $z=x+it$. 
With the mapping
$w=u+iv=\frac{N}{2\pi}\ln(z)$ the spatial direction $v$ becomes compact,
and at equal times ($u=0$) we obtain the desired result \cite{Cardy}. 
We will see
later that $\Del(r)$ is indeed a primary operator in a $c=1$ CFT.
}.

For the $\nu=\half$ FQHE, the corresponding quasihole propagator decays
exponentially in the bulk of the material (because of the gap). Low
energy excitations can propagate along the edge of the fluid, which is
contained by a confining potential. Such modes are called {\em edge}
states. Quasiholes at the edge have a propagator that decays
algebraically,
and with Ka\v{c}-Moody algebra techniques, Wen \cite{Wen}
showed that its exponent is $\half$ as well. 
As a remarkable aside: the {\em exclusion} statistics of the spinons
\cite{H91b},
which is $\frac{\pi}{2}$ (bosons: 0, fermions $\pi$), is the same as the
exchange statistics of the quasiholes in the $\nu =\half$ effect
\cite{Arovas}.

\subsection{ODLRO in the XY-model.}

In an effort to better understand our results
for the Heisenberg and ISE models, we
examined the overlap for the same insertion operator
acting on the $N-2$ site XY model groundstate with the
$N$ site  XY model groundstate.  The strikingly similar
result, here extended to a larger system size than
was possible for the Heisenberg and ISE models,  is
shown in fig. \ref{medec}.  Since the XY model can
be mapped onto spinless fermions, the ``order parameter''
can be recast in that language and a more detailed study made.
In that language the non-zero overlap is between the
 groundstate 
for $N/2$ spinless fermions,
with
either periodic of antiperiodic boundary conditions,
on $N$ sites 
(with positive
hopping integral so that their momentum 
in the groundstate are centered around
$\pi$) 
and the state obtained by adding two sites and
one fermion 
(in a superposition of being on
the two added sites with a relative minus sign for the two
different sites)
 to the
groundstate for the $\frac{N}{2}-1$ spinless fermions on
$N-2$ sites with the opposite boundary conditions
from the $N$ site case.
The change of boundary conditions is a
non-local operation in the fermion language, however
in the spin language it arises from a local operation, i.e.
the insertion of two additional spins in a singlet 
configuration.

Since the wavefunctions for the two fermion states are just Slater
determinants, we can bound the singlet insertion matrix element rigorously
from below.
The technical details can be found in 
Appendix B.
It is found there that \mbox{$|\langle 0,N|\Del(j)|0,N-2\rangle|$}
$>e^{-\frac{5}{2}}$ for $N\rightarrow\infty$.

It is somewhat remarkable that the ODLRO for the XY-model is the result
of three conspiring effects, every one of which by itself would lead to a
vanishing matrix element as  $N\rightarrow \infty$.

\begin{itemize}
\item $ |0,N\rangle$ had $\frac{N}{2}$ electrons, all in extended
states and $\Del(j) |0,N-2\rangle$ has $\frac{N}{2}-1$ electrons in
extended states too, but the last is in a localized {\em two} site state,
resulting in a decay $\propto\frac{1}{\sqrt{N}}$.
\item The extended states in the $N$-particle groundstate have
periodicity $N$, whereas the $(N-2)$-particle groundstate has periodicity
$N-2$, with a resulting orthogonality type catastrophe in the overlap.
\item The particles in the $N$-site groundstate have
periodic/antiperiodic boundary conditions, whereas the $(N-2)$-site version
has antiperiodic/periodic ones (depending on the parity of $\frac{N}{2})$.
\end{itemize}

For the XY-model too, we have done numerical calculations, studying the
decay of the singlet insertion as a function of the separation of the
singlet spins, see fig.\ (\ref{fig2a}). In this case we find an exponent
very close to $\frac{5}{8}$. In the next section we will give arguments
for this particular decay exponent from the point of view of the
Luttinger liquid picture, which describes the low energy excitations of these
spin models \cite{Haldane}.

Finally,
the XY model and the ISE model both have wavefunctions of
the Jastrow product form of the Calogero-Sutherland model
\cite{Suth71,Hald88}:
\begin{equation}
\Psi^{(0)}_N\left(n_1,\ldots,n_{\frac{N}{2}}\right) =
 \prod_{i=1}^{\frac{N}{2}}
(-)^{n_i}
\prod_{i<j}^{\frac{N}{2}} \sin^p\left(\frac{n_i -n_j}{N}\pi\right)
\label{eq:wfcn}
\end{equation}
For the XY model $p=1$ whereas for the ISE model $p=2$.
We have studied numerically the overlap for a number of
values of $p$
 and find that it remains finite in the limit of large system
size and is essentially constant for $1 \le p \le 2$. We
have also examined $0 \le p \le 1$ and find the overlap to be finite
in the large system size limit and monotonically decreasing with $p$
down to a value of $1/2$ at the point $p=0$ where the
wavefunction is that of Gutzwiller projected free bosons and the
can be computed exactly as $\frac{1}{\sqrt{2}} \sqrt{\frac{N}{N-1}}$.
Note that this implies $O(\frac{1}{N}))$ corrections in contrast to the
$O(\frac{1}{N^2})$ corrections we find for the XY, ISE and Heisenberg
models.  It is not clear when corrections of size 
$O(\frac{1}{N})$ first appear.

It may be tempting, based on the similarity of the wavefunctions
(\ref{eq:wfcn}) to relate the XY model ODLRO to a $\nu=1$ Read
order parameter, acting on the edge. However, the singlet decay exponent
$\frac{5}{8}$ is not consistent with the electron propagator exponent in
the  Quantum Hall situation.

\section{Singlet Insertion in the 1D Hubbard Model.}
\label{chain1d}

In the context of spin models, the singlet insertion operation may seem
somewhat esoteric, requiring the Hilbert space to change size. Now if we
imagine the sites to be physical {\em electrons}, then the singlet
insertion procedure corresponds to hopping two electrons in a singlet
configuration onto a chain. 

We follow the tradition of studying spin models in
the guise of the 1D Hubbard model in the limit of one particle/site, and
$U\rightarrow \infty$ \cite{HA87}, and consider 
the action of our order parameter in the context of
this simplest of interacting 1D itinerant electron models. 
Its Hamiltonian is given by:
\begin{equation}
H=t\sum_{is}(c_{i\sigma}^\dagger c_{i+1\sigma}+{\rm h.c.})
+U\sum_i n_{i\uparrow}n_{i\downarrow}.
\label{hubbarddef}
\end{equation}
The $i$'s are site-labels. The $U>0$ term discourages electrons to occupy
the same site. There is an additional non-academic reason
to study the effects of
singlet insertion in a Hubbard-type model. 
One of the mechanisms to explain the high transition
temperatures in the copper-oxide superconductors is built around the {\em
relevance} of the hopping of electron singlet pairs between the 2D layers
in these materials \cite{pwaconj}. We will attempt to decide on the
relevance of this operator in the {\em one}-dimensional Hubbard model for
$U\rightarrow\infty$, based on the ODLRO of our singlet insertion (see
\cite{Strong95} for possible ways to extend this procedure to two
dimensions). The inter-{\em chain} hopping operator acting at sites
$j,j+1$ is given by:
\begin{eqnarray}
\oc_S(j) &=& \half(c_{j\uparrow}^\dagger c_{j+1\downarrow}^\dagger -
c_{j\downarrow}^\dagger c_{j+1\uparrow}^\dagger )(
d_{j\uparrow} d_{j+1\downarrow} -d_{j\downarrow} d_{j+1\uparrow}) + {\rm
h.c.}\nonumber\\
&\equiv&\oc_1(j)\oc_2^\dagger(j)+{\rm h.c.},
\end{eqnarray}
where $c_{j\sigma}^\dagger $ are the electron creation operators for
chain 1, and the $d$'s act on chain 2. The relevance of this operator in the
renormalization group sense for
perturbing away form the uncoupled chain case,
is determined by its scaling dimension $d_s$ in $\langle 0
|\oc_S^\dagger(j) \oc_S(0)|0\rangle\sim j^{-2d_s}$ for 
$j\rightarrow\infty$. If
$d_s >2$, singlet pair hopping is irrelevant. $|0\rangle$ is the tensor
product groundstate of the two uncoupled Hubbard models. The $\oc_S$-$\oc_S$
correlator then factorizes into an $\langle\oc_1^\dagger(j)\oc_1(0)\rangle$ and
$\langle\oc_2^\dagger(j) \oc_2(0)\rangle$ part. 
Ogata and Shiba have pointed out
an additional simplification for large $U$ \cite{OgataShiba}: 
the groundstate of
(\ref{hubbarddef}), for $M$ electrons, factorizes into the convolution 
of a spin and a charge part:
\begin{equation}
\Psi_0(n_1,\ldots,n_M|\sigma_1,\ldots,\sigma_M)=
\Psi_{\rm SF}(n_1,\ldots,n_M)\chi(\sigma_1,\ldots,\sigma_M),
\label{oswave}
\end{equation}
where $\Psi_{\rm SF}$ is the determinantal groundstate 
wavefunction of $M$ free
spinless fermions, labeled by co-ordinates $\{ n_i\}$.
$\chi(\sigma_1,\ldots,\sigma_M)$ is the groundstate of the NNE Heisenberg
model (at least for $N=4m+2$, $m$ integer)
of $M$ spins, residing on the spinless fermions. Then $\langle
0_1|\oc_1^\dagger(j)\oc_1(0)|0_1\rangle$ reduces further to:
\begin{equation}
\langle \Psi_{\rm SF}|\psi_{j+1}\psi_{j}\psi_1^\dagger\psi_0^\dagger
(-)^{\sum_{i(<j)}n_i}|\Psi_{\rm SF}\rangle\cdot
\left|\langle\chi|\Del(j)^\dagger \Del(0)|\chi\rangle\right|
\label{randwalk}
\end{equation}
$\Del$ is the singlet insertion order parameter from the spin models 
which alternates with $j$,
i.e.\ keeps track of the number of fermions to the left of the insertion
site, whence the  $(-)^{\sum_{i(<j)}n_i}$ factor in the spinless 
fermion matrix element. $\psi_i,\psi_i^\dagger$ are spinless fermion
operators.  The second factor in eq.\ (\ref{randwalk}) is constant ($\sim
0.65$ as we saw before).

The matrix element decay eq.\ (\ref{randwalk}) then follows from the fact
that {\em free} fermion operators have dimension 1 and the alternating
term has dimension $\half$. We want to derive this a bit more extensively in
a bosonized language in the original Hubbard model. We will also include
a spin anisotropy term to accommodate the XY model limit later on.
In the continuum
limit the $c_{js}\rightarrow\psi_{R,s}(ja)+\psi_{L,s}(ja)$, the field
operators acting at the left and right Fermi points. $s=\pm \half$ is the
spin label, and $a$ the lattice spacing. 
In one dimension they can be rewritten in terms of boson fields
\cite{Haldane,Schulz93}:
\begin{eqnarray}
\psi_{\rule{0em}{0.4em}^R_L s}(j) & = & \lim_{\alpha\rightarrow 0}
\frac{1}{\sqrt{2\pi\alpha}}e^{\pm ik_F x -
i(\pm\frac{\phi_\rho+s\phi_\sigma}{\sqrt{2}}- \frac{\theta_\rho +
s\theta_\sigma}{\sqrt{2}})}\nonumber\\
& \equiv & e^{\pm ik_F  x-i\Theta_{\rule{0em}{.4em}^R_L s}(x)}.
\label{bosondef}
\end{eqnarray}
$\alpha$ is a cut-off scale. The fields $\phi_\mu$ and $\del_x
\theta_\mu$, $\mu=\rho,\sigma$ are canonically conjugate bosonic fields:
$[\phi_\mu(x),\del_x \theta_\nu(x')]=i\pi \delta_{\mu\nu}\delta(x-x')$. 
One also writes $\Pi_\mu=\pi\del_x \theta_\mu$.  The fields with label
$\sigma$ create/annihilate spin excitations (destroy up- and create
down-spin electrons and vice-versa); the $\rho$-fields relate to charged
excitations. A peculiarity of the 1D Hubbard model is that these fields
propagate independently: spin-charge separation, as is evident from the
Hamiltonian in the bosonic language:
\begin{eqnarray}
H&=&\int dx\, \left\{ \frac{\pi v_c K_\rho}{2} \Pi_\rho^2 +
\frac{v_c}{2\pi K_\sigma}(\del_x \phi_\rho)^2\right\} \nonumber\\
&& +\int dx\, \left\{ \frac{\pi v_s K_\sigma}{2} \Pi_\sigma^2 +
\frac{v_s}{2\pi K_\sigma} (\del_x \phi_\sigma)^2 + \hat{g}
\cos(\sqrt{8}\phi_\sigma)\right\}+\ldots
\end{eqnarray}
The ellipses refer to other irrelevant terms originating from the
non-linear electron dispersion relation 
and the underlying lattice. $v_c$ and
$v_s$ are the velocities of spin and charge excitations respectively.
$K_\rho$ and $K_\sigma$ are functions of the Hubbard parameters. All
correlation function exponents can be expressed in terms of these two.
 The spin
cosine term is irrelevant  if $0<U<\infty$ and the anisotropy $\Delta$
satisfies $-1<\Delta\leq 1$. The non-linear term
only acts to renormalize $K_\rho$ and $K_\sigma$: $K_\rho(U=0)=1$,
\mbox{$K_\rho(U\rightarrow\infty)$}$=\half$, and 
\mbox{$K_\sigma(\Delta,U\rightarrow\infty)$} $= 1-\frac{1}{\pi}
\arccos(\Delta)$ \cite{Affleck}. So at long wavelengths, we have two
uncoupled free linearly dispersing bosons with different velocities. The
electronic correlation functions can be deduced from the bosonic ones via
(\ref{bosondef}). 
The latter are easy to compute, since the fixed point Hamiltonian is quadratic.
For example \cite{Emery79}:
\begin{equation}
\langle e^{i\alpha \theta_{R\mu}(x,t)}e^{-i\beta\theta_{R\mu}(0,0)}\rangle
= \left(\frac{1}{(x-v_\mu t)^{\frac{K_\mu}{4} + \frac{1}{4K_\mu}+\half}}
\frac{1}{(x+v_\mu t)^{\frac{K_\mu}{4}+\frac{1}{4K_\mu}-\half}}
\right)^{\alpha\beta},
\end{equation}
Notice that for interacting electrons ($K_\mu\neq 1$) a right Fermi point
electron operator has an {\em anti}holomorphic part too!

In the following we will neglect the phase-factors, since we are just
interested in the scaling dimensions. Consider the case where the singlet
electrons are inserted, not as nearest neighbors, 
but at sites $x,x'$, and extracted
at sites $y,y'$: $\langle 0|\psi_\uparrow^\dagger(x)\psi_\downarrow^\dagger
(x') 
\psi_\uparrow(y)\psi_\downarrow(y')|0\rangle+\ldots$, where $1\ll |x-x'|,
|y-y'| \ll |x-y|$ in units of the lattice spacing $a$. In this limit we
can then apply the operator product expansion of fields $\psi^\dagger(x)$ and
$\psi^\dagger(x')$ 
and separately $\psi(y)$ and $\psi(y')$. In terms of the bosons:
\begin{eqnarray}
e^{i\theta_{R\uparrow}(xt)} e^{i\theta_{R\downarrow}(x't')}
&\sim & \left[ e^{\frac{i}{\sqrt{2}}\theta_{R\rho}(xt)}\cdot 
e^{\frac{i}{\sqrt{2}}\theta_{R\rho}(x't')}\right] 
\left[ e^{\frac{i}{\sqrt{2}}\theta_{R\sigma}(xt)}\cdot 
e^{-\frac{i}{\sqrt{2}}\theta_{R\sigma}(x't')}\right] \nonumber\\
&\stackrel{x',t'\rightarrow x,t}{\sim} & 
(\Delta x-v_c\Delta t)^{p_\rho+\frac{1}{4}} 
(\Delta x+v_c \Delta t)^{p_\rho-\frac{1}{4}}
:e^{i\sqrt{2}\theta_{R\rho}(xt)}:\nonumber\\
&&\times  \frac{1}{(\Delta x -v_s\Delta t)^{p_\sigma +\frac{1}{4}}
(\Delta x+v_s\Delta t)^{p_\sigma-\frac{1}{4}}}:\unity+\ldots:,
\label{fusion}
\end{eqnarray}
where $p_\rho= \frac{1}{8K_\rho}+\frac{K_\rho}{8}$ and
$p_\sigma=\frac{1}{8K_\sigma}+\frac{K_\sigma}{8}$. 
We used the
generalization of the Glauber formula: $e^A e^B = :e^{A+B} e^{\langle
0|AB+\frac{A^2+B^2}{2}|0\rangle}$ for normal ordering with respect to
the bosonic vacuum, and $A,B$ commuting with their commutators
\cite{Shankar91}.
There is a term with
identical exponents for $R\rightarrow L$, and less relevant terms,
containing mixtures of both. Notice that the equal-time 
singlet insertion decay exponent of fig.\ (\ref{fig2a}) can be read
off from (\ref{fusion}):
$2p_\sigma=\frac{K_\sigma}{4}+\frac{1}{4K_\sigma}=\half$ for $\Delta=1$
(NNE-model limit) and $\frac{5}{8}$ for $\Delta=0$ (XY-model limit).

The part of the electron singlet operator that yields the {\em spin}-singlet
correlator is $e^{\frac{i}{\sqrt{2}}\theta_{R\sigma}(x)}$. This
operator is semionic, in that commuting two of these through each other
results in a phase $e^{\frac{\pi i}{2}}$, for the isotropic Hubbard model. 
This is compatible with the
earlier observation that the singlet insertion is essentially a
two-spinon creation operator. \cite{ludwig} suggests a semionic operator
in the context of a CFT realization of the low energy part of the
ISE-model, bilinears of whose Fourier modes create 2 spinons as well. An
identification of the boson-exponential with this CFT operator seems
appropriate.

From eq.\ (\ref{fusion}) we learn that the {\em electron} singlet
insertion in bosonized form is
 $e^{i\sqrt{2}\theta_{R\rho}}$, with dimension $\frac{5}{4}$
and correlator decay $\propto (x-y)^{-\frac{5}{2}}$, as we found before
(see also \cite{Haldane,hubb_ll,Frahm}).
The expression for $\oc$ contains a charge-prefactor which {\em grows}
with separation between the singlet spins, since it has a negative
scaling dimension. This need not concern us, since this separation is
bounded from above by the distance to the singlet extraction, i.e., the
OPE is no longer valid when $x'\approx y,y'$. 

The decay exponent $\frac{5}{2}$ is insufficient to make singlet pair
hopping between chains relevant. From our spin model singlet insertion,
we find the culprit to be the operator $(-)^{\sum_{i(<j)}n_i}$, caused
by the alternation of the spin order parameter. It has been suggested
\cite{Strong95} that this can be remedied if it could be canceled
against a similar alternating factor in the {\em second} chain.

\section{Conclusion.}
\label{concl4}

We have shown that the operator that inserts a singlet into a 1D spin
chain acquires LRO in the ISE-, XY- and NNE-models. We provided numerical
evidence and suggested that its LRO might be expected based on its
interpretation as a two spinon insertion, as well as  via an analogy to Read's
order parameter in the $\nu=\half$ QHE (for NNE and ISE-models). The fact
that the order parameter breaks a $Z_2$ symmetry causes the more physical
electron singlet hopping operator in coupled 1D Hubbard chains to become
irrelevant. Bosonizing the anisotropic Hubbard chain allows us to
calculate the decay exponent of the spin singlet insertion operator or,
equivalently, the spinon propagator, as a function of the anisotropy.

The fact that this ODLRO is present in three quite different antiferromagnetic
isotropic 
spin-chains, raises the question whether there are models where it {\em
doesn't} occur. XY-, ISE- and NNE- models are all gapless and one might
imagine that that is a key feature to {\em this} particular ODLRO, but for
instance the Majumdar-Ghosh model with a gap above its
dimerized groundstate possesses the singlet-insertion ODLRO by definition.
This is an important issue since the usefulness of the order-parameter
depends on its ability to discriminate between different models.

\newpage

\section{Appendix}
\subsection{Appendix A: The Two-Spinon Matrix Element}

We compute the (normalized) overlap between the $N$-site  ISE groundstate
$|\Psi^{(0)}\rangle$ 
and the $N$-site two-spinon wavefunction, localized at sites $\alpha$ and 0,
$|\Psi^{(2)}(\alpha)\rangle$.
The two-spinon wavefunction is an adiabatic deformation of the $(N-2)$-site
groundstate plus a singlet. 
Because of translational invariance we can fix $\beta$ to be 0. 
What we need to compute is:
\begin{equation}
\frac{\langle\Psi^{(0)} | \Psi^{(2)}(\alpha)\rangle}{\sqrt{
\langle\Psi^{(0)} |\Psi^{(0)}\rangle \langle\Psi^{(2)}(\alpha) |
\Psi^{(2)}(\alpha)\rangle}} =
\frac{\langle\Psi^{(0)} |\Psi^{(2)}(\alpha)\rangle}{
\langle\Psi^{(0)} |\Psi^{(0)}\rangle}
\left(\frac{\langle\Psi^{(2)}(\alpha) |\Psi^{(2)}(\alpha)\rangle}{
\langle\Psi^{(0)}
|\Psi^{(0)}\rangle}\right)^{-\frac{1}{2}}\equiv A\cdot B^{-\half}
\end{equation}
First we will determine $\langle\Psi^{(0)}|\Psi^2(\alpha)\rangle$. Setting
$M=\frac{N}{2}-1$:
\begin{eqnarray}
 \langle\Psi^{(0)}|\Psi^{(2)}(\alpha)\rangle & = & 
{\scriptstyle\left( \frac{N}{2} \right)}! 
 ~\sum_{ n_1,\ldots,n_M }\frac{1}{\sqrt{2}}
\Psi^{(2)}(n_1,\ldots,n_M|0,\alpha)\times\nonumber\\
&&\left\{\Psi^{(0)}(n_1,\ldots,n_M,0) - 
\Psi^{(0)}(n_1,\ldots,n_M,\alpha)\right\}^* \nonumber\\
&=&\frac{1}{\sqrt{2}}{\scriptstyle \left( \frac{N}{2} \right)} !~
\sigma(\alpha)\left(1-(-)^\alpha\right).
\label{sigmadef}
\end{eqnarray}
Here
\begin{eqnarray}
\sigma(\alpha) &=&\sum_{n_1,\ldots,n_M}
\left(\prod_{i<j}\sin^4\left(\frac{n_i-n_j}{N}\pi\right)\right)\nonumber\\
&\times&\prod_{i=1}^{M}\sin^3\left(\frac{\pi n_i}{N}\right)
\prod_{i=1}^{M}\sin\left(\frac{n_i-\alpha}{N}\pi\right).
\end{eqnarray}
In eq.\ (\ref{sigmadef}) we shifted all the $n_i$ by $\alpha$ in the second
term to bring both terms in the same form. 

Since $\Psi^{(0)}$ is a singlet, we know that if the two spinons in
$\Psi^{(2)}$
were in the $S_z=0$ {\em triplet} state the overlap should be zero, 
i.e.\ 
$\langle\Psi^{(0)}|\Psi^{(2)}(\alpha),{\rm triplet}\rangle \propto
\sigma(\alpha)\left(1+(-)^\alpha\right) = 0$ for all $\alpha$.
Therefore $\sigma(\alpha)$ vanishes for all {\em even} $\alpha$. 
At the same
time we see from eq.\ (\ref{sigmadef}) that
 $\sigma(\alpha)$ is a polynomial
in $\cos\left(\frac{\pi\alpha}{N}\right)$.
This is easily checked by
expanding the $\sin\left(\frac{n_i-\alpha}{N}\pi\right)$ and noting that
$\sigma(\alpha)$ is even in $\alpha$. We conclude
immediately ~\cite{gradryz}:
\begin{eqnarray}
\sigma(\alpha)&\propto &\prod_{j=1}^{\frac{N}{2}}
\left(\cos\left(\frac{\pi\alpha}{N}\right)-\cos\left(\frac{2\pi
j}{N}\right)\right)\nonumber\\
&\propto & \frac{\sin\left(\frac{\pi\alpha}{2}\right)}{
\sin\left(\frac{\pi\alpha}{N}\right)}.
\end{eqnarray}
which holds, even when $\alpha$ is not integer.
This gives us the $\alpha$ dependence in $A$. 
To get the normalization
correctly, let us write out $A$ ($\alpha$ odd):
\begin{eqnarray}
A&=&\frac{1}{\sqrt{2}}\frac{\left(\frac{N}{2}\right)!\cdot
2\cdot\sum_{n_1,\ldots,n_M}\prod_{i<j}
\sin^4\left(\frac{n_i-n_j}{N}\pi\right) 
\prod_i\sin^3\left(\frac{n_i}{N}\pi\right)
\sin\left(\frac{n_i-\alpha}{N}\pi\right)}{\left(\frac{N}{2}\right)!
\cdot
\sum_{n_1,\ldots,n_{M+1}}\prod_{i<j}
\sin^4\left(\frac{n_i-n_j}{N}\pi\right) }\nonumber\\
&=&\frac{2\sum_{n_1,\ldots,n_M}\prod_{i<j}
\sin^4\left(\frac{n_i-n_j}{N}\pi\right) 
\prod_i\sin^3\left(\frac{n_i}{N}\pi\right)
\sin\left(\frac{n_i-\alpha}{N}\pi\right)}{
\sqrt{2} N\sum_{n_1,\ldots,n_{M}}\prod_{i<j}
\sin^4\left(\frac{n_i-n_j}{N}\pi\right)
\prod_i\sin^4\left(\frac{\pi n_i}{N}\right) }.
\end{eqnarray}
In the last step we changed the summation variables $n_i \rightarrow
n_i+n_{M+1};i=1,\ldots,M$ in the denominator and did the trivial 
$n_{M+1}$ summation.
Despite the fact that physically $\alpha$ is actually an integer, we can
formally take the limit $\alpha\rightarrow 0$ in the last expression and
obtain: $A(\alpha=0)=\frac{\sqrt{2}}{N}$. Therefore:
\begin{equation}
\label{adef}
A=\frac{4}{\sqrt{2}N^2}\frac{\sin (\frac{\pi\alpha}{2})}{
\sin(\frac{\pi\alpha}{N})}
\end{equation}

Now we turn our attention to the second piece. 
Only its numerator is $\alpha$ dependent:
\begin{eqnarray}
\lefteqn{ \langle \Psi^{(2)}(\alpha)|\Psi^{(2)}(\alpha)\rangle =}\nonumber\\
&& \left(\frac{N}{2} -1\right)! ~\sum_{\{ n_1,\ldots,n_M\} }
\left(\Psi^{(2)}(n_1,\ldots,n_M|0,\alpha)\right)^2\nonumber\\
&&= 2\left(\frac{N}{2} -1\right)! ~\sum_{n_1,\ldots,n_M} 
\prod_{i<j}\sin^4\left(\frac{n_i-n_j}{N}\pi\right)\times\nonumber\\
&&\hspace{.15truein}\prod_{i=1}^M\sin^2\left(\frac{\pi
n_i}{N}\right)\sin^2\left(\frac{n_i-\alpha}{N}\pi\right).
\end{eqnarray}
But this sum can be recognized as the $\langle S^z(\alpha) S^z(0)\rangle$
static correlation function in the ISE
model~\cite{Hald88}, or---after an (exact)
conversion of the sums to integrals (see chapter \ref{dsf})---as 
the one-particle density matrix in the
Calogero-Sutherland model at half filling~\cite{Suth71,Hald88}.  
The result for large
$N$ 
is proportional to
$\frac{{\rm Si}(\pi\alpha)}{\pi\alpha}$, where
${\rm Si}(x)$ is the sine-integral function $\int_0^x dx'\,
\frac{\sin(x')}{x'} $.  
To normalize $B$ properly we write it
out:
\begin{eqnarray}
B&=& \frac{\left(\frac{N}{2}-1\right)!\cdot
\sum_{n_1,\ldots,n_M}\prod_{i<j}
\sin^4\left(\frac{n_i-n_j}{N}\pi\right) 
\prod_i\sin^2\left(\frac{n_i}{N}\pi\right)
\sin^2\left(\frac{n_i-\alpha}{N}\pi\right)}{
\left(\frac{N}{2}\right)!\cdot
\sum_{n_1,\ldots,n_{M+1}}\prod_{i<j}
\sin^4\left(\frac{n_i-n_j}{N}\pi\right) }\nonumber\\
&=&2\cdot\frac{\sum_{n_1,\ldots,n_M}\prod_{i<j}
\sin^4\left(\frac{n_i-n_j}{N}\pi\right) 
\prod_i\sin^2\left(\frac{n_i}{N}\pi\right)
\sin^2\left(\frac{n_i-\alpha}{N}\pi\right)}{
N^2\sum_{n_1,\ldots,n_{M}}\prod_{i<j}
\sin^4\left(\frac{n_i-n_j}{N}\pi\right)
\prod_i\sin^4\left(\frac{\pi n_i}{N}\right) }.
\end{eqnarray}
In the last step we shifted the summation variables, as we did previously
for $A$,
 to simplify the denominator.
We now recognize that for $\alpha=0$, $B=2/N^2$. 
As for $A$, the normalization can be obtained from the
case $\alpha = 0$ and we find:
\begin{equation}
B=\frac{2}{N^2}
\frac{{\rm Si}(\pi\alpha )}{\pi\alpha},
\end{equation}
The entire expression for the overlap, with normalization, then becomes:
\begin{equation}
\frac{2}{N}\frac{\sin\left(\frac{\pi\alpha}{2}\right)}{ 
\sin(\frac{\pi\alpha}{N})}
\sqrt{\frac{\pi\alpha}{{\rm Si}(\pi\alpha)}}.
\end{equation}

\subsection{Appendix B: Lower Bound on the XY ODLRO}

We want to show that $\langle 0,N|\Del(j)|0,N-2\rangle$ is finite when
$N\rightarrow\infty$.
We begin, following Anderson's treatment of
the orthogonality catastrophe \cite{pwaorthog},
 by writing the overlap as the $\frac{N}{2}\times\frac{N}{2}$ 
determinant of 
of the matrix formed from the inner products of the
single particle wavefunctions comprising the two states:
\begin{equation}
\label{eq:mat_def}
{\rm overlap} = |M_{i,j}| = \left|\langle \Psi_i^{(1)} | \Psi^{(2)}_j \rangle
\right|
\end{equation}
where, for $\frac{N}{2}$ an odd integer
and periodic boundary conditions for the
$N$ site case:
\begin{equation}
\label{eq:psi}
| \Psi_k^{(a)} \rangle = \sum_{n=1}^{N+2(1-a)}
\frac{(z_{k,a})^n}{\sqrt{N+2(1-a)}}   | n \rangle
\end{equation}
except that
\begin{equation}
\label{eq:psi2}
| \Psi_{\frac{N}{2}}^{(2)} \rangle =  
\frac{1}{\sqrt{2}}(| N-1 \rangle - | N \rangle)
\end{equation}
and 
\begin{eqnarray}
\langle \Psi_{i \neq \frac{N}{2}}^{(2)} | N-1 \rangle &  = & 0 \nonumber
\\
\langle \Psi_{i \neq \frac{N}{2}}^{(2)} | N \rangle &  = & 0 
\end{eqnarray}
The $z$'s are given by:
\begin{eqnarray}
z_{k,1} & = & \exp(\frac{2\pi i}{N}(k+\frac{N}{4}-\frac{1}{2}))
\nonumber\\
z_{k,2} & = & \exp(\frac{2\pi i}{N-2}(k+\frac{N}{4}-1))
\label{eq:kdef}
\end{eqnarray}
The entries in the matrix are then given 
for $j \neq \frac{N}{2}$ by:
\begin{equation}
M_{i,j} = z_{j,2}\frac{1 +  z_{i,1}^2}
{\sqrt{N(N-2)}(z_{i,1}-z_{j,2})} 
\end{equation}
and by:
\begin{equation}
M_{i,\frac{N}{2}} = \frac{z_{i,1} - 1}
{\sqrt{2N}}\cdot (z_{i,1})^{-N}.
\end{equation}
The cofactors associated to the elements
$M_{i,\frac{N}{2}}$ are determinants of matrices of the form
$X_{i,j} = \frac{f_i g_j}{A_i -B_j}$, where
$f_i$ and $A_i$ depend only on $i$,
while $g_j$  and $B_j$ depend only on $j$. Such determinants can 
be rewritten using:
\begin{equation}
|X| = (\prod_{i} f_i\prod_j g_j)
 \frac{\prod_{i < j}(A_j - A_i)(B_i-B_j)}
{\prod_{i,j} (A_i - B_j)}
\end{equation}
Neglecting phases multiplying it, 
this leads to a rewriting of the whole determinant as:
\begin{eqnarray}
|M|& = & \frac{\prod_i (1 + z_{i,1}^2)}{\sqrt{2N} [N(N-2)]^{\frac{N-2}{4}}}
\frac{[\prod_{i < j}(z_{i,1}-z_{j,1})][\prod_{i < j}(z_{i,2}-z_{j,2})]}
{\prod_{i,j}(z_{i,1}-z_{j,2})} \nonumber\\
&&\times\sum_m  \frac{z_{m,1}-1}{z_{m,1}^2 +1}
\frac{\prod_{i} (z_{m,1} - z_{i,2})}{\prod_{i (\neq m)} (z_{m,1} - z_{i,1})}
\end{eqnarray}

We are unable to compute this exactly, but we have shown by
careful consideration of the infinite
products that to leading order in $\frac{1}{N}$ the determinant is
bounded below by $ e^{-\frac{5}{2}}$ 
and must therefore remain
finite in the limit of $N \rightarrow \infty$. 
To illustrate, we examine a single term in the
sum over $m$, $T_m$, defined by:
\begin{equation}
\label{eq:prod_def}
T_m = \frac{z_{m,1}-1}{z_{m,1}^2 +1}
\frac{ \prod_{i = 1}^{\frac{N-2}{2}} (z_{m,1} - z_{i,2})}
{\prod_{i = 1,i \neq m}^{\frac{N}{2}} (z_{m,1} - z_{i,1})}
\end{equation}
using the fact that the $z_{i,1}$ are roots of unity
and $z_{i,2}$ of (-1) this can be rewritten as:
\begin{equation}
T_m = \frac{z_{m,1}- 1}{N z_{m,1}} 
\frac{\prod_{i} (z_{m,1} + z_{i,1})}
{\prod_{j} (z_{m,1} + z_{j,2})}
\end{equation}
Removing the term in the numerator given by
$z_{m,1} -1$ and relabeling the remaining $z_{i,1}$ so that
they are in one to one correspondence with the $z_{j,2}$ and
$j$ runs from $1$ to $\frac{N-2}{2}$ we get
\begin{equation}
T_m = \frac{(z_{m,1}-1)^2}{N z_{m,1}}
\prod_j \frac{z_{m,1} + z_{j,1}}
{z_{m,1} + z_{j,2}}
\end{equation}
Since both sets of $z$'s come in complex conjugate pairs or are
real, for every $m$ there is an n such that $T_m = T_n^{\star}$.
Therefore, only the real part of $T_m$ matters and, since the
product can be shown to be real, we may replace
$\frac{(z_{m,1} - 1)^2}{z_{m,1}}$ by  $2$. The prefactor of the
product is then of uniform sign and the sum can be bounded by
replacing the product with a lower bound.
The smallest individual term in the product can be calculated and
is given by $1 - \frac{2}{N} + O(\frac{1}{N^{2}})$ 
so that the product is bounded
below by $\frac{1}{e}$ as is $T_m$.  The prefactor of the sum over $m$
is also real and bounded below by $e^{-\frac{3}{2}}$ by similar arguments
so that the whole determinant is real and bounded below by
$e^{-\frac{5}{2}}$.


\comment{
%
%
\begin{figure}
\caption{ Calculated Overlaps as Functions of System Size}
Shown are the calculated overlaps between the $N$ spin groundstates of
the $1/r^2$, nearest neighbor Heisenberg and XY models
and the states obtained by inserting a nearest
neighbor singlet pair of spins into the $N-2$ spin groundstates of those
models.
 \label{fig:size}
 \end{figure}

 \begin{figure}
 \caption{Decay of the Overlap with Separation}
The overlaps between the $N$ spin groundstates
and the states obtained from the $N-2$ spin groundstates
by inserting two spins in a singlet configuration
separateed from each other by $\frac{N-2}{2}$.
The XY results are from exact numerical results,
while those of the $ISE$ model are  Monte Carlo.
 \label{fig:sep}
 \end{figure}

\part*{}
\renewcommand{\thechapter}{}
\def\chapvar{}
\chapter{Conclusion}
\label{conclusion}
\markright{Conclusion}

We have seen in chapters \ref{chapter1} and \ref{invar} that it is possible to
show full integrability of the ISE-model, both in the sense of the
existence of a transfer matrix that satisfies the Yang-Baxter fundamental
commutation relations (just as for all other integrable spin models) and of the
presence of a set of constants of the motion, constructed by taking the static
limit of a dynamical deformation of the spin model. We found that to
every eigenvalue of the ISE-Hamiltonian belongs a set of degenerate 
wavefunctions, 
that derive from acting on a polynomial-type Yangian Highest Weight
State with the Yangian algebra. The Yangian transfer
matrix representations told us unambiguously the spin content of this Yangian
representation, that is, the state counting rules of the model's
quasiparticles: the spinons. Those counting rules are non-trivial, leading
to the fractional statistics nature of those particles, a phenomenon
presaged  by the fact that for two Yangian representations $\T_1$ and $\T_2$,
\mbox{$\T_1\!\otimes\! \T_2\, \mbox{$\simeq \hspace{-0.14truein} /$}\,
\T_2\!\otimes\T_1$}, in general (lack of cocomutativity). 

As far as the integrals of motion are concerned, we saw that such a set,
that contains the Hamiltonian $H_{\rm ISE}$ could indeed be constructed, but
this required a  somewhat unnatural deviation through the dynamical
deformation of the static spin model. A more direct approach should be
possible once we can solve the model
involving the introduction of a second quantum parameter, deforming it in
the direction of the  {\em hyperbolic} models, i.e.\ introducing an exchange
integral in $H_{\rm ISE}$ 
which is an elliptic Weierstrass function $P(x|\omega_1,\omega_2)$
\cite{H94,I90}. Since the NNE-model is obtained in the limit $\omega_1=iN;\;
\omega_2\rightarrow 0$, and the ISE-model when $\omega_1=N;\;
\omega_2\rightarrow \infty$, we could interpolate
between the traditional Nearest
Neighbor constants of motion to the more esoteric ISE ones. 

In chapter \ref{dsf} we stated the remarkably simple, empirically found, 
selection rules for excitations that
can be generated by action of a local spin operator. The rule essentially
states that the only Yangian multiplets that are connected via $S^a_i,\;
a=1,2,3$ are essentially those whose motifs differ little: 
the number of their Drinfel'd zeroes in
any range $[a,b],\; 1\leq a < b <N $ differs by less than two.
This rule has only been proved in the special case of YHWS in both ket and
bra of the $\svec_i$ matrix element, via Jack polynomial Clebsch-Gordan
coefficients \cite{Ha94}. The question of the extension  to the case of non-YHWS matrix 
elements is still open.

As a corollary to the Yangian selection rule, 
$S^a_i$ acting on the groundstate in a given
magnetic field, changes the number of spinons by no more than two, and can
add up to two magnons (which move in opposite directions).  
In low fields, the dominant contribution to the spectral function comes
from the
addition of two spinons, in high fields, from the addition of one magnon.
These selection rules
predict the regions in the energy-momentum diagram that can be accessed in a
neutron scattering experiment. The actual intensity of the inelastic
scattering cannot be determined for all possible excitations. Most notably,
(less polarized) excitations further down in a Yangian multiplet are not
available.  One would have to know what the permutation algebra of Yangian
lowering operators with local spin operators would be, to be able to compute a
typical matrix element  like $\langle\Gamma|T^+(\lambda_0)\svec
|0\rangle$. If one believes that
contributions from states 
further down in a Yangian multiplet are related to those
involving their parent YHWS $|\Gamma\rangle$---as is the case for
excitations of types $(i),\; (iv)$ and $(v)^b$ in table \ref{sumtable}---it
seems reasonable to insert diagonal elements 
of the transfer matrix between $\svec_i$ and $T^+$ in \mbox{$\langle \Gamma
|T^+(\lambda_0)\rule{0em}{1ex}^{T^{aa}(u)}_{\rule{.8em}{0em}\wedge} 
\svec_i|0\rangle$}.
Since $|0\rangle$ is known analytically, one could choose $T^{aa}$
judiciously, such that its action on $\svec_i|0\rangle$ is manageable.
Commuting it to the left through $T^+(\lambda_0)$ will fix the matrix element.

Chapter \ref{chapter4} presented evidence for the surprising existence of a
new from of ODLRO (off diagonal long range order) in the groundstate of one
dimensional antiferromagnetic spin chains, that consists of inserting a
singlet into the groundstate, and extracting it somewhere else. The ODLRO is
present in short ranged XY and NNE and long ranged ISE spin models, with
virtually identical values for the order parameter, which alternates
in sign depending on where in the chain it is applied. 
This raises the question which
models would not have this ODLRO. The ODLRO could be seen to exist, 
at least in the
ISE case, by the fact that it is related to the spinon propagator, which can
be shown analytically to have a finite piece in the thermodynamic limit.
The spinon propagator bears resemblance to the Read order parameter in the
$\nu=\half$  Hall effect, which is known to have true long range order. The
singlet insertion operation was applied to coupled
one dimensional Hubbard chains, with large on-site repulsion. The
corresponding operator makes a singlet pair of electrons hop from one chain to
the other. That operator does not have LRO, in fact it is not even relevant
(its correlator decays too fast), because the spin model order parameter
breaks the $Z_2$ symmetry of
translating by a lattice site.
Relevancy could be restored if the chains are properly correlated,
such that the symmetry breaking alternation term  in both
chains cancels.

The big question here is if, and how this can be extended to two dimensions. 
Numerically,
things get difficult very quickly as a function of the system size, and
there is no two dimensional ISE model that we may borrow intuition from. The
issue of how to insert the singlet in a natural way into a 2D lattice is
unclear as well. One could also consider generalizations to models with
higher spin and $SU(n)$ spins.

\end{document}